\documentclass[a4paper, traditabstract]{aa} 

\usepackage{graphicx}
\usepackage{txfonts}
\usepackage[dvipsnames,svgnames,x11names]{xcolor}

\usepackage{microtype}
\usepackage{url}
\usepackage{booktabs}
\usepackage{commath}
\usepackage[bookmarks, colorlinks, breaklinks, ]{hyperref}  
\hypersetup{linkcolor=MidnightBlue,citecolor=MidnightBlue,filecolor=black,urlcolor=MidnightBlue} 
\usepackage{amsmath,amsfonts,amssymb}
\usepackage{fixltx2e}
\usepackage{caption,subcaption}
\usepackage{natbib}
\usepackage{tabularx}
\usepackage[export]{adjustbox}
\usepackage{mathbbol}
\usepackage{placeins}

\bibpunct{(}{)}{;}{a}{}{,}

\newcommand{\Planck}{{\em Planck}}
\newcommand{\XMM}{{\em XMM-Newton}}
\newcommand{\Chandra}{{\em Chandra}}
\newfont{\gwpfont}{cmssq8 scaled 1000}
\newcommand{\rexcess}{{\gwpfont REXCESS}}

\newcommand{\Rfive}{R_{\rm 500}}
\newcommand{\Mfive}{M_{500}}
\newcommand{\Pfive}{P_{e,500}}
\newcommand{\thetaf}{\theta_{\rm 500}}
\def\msol {{\rm M_{\odot}}}
\def\keV {\rm keV}

\newcommand{\thetas}{\theta_{\rm s}}
\newcommand{\yo}{y_{\rm o}}     
\newcommand{\fnu}{f_\nu}                
\newcommand{\Tth}{T_{\thetas}}
\newcommand{\tth}{t_{\thetas}}

\newcommand{\Om}{\Omega_\textrm{m}}
\newcommand{\Ol}{\Omega_\Lambda}

\begin{document}

   \title{Joint measurement of the galaxy cluster pressure profile\\ with \Planck\ and SPT-SZ}
   \author{J.-B. Melin
          \inst{1}
          \and G. W. Pratt
           \inst{2}
                  }

   \institute{Université Paris-Saclay, CEA, Département de Physique des Particules, 91191, Gif-sur-Yvette, France\\
                 \email{jean-baptiste.melin@cea.fr}
                  \and Université Paris-Saclay, Université Paris Cité, CEA, CNRS, AIM, 91191, Gif-sur-Yvette, France
             }

   \date{Received ...; accepted ...}

% \abstract{}{}{}{}{} 
% 5 {} token are mandatory
 
  \abstract
  {We measured the average Compton profile of 461 clusters detected jointly by the South Pole Telescope (SPT) and \Planck. The number of clusters included in this analysis is about one order of magnitude larger than in previous analyses. We propose an innovative method developed in Fourier space to combine optimally the \Planck\ and SPT-SZ data, allowing us to perform a clean deconvolution of the point spread and transfer functions while simultaneously rescaling by the characteristic radial scale $R_{\rm 500}$ with respect to the critical density. The method additionally corrects for the selection bias of SPT clusters in the SPT-SZ data. We undertake a generalised Navarro-Frenk-White (gNFW) fit to the profile with only one parameter fixed, allowing us to constrain the other four parameters with excellent precision.
The best-fitting profile is in good agreement with the universal pressure profile based on \rexcess\ in the inner region and with the Planck intermediate paper V profile based on \Planck\ and the \XMM\ archive in the outer region. We investigate trends with redshift and mass, finding no indication of redshift evolution but detecting a significant difference in the pressure profile of the low- versus high-mass subsamples, in the sense that the low mass subsample has a profile that is more centrally peaked than that of the high mass subsample.
We also scaled the average Compton profile by the mean Universe density ($R_{\rm 200m}$) and provide the best-fitting gNFW profile. Using the profiles scaled by both the critical ($R_{\rm 500}$) and the mean Universe density ($R_{\rm 200m}$), we studied the outskirt regions by reconstructing the average Compton parameter profile in real space. These profiles show multiple pressure drops at $\theta>2\theta_{\rm 500}$, but these cannot clearly be identified with the accretion shocks predicted by hydrodynamical simulations. This is most probably due to our having reached the noise floor in the outer parts of the average profile with the current data sets.}


   \maketitle
%
%-------------------------------------------------------------------

\section{Introduction}

The Sunyaev-Zeldovich \citep[SZ,][]{sz1970,sz1972} effect is the inverse Compton scattering of cosmic microwave background (CMB) photons by hot electrons in the Universe. It allows the electronic pressure of the hot gas in the cosmic web to be probed, as seen for example in the all-sky map obtained by \Planck\ ~\citep{planck_szmap2016}. The SZ effect depends on the Compton parameter

\begin{equation}
\label{eq:basicsz}
y={\sigma_{\rm T} \over m_{\rm e} c^2} \int_{los} P_e dl
\end{equation}
where $\sigma_{\rm T}$ is the Thomson cross-section, $m_{\rm e}$ is the electron rest mass, $c$ is the speed of light and $P_e=n_e \times T_e$ is the electronic pressure (product of electronic density and temperature). The integral is performed along the line-of-sight ($los$).

The cosmic web is composed of filaments, sheets, and nodes. Clusters of galaxies are located at the nodes. They constitute the densest regions of the cosmic web, dominated by gravitational processes. As a consequence, they do not show a preferred spatial scale and are expected to be self similar: at fixed cosmic time, any cluster can be matched to another according to a single parameter (the mass). This property allows for statistical analyses of clusters after rescaling, which we take advantage of in this work. In this article we use as our main parameter the cluster mass $\Mfive$, corresponding to the mass enclosed in a sphere of radius $\Rfive$, whose mean density is 500 times the critical density of the Universe at the cluster redshift $z$. We note $\thetaf=\Rfive/D_{\rm ang}(z)$ the angular radius of the cluster ($D_{\rm ang}(z)$ being the angular distance of the cluster). We also use $M_{\rm 200m}$, $R_{\rm 200m}$\, and $\theta_{\rm 200m}$, the equivalent quantities defined with respect to 200 times the mean density of the Universe.

Galaxy clusters are filled with a hot, gaseous,  intracluster medium (ICM), the radial pressure profile of which decreases from the centre (radial coordinate $r=0$) to the outskirts ($r \geqslant \Rfive$, $\Rfive$). The SZ effect is ideal to probe the pressure profile $P_e(r)$ of the ICM in galaxy clusters, and hence constrain models of structure formation for different cosmological scenarios.  
First measurements of pressure profiles based on X-ray data were performed by \cite{nagai2007} and \cite{arnaud2010} using \Chandra\ and \XMM\ respectively on relatively small cluster samples (a few to a few tens of objects). \cite{nagai2007} fitted their data with a generalised Navarro-Frenk-White (gNFW) profile \cite[Eq.~\ref{eq:gnfw500} hereafter, see also][]{zhao1996} which has been widely adopted by the cluster community since then.
The goal of these studies was to support the imminent first SZ observations by providing a generic profile to improve the efficiency of cluster detection algorithms, to help in the analysis and interpretation of SZ data, and to improve in the theoretical modelling of the ICM \citep[Appendix A of][]{nagai2007}. The \cite{nagai2007} and \cite{arnaud2010} studies were extended to the low mass scale by \cite{sun2011} using galaxy groups and \Chandra\ data.
These three data sets are compatible. In the following, we use the pressure profile from~\cite{arnaud2010} based on the \rexcess\ sample~\citep{boehringer2007} to compare with our results. We refer to this as the universal pressure profile (UPP). We also use the profiles extracted from the \rexcess\ Cool-core and \rexcess\ Morphologically disturbed subsamples.

First SZ profiles were observed with the Sunyaev-Zel’dovich Array (\citealp[][based on three clusters]{mroczkowski2009}; \citealp[][25 clusters]{bonamente2012}), the South Pole Telescope~\citep[][15 clusters]{plagge2010}, and the Atacama Cosmology Telescope~\citep[][nine clusters]{sehgal2011}. \cite{pipv} studied the average SZ profile of 62 clusters detected by \Planck\ which were part of the \XMM\ archive, and \cite{sayers2013} the average profile of 45 X-ray selected clusters observed with Bolocam. These six studies all found profiles in good agreement with the UPP, but the latter two suggested some possible deviations at small ($r<0.1\Rfive$) and large ($r>\Rfive$) radii. In particular, the average profiles from \cite{pipv} (hereafter PIPV) and \cite{sayers2013} showed higher pressure than the UPP at $r>\Rfive$.

The advent of high resolution SZ observations combined with \Planck\ and/or \XMM\ has allowed for further progress. \cite{ruppin2017} studied the SZ profile of PSZ1~G045.85+57.71 with NIKA and \Planck, and \cite{ruppin2018}  studied PSZ2~G144.83+25.11 with NIKA2 and \Planck, demonstrating the high efficiency of the NIKA and NIKA2 Kinetic Inductance Detector   experiments to map profiles at high ($\sim$12~arcsec) resolution.  Recently, attention has focused on the synergy in spatial scales between different experiments for the measurement of SZ profiles. \cite{romero2017} combined data from MUSTANG and Bolocam for 14 cluster profiles, and \cite{romero2018} combined the data from MUSTANG, NIKA, Bolocam, and \Planck\ to study the profile of CLJ~1226.9+3332 at redshift $z=0.89$. \cite{oppizzi2022} performed a measurement of the pressure profiles of six clusters observed with \Planck\ and SPT-SZ. Similarly \cite{pointecouteau2021} used the \Planck\ and ACT data jointly to measure the average pressure profile of 31 clusters, demonstrating that space-based and ground-based data can be efficiently combined to improve constraints on the cluster pressure profile. Finally, \cite{sayers2023} combined Bolocam, \Planck\ and \Chandra\ data to study the redshift evolution and mass dependence of the average pressure profile of 40 clusters. 

The quest for high resolution individual SZ profile observations at high redshift is now ongoing.  Using NIKA2, \cite{keruzore2020} presented the profile of ACT-CL~J0215.4+0030 at $z\sim0.9$, and \cite{ruppin2020} combined NIKA2 and \Chandra\ to study MOO~J1142+1527 at $z=1.2$. These high redshift SZ profiles are in agreement with the UPP, indicating probable self similar evolution of the average cluster pressure profile with time. However, the studies with MUSTANG-2 and \Chandra\ of IDCS~J1426.5+3508 at $z=1.75$ \citep{andreon2021}, and  JKCS~041 at $z=1.8$ with MUSTANG-2 \citep{andreon2023} have indicated disagreement between the pressure profiles of these high $z$ clusters and the UPP.

In parallel to this observational effort, impressive progress has been made to understand the impact of astrophysical processes on the pressure profile of galaxy clusters using numerical  simulations~\citep{molnar2009,shaw2010,battaglia2012,mccarthy2014}.
\cite{he2021} recently combined simulations with \XMM\ and \Chandra\ data to better understand the hydrostatic X-ray mass bias and its impact on the average pressure profile.
Other recent simulation studies focused on cluster outskirts~\citep{aung2021,baxter2021,zhang2021}. These have showed that a pressure shock is expected at $2<r<2.5 R_{\rm 200m}$ in the average pressure profile of relaxed clusters.

The observation of cluster outskirts at large radius in SZ is of growing interest~\citep{ghirardini2018,hurier2019}. \cite{pratt2021} stacked the \Planck\ y-maps of ten nearby galaxy groups and studied the resulting average profile, finding a $~3\sigma$ feature at $~2R_{500}$ that they attributed to a signature of internal accretion shocks. Recently, \cite{anbajagane2022} studied the stacked profile of the SPT-SZ clusters in the combined SPT-SZ and \Planck\ y-maps \citep{bleem2022} and found a pressure deficit at $1.1R_{\rm 200m}$ ($3.1\sigma$) and a decrease in pressure at $4.6R_{\rm 200m}$ ($2.0\sigma$). They attributed these features to a shock-induced thermal non-equilibrium between electrons and ions for the former, and accretion shocks for the latter.

In this work, we average the SZ profiles of 461 South Pole Telescope (SPT) clusters in the \Planck\ and SPT-SZ data sets. We compare the average profile extracted from \Planck\ only, SPT-SZ only, and from the two data sets jointly, and provide the best-fitting gNFW model fit. We then compare our average profile to previous work, study cluster cores, and investigate possible trends with mass and redshift. We then perform the averaging of the SZ profiles in rescaling with $R_{\rm 200m}$, in a first attempt to obtain constraints on the profile in the cluster outskirts. In Section~\ref{sec:datasets}, we present the data sets we use for our study. We detail our analysis method in Section~\ref{sec:pp_harmo}. We present our results in Section~\ref{sec:results} and discuss them in Section~\ref{sec:discussion}. In Section~\ref{sec:sum}, we summarise our results and conclude. Throughout the article we assume a flat $\Lambda$CDM cosmology with $H_0=70$~km$^{-1}$s$^{-1}$Mpc$^{-1}$ and $\Om=1-\Ol=0.3$.

\section{Data sets}
\label{sec:datasets}

We present the cluster sample used in the analysis in Section~\ref{sec:clusamp} and the data sets in Section~\ref{sec:plck} and~\ref{sec:sptsz}. The \Planck\ and SPT-SZ data sets along with their formatting are the same as in~\cite{melin2021}. We provide the main characteristics below, but encourage the reader to consult the above article for more detailed information.

\subsection{Cluster sample}
\label{sec:clusamp}

We used the fiducial SPT-SZ 2500d cluster catalogue published by the SPT collaboration~\citep{bleem2015}. It contains 677 candidates, out of which 620 are outside the SPT-SZ boundary mask provided with the public data\footnote{\url{https://lambda.gsfc.nasa.gov/product/spt/spt_prod_table.cfm}}. We selected among them the 477 detections with redshift. We discarded five objects because they are on the boundary of the SPT-SZ footprint so we cannot extract the background statistics: less than 50\% of the pixels are observed in the tangental maps centered on these clusters. Finally, we removed eleven clusters with $Y^{0.75}\, {\rm SPT} \geqslant 2 \times 10^{-4} \, {\rm arcmin}^2$, for which we showed that the filter transfer function (FTF) is not accurate~\citep[Fig. B.3. of][]{melin2021}. Our final sample contains 461 clusters. $\Mfive$ is provided with the catalogue. Throughout this work, we renormalise the values of $\Mfive$ by a factor 0.8 to make them compatible with the \cite{arnaud2010} mass definition, as was done in~\cite{tarrio2019} and~\cite{melin2021}. All the quantities derived from the mass (e.g. $\Rfive$) are based on these rescaled masses. We only used the original SPT masses at four specific places in the article: in Section~\ref{sec:ilc} for the temperature-mass relation; in Section~\ref{sec:malmquist} for the ‘unbiased SPT-significance’-mass relation;  in Table~\ref{tab:bestfits500} for naming the low-mass/high-mass subsamples; and in  Section~\ref{sec:th200m} for the scaling with $\theta_{\rm 200m}$. These choices are explained in each respective Section.

\subsection{\Planck}
\label{sec:plck}

We use the \Planck\ full mission maps from the High Frequency Instrument. The instrument observed the all sky in six frequency bands: 100, 143, 217, 353, 545, and 857~GHz. We assume that the beams are Gaussian with full width at half maximum (FWHM) of 9.659, 7.220, 4.900, 4.916, 4.675, 4.216~arcmin respectively as done in~\cite{PSZ2}. The maps are provided in galactic coordinates at HEALPix resolution $N_{side}=2048$ (1\farcm72~pixel$^{-1}$) in the \Planck\ Legacy Archive\footnote{\url{https://pla.esac.esa.int}} with their associated products, including the frequency response of each band. We upgraded the maps to  $N_{side}=8192$ (0\farcm43~pixel$^{-1}$) by zero padding in harmonic space, that is, by adding new modes at high multipole with zero power. We then rotated the coordinate system from Galactic to equatorial, to match the SPT-SZ data format.
 
\subsection{SPT-SZ}
\label{sec:sptsz}

We work with the "SPT Only Data Maps". SPT-SZ observed in three frequency bands: 90, 150 and 220~GHz. The maps are all provided at Gaussian resolution with FWHM of 1.75~arcmin in HEALPix format at resolution $N_{side}=8192$ (0\farcm43~pixel$^{-1}$) and in equatorial coordinates. The FTF are provided in the public archive with the maps. We retrieved the frequency responses of the instrument from Fig.~10 of~\cite{chown2018} using the WebPlotDigitizer\footnote{\url{https://automeris.io/WebPlotDigitizer/}}.

\subsection{Consistency between \Planck\ and SPT-SZ} 

\citet{melin2021} performed extensive consistency tests between \Planck\ and SPT-SZ; we summarise here the main results. Fluxes of SPT-SZ clusters extracted from \Planck maps are fully compatible with published SPT fluxes. This result is discussed in Sect.~5.1 and illustrated in Fig.~1 of~\cite{melin2021}. We note that Fig.~1 of~\cite{melin2021} shows that the SPT-SZ fluxes are subject to Malmquist bias, as expected. It is therefore important to correct for this bias when performing a joint \Planck\ and SPT-SZ analysis of SPT-SZ selected clusters as we do in Sect.~\ref{sec:malmquist}.

The signal-to-noise ratio of SPT-SZ clusters extracted from public SPT-SZ maps is 10\% lower than the signal-to-noise ratio published by the SPT collaboration. This is expected because the public SPT-SZ maps are shallower than the private SPT-SZ maps. The cluster sizes extracted from the public SPT-SZ maps are in good agreement with the cluster sizes published by the SPT collaboration. Finally, cluster fluxes extracted from SPT-SZ public maps are also fully compatible with the flux published by the SPT-SZ collaboration except for clusters with $Y^{0.75}\, {\rm SPT} \geqslant 2 \times 10^{-4} \, {\rm arcmin}^2$, due to inaccuracy of the FTF. This is the reason for having excluded eleven clusters from our sample in Sect.~\ref{sec:clusamp}. All of these consistency tests are presented in Appendix~B. of~\cite{melin2021}.

\section{Pressure profile in harmonic space}
\label{sec:pp_harmo}

We give important conventions for the cluster profile rescaling and fitting in Section~\ref{sec:conventions}. We discuss the advantages of working in harmonic space in Section~\ref{sec:av_harmo}. We then introduce the internal linear combination to extract SZ maps in Section~\ref{sec:ilc}. In Section~\ref{sec:malmquist}, we detail the Malmquist bias correction of the SPT-SZ signal. We describe in Section~\ref{sec:renorm} the averaging procedure of the SZ maps. Then we present the likelihood used to perform the gNFW fits in Section~\ref{sec:gnfwfits}.

\subsection{Conventions}
\label{sec:conventions}

Following conventions in~\cite{arnaud2010}, we defined the cluster scaled pressure profile as

\begin{equation}
\label{eq:px}
p(x)={P_e(r) \over \Pfive}
\end{equation}
with the scaled radius
\begin{equation}
\label{eq:x}
x={r \over \Rfive}
\end{equation}
and the characteristic electronic pressure
\begin{equation}
\label{eq:p500e}
\Pfive = {\mu \over \mu_e} P_{g,500}.
\end{equation}
Here $\mu=0.6$ is the mean molecular weight and $\mu_e={5 \over 1+{2/\mu}} \approx 1.15$ the mean molecular weight per free electron, and
\begin{equation}
\label{eq:p500g}
P_{g,500}=  1.65 \times10^{-3} \, E(z)^{8/3} \, \left[ {\Mfive \over {3\times10^{14}\,{\rm h^{-1}_{70}}\,\msol}}\right]^{2/3}~~{\rm h_{70}^{2} \,\keV \, cm^{-3}}
\end{equation}
is the characteristic gas pressure (\citealp[Eq. 3 of][]{nagai2007}; \citealp[Eq. 5 of][]{arnaud2010}), with $E(z)$ the Hubble parameter at redshift $z$ normalised to its present value and $h_{70}$ the Hubble constant normalised to $70 \, {\rm km/s/Mpc}$.
\cite{arnaud2010} noticed that $p(x)$ depends on $\Mfive$ and proposed to define (their Eq.~8)
\begin{equation}
\label{eq:bpx}
p(x,\Mfive)=\mathbb{p}(x) \,\left[\frac{\Mfive}{3 \times10^{14}\,{\rm h^{-1}_{70}}\,\msol}\right]^{0.12}, 
\end{equation}
where $\mathbb{p}(x)$ is the average scaled profile (gNFW) that we constrain in this work:
\begin{equation}
\label{eq:gnfw500}
\mathbb{p}(x) = \frac{P_{0} } { (c_{500}x)^{\gamma}\left[1+(c_{500}x)^\alpha\right]^{(\beta-\gamma)/\alpha} }.
\end{equation}
Above, $P_0$ is the normalisation (independent of the mass), $c_{500}=\Rfive/r_s$ is the concentration parameter, $r_s$ is the scale radius, and $\alpha$, $\beta$ and $\gamma$ are the intermediate ($r \sim r_s$), outer ($r >> r_s$) and inner ($r<<r_s$) slopes of the profile, respectively. The average scaled profile $\mathbb{p}(x)$ thus has five parameters: $\Theta=[P_0,c_{500},\gamma,\alpha,\beta]$.

We can now define the characteristic Compton parameter
\begin{equation}
\label{eq:def_y500}
y_{500} = {\sigma_{\rm T} \over m_{\rm e} c^2}  \times (2 R_{500})  \times \Pfive \times \left[\frac{\Mfive}{3 \times10^{14}\,{\rm h^{-1}_{70}}\,\msol}\right]^{0.12}.
\end{equation}
In Sect.~\ref{sec:renorm}, we renormalise the Compton parameter map of each cluster by the above quantity.

\subsection{Advantages of working in harmonic space}
\label{sec:av_harmo}

The observed clusters were convolved by the instrumental point spread functions (PSF) of the experiments, and also by the FTF for SPT-SZ. The averaging process requires a radial rescaling of  the profiles by $\thetaf$, with $\thetaf$ ranging from 0.88 to 16.5~arcmin. The effect of the PSF and FTF will therefore be different on the rescaled pressure profile from one cluster to another.
To undertake the cleanest possible PSF and FTF deconvolution across the large range of cluster sizes, we performed our analysis in Fourier space. The PSF and FTF convolutions in real space translate into multiplications in Fourier space, making the deconvolutions easier. We adopted the flat sky approximation which allows us to associate the amplitude of a Fourier mode $\lVert \vec{k} \rVert$ to the multipole $l$ in spherical harmonics. The conversion from $\lVert \vec{k} \rVert$ to $l$ has the advantage to present results independent of the specific characteristics of the tangential maps.
As a consequence, the measurements are presented as $y_l/y_{500}/\thetaf^2$ versus $l\thetaf$ (harmonic space, $l$ being the multipole) instead of $y(\theta)/y_{500}$ versus $\theta/\thetaf$ (real space, $\theta$ being the angular radius). Although the measurement is actually performed in Fourier space and then converted to harmonic space, we will use the term `harmonic space' for the remainder of the article.

Working in harmonic space has another major advantage. To a first approximation, the instrumental and astrophysical noise in harmonic space can be considered as independent from one multipole to another. This is rigorously the case for instrumental white noise and unlensed primary CMB anisotropies. However, real data can contain some correlated instrumental noise in Fourier space, and also some lensing of the primary anisotropies. In the following, we therefore assume that the measurements in harmonic space are correlated from one multipole to another, and we use a covariance matrix $\mathcal{S}$, measured on the data with bootstrap resampling, to take this correlation into account. In particular, we used this matrix to write the likelihood, detailed in Section~\ref{sec:gnfwfits}. Full details on the calculation of $\mathcal{S}$ are given in  Appendix~\ref{app:harmonoise}.

The method to extract the average cluster pressure profile is now presented in detail in the following Sections. We first extracted an SZ map for each cluster using an internal linear combination (Section~\ref{sec:ilc}). An important aspect, detailed in Section~\ref{sec:malmquist}, is to correct for the Malmquist bias of the SPT clusters in the SPT-SZ maps. We then renormalised, rescaled, and  azimuthally averaged each map, and finally bin averaged them (Section~\ref{sec:renorm}).

\subsection{Internal linear combination}
\label{sec:ilc}

We followed closely the method developed in~\cite{melin2021} for extracting clusters from \Planck\ and SPT-SZ data jointly, with two main differences. First, we did not use a matched multi-filter (which assumes a cluster profile) but instead applied an internal linear combination (ILC) to reconstruct a Compton parameter y-map for each cluster. Secondly, we did not apply the method blindly, but centred the extraction at the location of the SPT-SZ clusters. We recall briefly below how we prepare the data and then provide details on the ILC.

We first cut $10 \deg \times 10 \deg$ tangential maps centred on the 461 clusters of our sample for each \Planck\ and SPT-SZ map, yielding nine frequency maps in total for each cluster with pixel size of about $0.43\, {\rm arcmin}$. As in~\cite{melin2021}, we started by detecting and masking bright point sources ($S/N>8$) with a single frequency matched filter, applied to individual channel maps. Map areas corresponding to masked point sources were set to zero. This step is important to avoid contamination of the cluster signal by bright sources, and to lower the astrophysical noise in the data. We note that fourteen out of the 461 clusters are overlapping by less than $3 \times \thetaf$. We left these in the analysis, but checked their impact on the average profile (see Sect.~\ref{sec:fullsamp}). We write the tangental maps as
\begin{equation}
\label{eq:datamodel}
\vec{m}(\vec{x}) =  \yo\vec{\tth}(\vec{x}) + \vec{n}(\vec{x})
\end{equation}
where $\vec{m}(\vec{x})$ is the column vector corresponding to the tangental maps, $\yo\vec{\tth}(\vec{x})$ the cluster component, and $\vec{n}(\vec{x})$ the instrumental and astrophysical noise.
The quantity $\yo\vec{\tth}(\vec{x})$ can be decomposed into the central Compton parameter $\yo$, and the cluster profile $\vec{\tth}(\vec{x})$, whose $i^{th}$ component is \mbox{$\fnu(\nu_i) [b_i\ast \Tth](\vec{x})$}. The quantity $\Tth$ is the two dimensional SZ profile\footnote{That is, the three dimensional SZ profile integrated along the line-of-sight.}, convolved by $b_i$ (defined hereafter) and multiplied by $\fnu$, the SZ frequency spectrum integrated over the frequency response. We included the relativistic corrections to the calculation of $\fnu$ using the formula from~\cite{itoh1998} and the cluster temperature-mass relation\footnote{We note that we have not recalibrated the masses $\Mfive$ by 0.8 to compute the temperature because the relation has been derived with the SPT mass definition.} from Eqn.~19 of~\cite{bulbul2019}. The quantity $b_i$ is the PSF for \Planck\ frequencies, and the PSF convolved by the FTF for SPT-SZ frequencies, respectively. As stated above, we used the flat sky approximation. We adopted the fast Fourier transform to perform all the calculations, and only translated to harmonic space as a final step. Equation~\ref{eq:datamodel} can be written in Fourier space as
\begin{equation}
\label{eq:datamodel_fourier}
\vec{m}(\vec{k}) =  \vec{a}(\vec{k})\yo\Tth(\vec{k}) + \vec{n}(\vec{k})
\end{equation}
with $\vec{a}(\vec{k})$ the mixing vector whose  $i^{th}$ component is \mbox{$\fnu(\nu_i) b_i(\vec{k})$}. The ILC provides an estimator of $\yo\Tth$~\citep[e.g.][]{remazeilles2011}
\begin{equation}
\label{eq:ilc}
\widehat{\yo\Tth}(\vec{k}) =  \left[ \vec{a}^t(\vec{k}) \cdot \vec{P}^{-1} \cdot \vec{a}(\vec{k}) \right ]^{-1} \, \vec{a}^t(\vec{k}) \cdot \vec{m}(\vec{k}),
\end{equation}%
with $\vec{P}(\vec{k}) = \langle \vec{n}^*(\vec{k}) \vec{n}(\vec{k}) \rangle$ being the power spectrum matrix of the noise across frequency channels. In practice $\vec{P}(\vec{k})$ was estimated directly from the maps by masking a circular region of $4\thetaf$ radius to exclude the studied cluster. The error on the estimator is given by
\begin{equation}
\label{eq:ilc_error}
\sigma(\vec{k}) =  \left[ \vec{a}^t(\vec{k}) \cdot \vec{P}^{-1} \cdot \vec{a}(\vec{k}) \right ]^{-1/2}.
\end{equation}
We then divided the profiles $\widehat{\yo\Tth}(\vec{k})$ and associated error $\sigma(\vec{k})$ by the pixel window function $w(k)$ to correct our profile for the $0.43\, {\rm arcmin}$ pixelisation. We rejected modes $\vec{k}$ corresponding to multipoles $l>10,000$ in SPT-SZ data, for which the FTF is not provided~\citep{chown2018}.

\subsection{Malmquist bias correction of the SPT-SZ signal}
\label{sec:malmquist}

An important point of the analysis is to correct for the Malmquist bias of the SZ signal in the SPT-SZ data. Our cluster sample being selected in the SPT-SZ data, we expect the signal to be overestimated in the SPT-SZ maps, especially at low detection significance. We calculated a correction factor for each cluster to be applied to the SPT-SZ maps before combining them in the ILC. We did not apply a correction to the \Planck\ data, since the sample has not been selected from it.
We computed the correction factor using the formalism in~\cite{bleem2015}. We adopted the `unbiased SPT-significance'-mass relation
\begin{equation}
\label{eq:zetamass}
\zeta =  A_{\rm SZ} \left ( {\Mfive \over 3 \times 10^{14} \, \msol \, h^{-1} } \right )^{B_{\rm SZ}}  \left ( {E(z) \over E(0.6)} \right )^{C_{\rm SZ}}
\end{equation}
with $A_{\rm SZ}=4.14$, $B_{\rm SZ}=1.44$, $C_{\rm SZ}=0.59$, and a log-normal scatter, $D_{\rm SZ}=0.22$. $h$ is the Hubble constant normalised to 100~km/s/Mpc. The significance $\xi$ is related to the unbiased significance $\zeta$ via
\begin{equation}
\label{eq:zetaxi}
\zeta =  \sqrt{ \langle \xi \rangle^2-3}
\end{equation}
for $\zeta>2$. For each published SPT-SZ cluster Malmquist bias corrected mass $\Mfive$, and significance $\xi$, we drew N=1,000,000 values of $\zeta_{\rm true}$ following Eqn.~\ref{eq:zetamass} including the log-normal scatter. We then computed $\xi_{\rm true}=\sqrt{\zeta_{\rm true}^2+3}$ for the N values. We added a noise with normal distribution $\mathcal{N}(0,1)$ to $\xi_{\rm true}$ to obtain observed significances $\xi_{\rm obs}$. We selected  the values that verify $\xi_{\rm obs}>4.5$ corresponding to the selection threshold of the SPT-SZ cluster catalogue, and $\zeta_{\rm true}>2$ corresponding to the validity range of Eq.~\ref{eq:zetaxi}. We computed the average of $\xi_{\rm obs}$ in bins of $\xi_{\rm true}$ for the selected values. We thus obtained a $\xi_{\rm obs}-\xi_{\rm true}$ relation for the considered cluster. We interpolated this relation at the published observed significance $\xi$ to obtain the average true significance $\xi_{\rm t}$. We adopted the ratio $\xi_{\rm t}/\xi$ as the correction factor to be applied to each SPT-SZ map. The correction is negligible for $\xi>7$ clusters but is important for lower significance clusters, reaching $0.4$ at the detection threshold $\xi=4.5$. We note that we have not recalibrated the masses $\Mfive$ by 0.8 to compute the Malmquist bias correction factor, since Eqn.~\ref{eq:zetamass} has been derived by the SPT collaboration with their mass definition.

\begin{figure}
\centering
\includegraphics[width=\hsize]{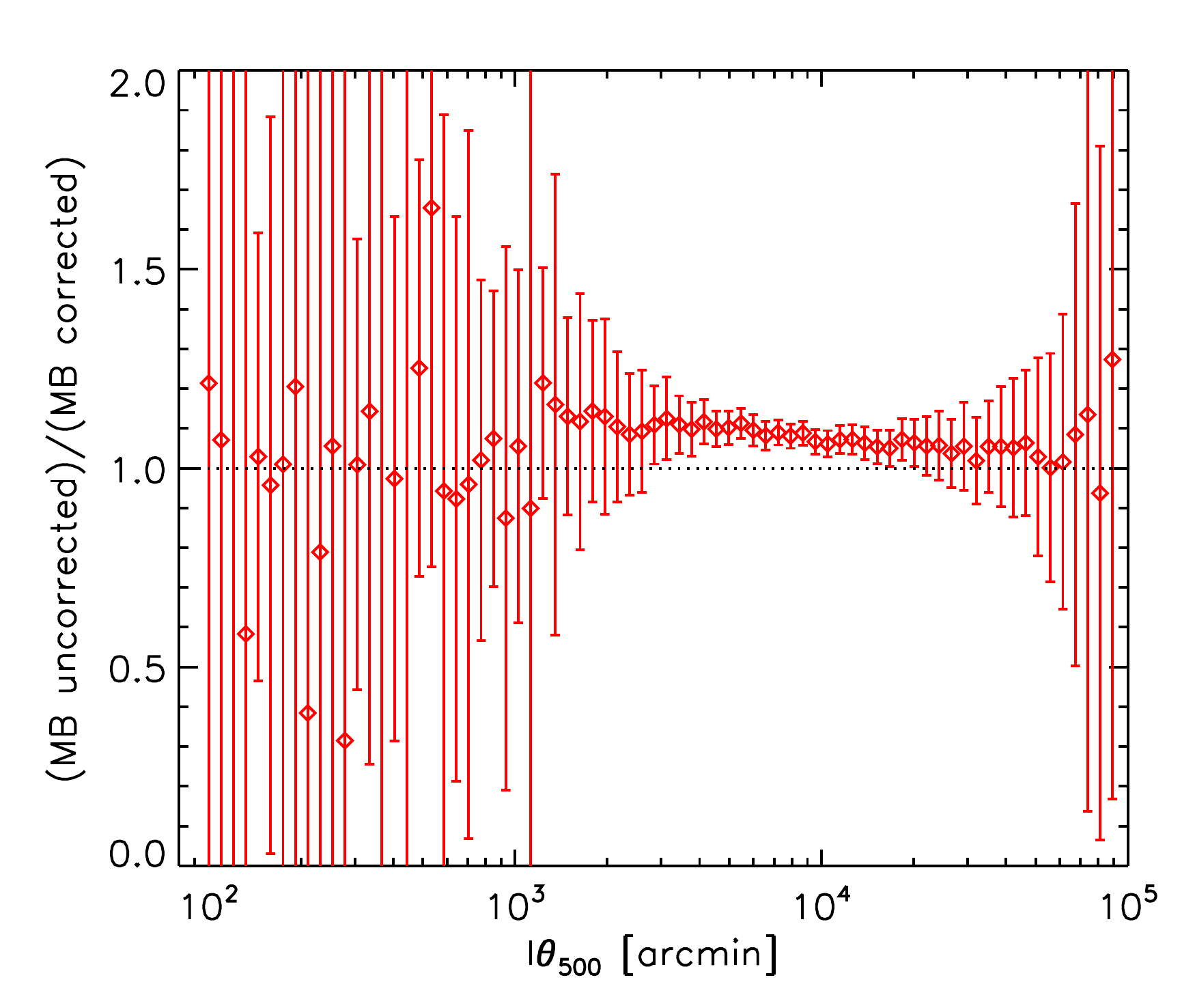}
\caption{\footnotesize Effect of the Malmquist bias (MB) correction on the measured average SPT-SZ  rescaled harmonic transform Compton parameter profile. The correction lowers the SPT profile by about 10\%, which is larger than the statistical error in the range $3 \times 10^3 \, {\rm arcmin}<{l\thetaf}<2 \times 10^4 \, {\rm arcmin}$.}
\label{fig:mbcorrection}
\end{figure}

The impact of the Malmquist bias correction on the average SPT-SZ rescaled harmonic transform Compton parameter profile is shown in Fig.~\ref{fig:mbcorrection}. The uncorrected SPT-SZ profile is about 10\% higher than the corrected profile. The correction is larger than the statistical error in the range $3 \times 10^3 \, {\rm arcmin}<{l\thetaf}<2 \times 10^4 \, {\rm arcmin}$ (see Sect.~\ref{sec:renorm} below for definition of ${l\thetaf}$).
%--------
%--------

%--------
% pixels vs clusters
%
\begin{figure}
\centering
\includegraphics[width=\hsize]{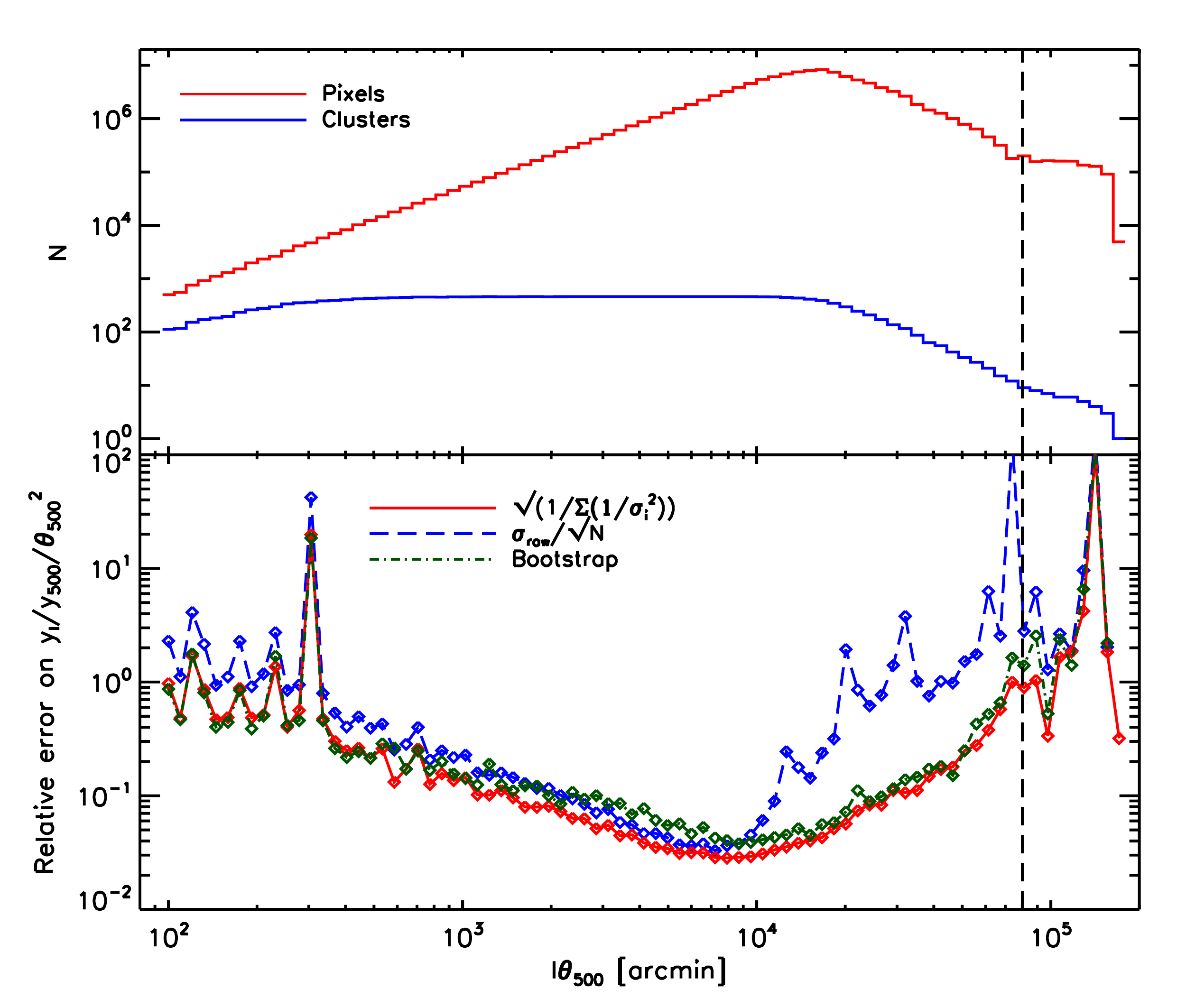}
\caption{\footnotesize {\it Top:} Number of pixels and clusters contributing to measurement of the harmonic transform Compton parameter profile in a given $l\thetaf$ bin. See text in Section.~\ref{sec:gnfwfits} for a description of the two histograms and their interplay. {\it Bottom:} Relative error on the profile. The solid red line is the inverse variance weighted errors used throughout the analysis (shown as red error bars in Fig.~\ref{fig:press_prof_all} and~\ref{fig:press_prof_fit}). The dotted-dashed green line corresponds to bootstrap errors. The dashed blue line is the raw scatter in a given $l\thetaf$ bin divided by the square root of the number of clusters contributing to this bin.}
\label{fig:bootstrap}
\end{figure}
%--------

\subsection{Averaging procedure}
\label{sec:renorm}

We renormalised each Fourier map obtained after the ILC $\hat{y}(\vec{k})=\widehat{\yo\Tth}(\vec{k})/w(k)$ by dividing it by $\thetaf^2 y_{500}$. We then bin averaged it in $\lVert \vec{k} \rVert \thetaf$ bins. This procedure performs the spatial rescaling of the profile in $\thetaf$ and the azimutal averaging at the same time. The averaging in bins was performed with inverse variance weights ($\sigma^2(\vec{k})/w^2(k)/\thetaf^4/y^2_{500}$). The profiles binned in $\lVert \vec{k} \rVert \thetaf$ were averaged together also using inverse variance weights. The final measurement is an estimate of the average of the two dimensional SZ profile of the cluster sample and its associated errors. This measurement is obtained in Fourier space $y_k/y_{500}/\thetaf^2$ versus $k \thetaf$ and we converted it to harmonic space $y_l/y_{500}/\thetaf^2$ versus $l\thetaf$. Errors on ${\tilde y}_l=y_l/y_{500}/\thetaf^2$ are the inverse weighted average errors. We denote these ${\tilde \sigma}_l$. We chose to adopt a logarithmic binning with 25 points per decade. 

The link between the profile shape in harmonic space and its shape in real space is not intuitive for all. In Appendix~\ref{app:realfour}, we provide a visual representation of the correspondance between three dimensional pressure profiles in real space, the two dimensional projected SZ profiles in real space, and the two dimensional projected SZ profiles in harmonic space, to help the reader pass from one space to the other.

%%%------
\begin{figure*}
   \sidecaption
   \includegraphics[width=0.7\hsize]{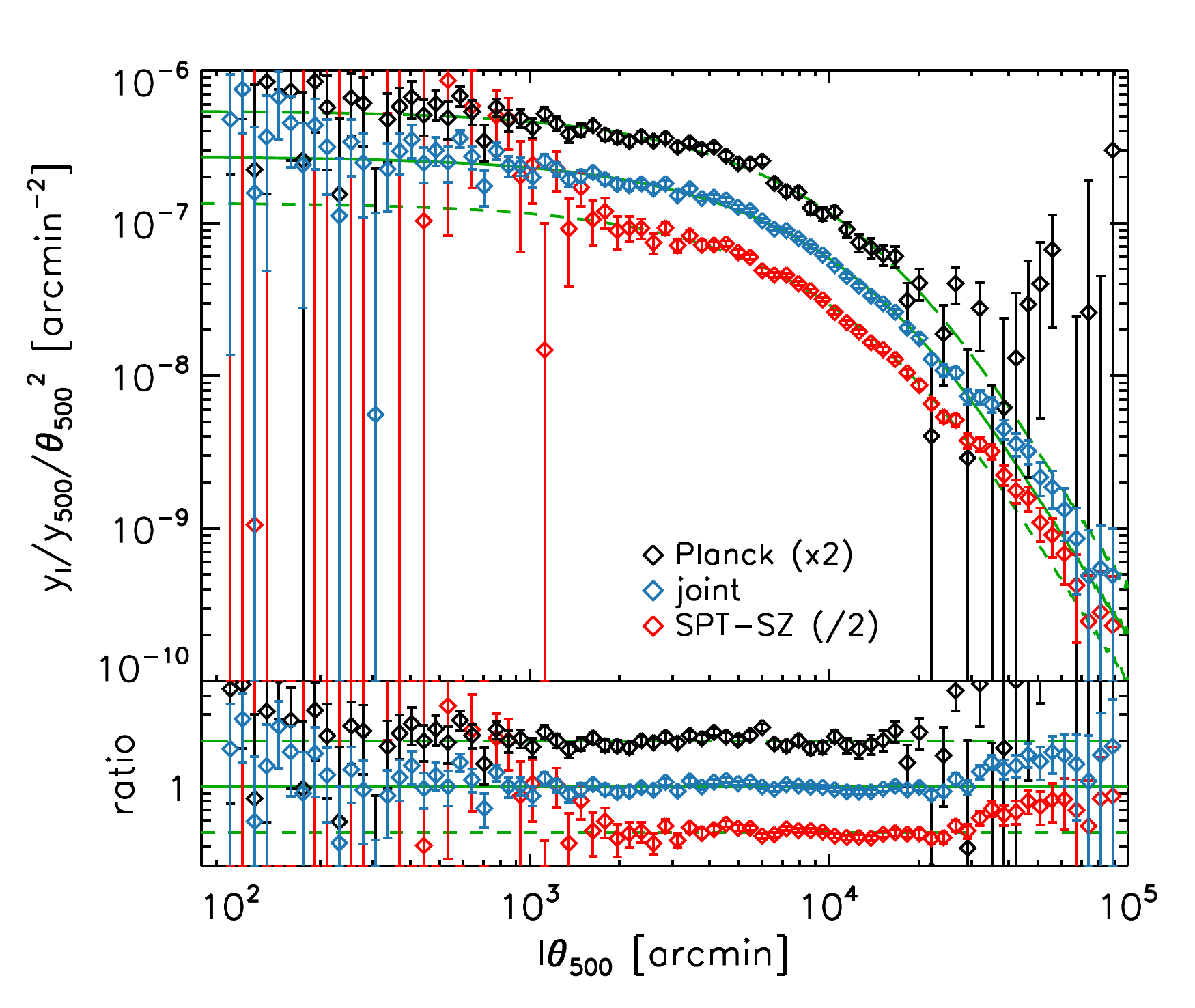}
      \caption{\footnotesize Average rescaled harmonic transform Compton parameter profile measurements from \Planck\ only, SPT-SZ only and joint (\Planck+SPT-SZ) in black, red and blue respectively. We multiplied (resp. divided) the \Planck\ (resp. SPT-SZ) measurements by a factor two for clarity. The joint analysis allows to take advantage of the two data sets at all scales. The solid green line is the PIPV profile. The long dashed (resp. short dashed) green line is the PIPV profile multiplied (resp. divided) by a factor of two.}
         \label{fig:press_prof_all}
\end{figure*}
%%%------

\begin{table*}
        \centering
        \caption{\footnotesize Best-fitting parameters (given by the maximum of the likelihood) for the full sample when the profiles are rescaled with $\thetaf$. Errors are the 16 and 84 percentiles. In the fitting procedure, the gNFW parameter $\alpha$ is fixed to the UPP value of $\alpha=1.05$. In this Table, $\Mfive$ is the original SPT mass, uncorrected for the 0.8 factor (see Sect.~\ref{sec:clusamp} for details). $\log$ is the decimal logarithm. The final column gives the reduced $\chi^2$ for the corresponding fit. }
	\label{tab:bestfits500}
	\begin{tabular}{c c c c c c c}
\toprule
\toprule
	 & $\log(P_0)$ & $c_{500}$ & $\gamma$ &  $\beta$ & $\alpha$ & $\chi^2/{\rm d.o.f.}$\\
\midrule
    {\bf All} & ${\bf 0.23} \substack{{\bf +0.16}\\ {\bf -0.21}}$ & ${\bf 0.61} \substack{{\bf +0.14} \\ \bf {-0.17}}$ & ${\bf 0.71} \substack{{\bf +0.09} \\ {\bf -0.06}}$ & ${\bf 6.32} \substack{{\bf +0.88} \\ {\bf -0.91}}$ & {\bf 1.05} & {\bf 1.37} \\
\\
          $z < 0.6$ & $ 0.59 \substack{+0.17 \\ -0.20}$ & $1.07 \substack{+0.22 \\ -0.34}$ & $0.55 \substack{+0.12 \\ -0.08}$ & $4.90 \substack{+0.49 \\ -0.53}$ & 1.05 & 1.19 \\
%          \hline
           $z\geqslant0.6$ & $1.11 \substack{+0.13 \\ -0.63}$ & $1.46 \substack{+0.02 \\ -0.82}$ & $0.19 \substack{+0.53 \\ -0.16}$ & $5.06 \substack{+0.89 \\ -0.61}$ & 1.05 & 1.37\\

%          \hline
\\
           $\Mfive<3.6\times10^{14} M_\sun$ & $-0.74 \substack{+0.49 \\ -0.13}$ & $0.03 \substack{+0.03 \\ -0.01}$ & $0.62 \substack{+0.09 \\ -0.15}$ & $82.83 \substack{+5.8\\ -63.93}$ & 1.05 & 1.56\\

%          \hline
          $\Mfive \geqslant 3.6\times10^{14} M_\sun$ & $0.45 \substack{+0.19 \\ -0.19}$ & $0.82 \substack{+0.19 \\ -0.22}$ & $0.62 \substack{+0.09 \\ -0.10}$ & $5.55 \substack{+0.57 \\ -0.61}$ & 1.05 & 1.59 \\
\bottomrule
\end{tabular}
\end{table*}

\subsection{Likelihood for gNFW profile fits}
\label{sec:gnfwfits}

We write the log likelihood as $\ln \mathcal{L}(\Theta) = -{\chi^2(\Theta) \over 2}$ with
\begin{equation}
\label{eq:chi2}
 \chi^2(\Theta) = \sum_{{l\thetaf}_{\rm bin},{l'\thetaf}_{\rm bin}}^{<{l\thetaf}_{\rm max}} {\left [ {\tilde y}_l - {\tilde y}_{{\rm mod},l}(\Theta) \right] \mathcal{S}_{ll'}^{-1} \left [ {\tilde y}_{l'} - {\tilde y}_{{\rm mod},l'}(\Theta) \right]}
\end{equation}
where $\Theta$ is the set of the five gNFW profile parameters. The quantity ${\tilde y}_{{\rm mod},l}$ is the model, that is, the harmonic transform of the SZ profile constructed from the integrated gNFW profile along the line-of-sight. $\mathcal{S}$ is the covariance matrix, estimated as detailed in Appendix~\ref{app:harmonoise}. The quantity ${\tilde y}_l$ is the average profile measurement obtained as described in the previous Section. The sum is performed over the binned multipoles ${l\thetaf}_{\rm bin}, {l'\thetaf}_{\rm bin}$ lower than ${l\thetaf}_{\rm max}$. 

%%%--------

%%%------
\begin{figure*}
   \sidecaption
   \includegraphics[width=0.7\hsize]{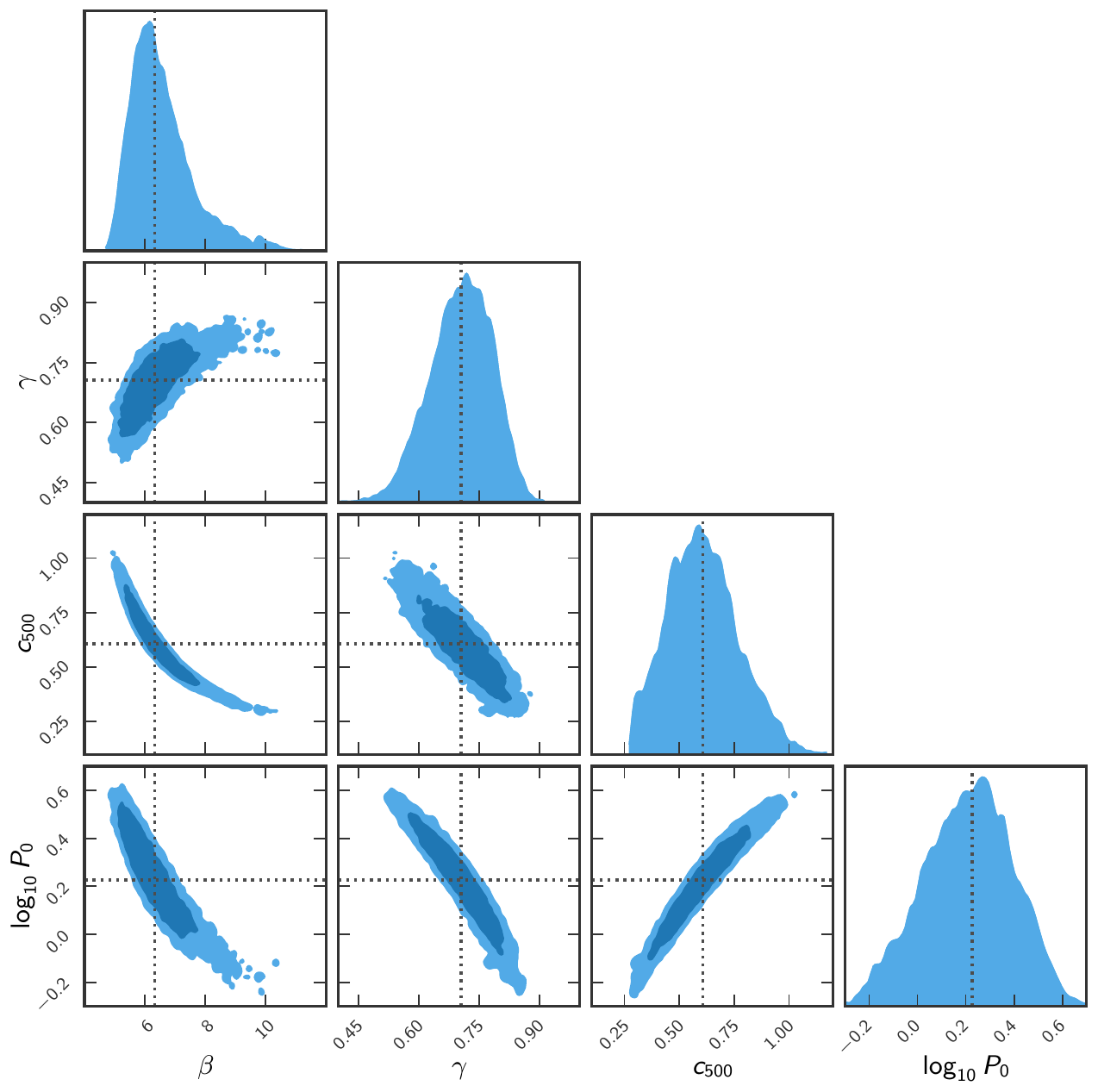}
      \caption{\footnotesize Marginalised posterior likelihood for the parameters of the best-fitting gNFW model (Eqn.~\ref{eq:gnfw500}) with fixed $\alpha=1.05$. Contours represent the 68\% and 95\% confidence regions. Blue contours show the results for the fit to the full data set; the dotted blue lines show the best-fitting value of each parameter as reported in Table~\ref{tab:bestfits500}.}
         \label{fig:gnfw_parameters500}
\end{figure*}
%%%------

%%%--------
%--------
% cf REXCESS, PIPV
%
\begin{figure*} %[!h]
\centering
\includegraphics[width=0.49\hsize]{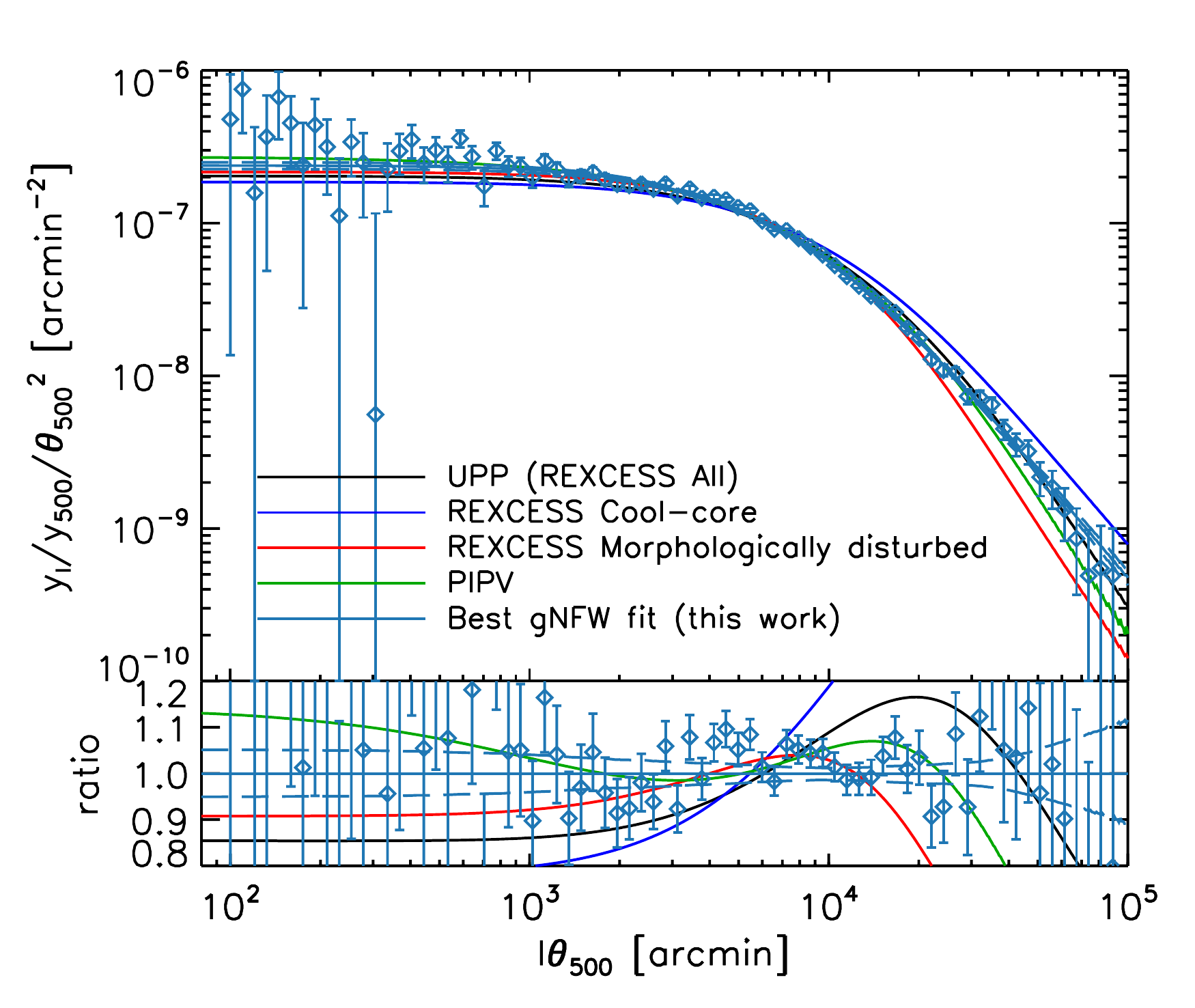}
\includegraphics[width=0.49\hsize]{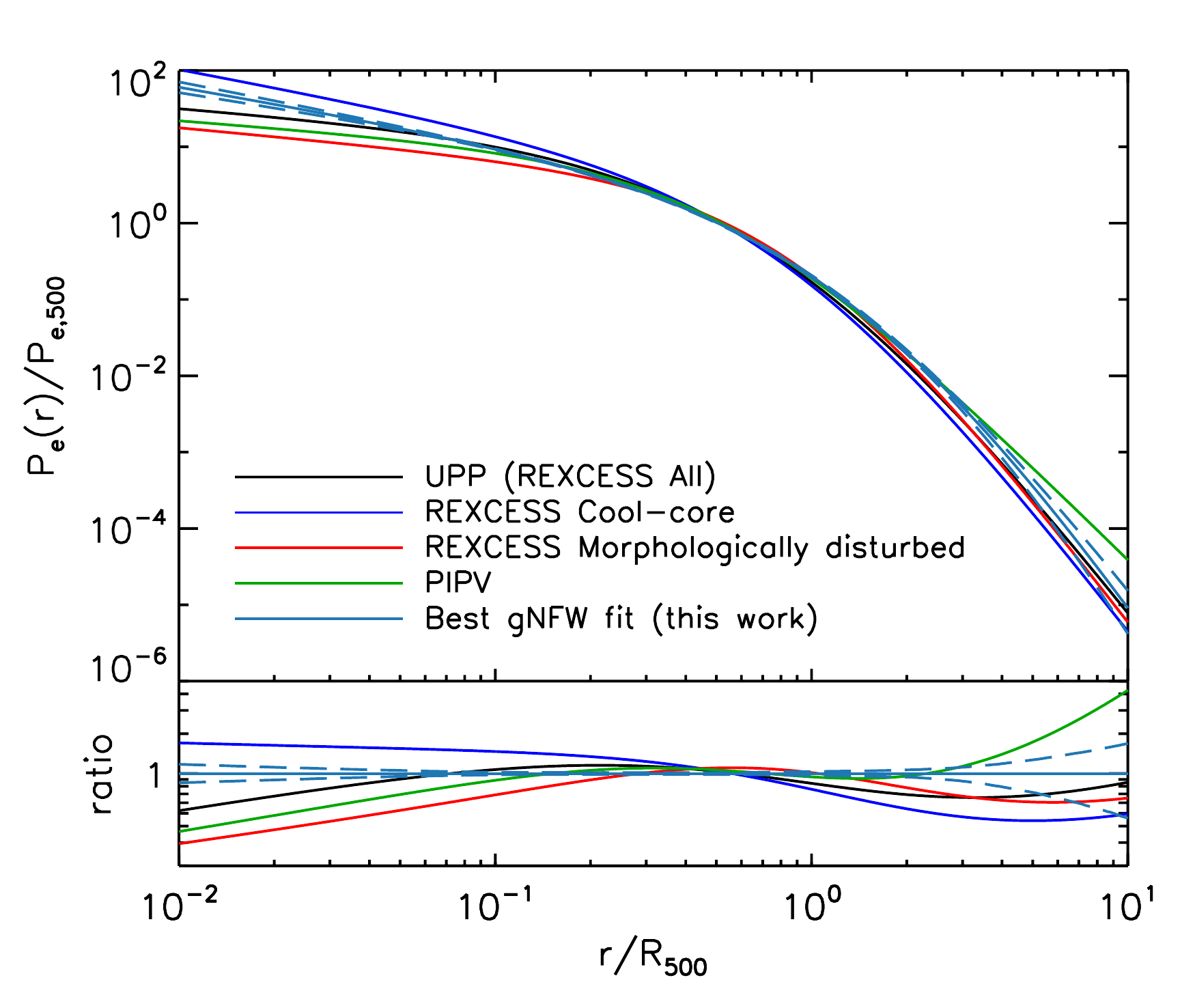}
\caption{\footnotesize Average pressure profile. {\it Left:}  Average rescaled harmonic transform Compton parameter joint measurement and best fitting gNFW model, with the  \rexcess\ and PIPV profiles overplotted for comparison. The best fit (solid blue line) is close to the UPP at large $l \thetaf$ (the central regions), and to the PIPV profile at small $l \thetaf$ (the outer regions). {\it Right:} Corresponding 3D pressure profiles. In both panels the dashed blue lines delineate the 68\% C.L. uncertainties of our best fitting gNFW profile.}
\label{fig:press_prof_fit}
\end{figure*}
%--------

For the fit to the full sample, we fixed ${l\thetaf}_{\rm max}=8\times 10^4 \, {\rm arcmin}$, a value chosen to avoid the inclusion of systematics at high ${l\thetaf}$. Figure~\ref{fig:bootstrap} (top) shows the number of pixels (resp. clusters) contributing to the measurement in a given ${l\thetaf}_{bin}$ as the red (resp. blue) histogram. At low ${l\thetaf}$ values (around $10^2 \, {\rm arcmin}$), the number of contributing clusters is around 100. As ${l\thetaf}$ increases, the number of contributing clusters increases together with the number of pixels. The number of pixels increases both because of the increasing number of clusters in each bin, and because of the logarithmic binning in ${l\thetaf}$ (the annulus contains more pixels when ${l\thetaf}$ increases). Around ${l\thetaf}=10^3{\rm arcmin}$, all the clusters of the sample contribute, which explains the flattening of the histogram of clusters. However, the number of pixels continues to increase due to the logarithmic binning. Around  ${l\thetaf}=2 \times 10^4 \, {\rm arcmin}$, the number of clusters decreases along with the number of pixels. At ${l\thetaf}=8 \times 10^4 \, {\rm arcmin}$ (marked as the dashed black vertical line in Figure~\ref{fig:bootstrap}), the number of clusters continues to decrease but the number of pixels increases again. This effectively means that the reconstructed profile is based on more pixels but fewer clusters. The information contained in the average is thus subject to possible systematic effects in the remaining about ten individual cluster profiles included in the averaging procedure. To minimise the risk of including systematic effects, we therefore exclude ${l\thetaf}$ bins above ${l\thetaf}_{\rm max}=8 \times 10^4 \, {\rm arcmin}$ in our likelihood for the full sample. Using the same criteria to fix ${l\thetaf}_{\rm max}$ for the subsamples, we find the same value of ${l\thetaf}_{\rm max}=8 \times 10^4 \, {\rm arcmin}$ for the high~mass and low~$z$ subsamples. 

We maximised the likelihood using Bayesian Monte Carlo Markov chain (MCMC) sampling. We used the \texttt{emcee} package~\citep{foreman2013} to marginalise over the gNFW parameters, adopting the same flat priors for all considered samples and subsamples except for the low mass subsample: ${\mathcal U}_{[-3,2]}$ for $\log(P_0)$, ${\mathcal U}_{[0.01,3]}$ for $c_{500}$ (or $c_{200}$), ${\mathcal U}_{[0.01,2.5]}$ for $\gamma$ and ${\mathcal U}_{[1,20]}$ for $\beta$. For the low mass sample, we also adopted flat priors but modified the upper bounds adopting ${\mathcal U}_{[-3,1.5]}$ for $\log(P_0)$, ${\mathcal U}_{[0.01,2]}$ for $c_{500}$, ${\mathcal U}_{[0.01,1.5]}$ for $\gamma$ and ${\mathcal U}_{[1,75]}$ for $\beta$. The \citet{arnaud2010} UPP parameters were used as a starting point of the fit, shortening the burn-in phase, and the final converged chains contained $\sim150\,000$ iterations, which were subsequently thinned by a factor of five.

\section{Results}
\label{sec:results}

We present results for the full cluster sample in Section~\ref{sec:fullsamp}. In Section~\ref{sec:bestfit}, we provide the best gNFW fit. In Section~\ref{sec:intrinsic}, we study the intrinsic scatter around the profile.

\subsection{Full sample profile}
\label{sec:fullsamp}

The blue diamonds in Fig.~\ref{fig:press_prof_all} show the average rescaled Compton parameter profile joint measurement in harmonic space for the full sample (461 clusters). We overplot the measurement based on \Planck\ data only with black diamonds, and that based on SPT-SZ data only with red diamonds. For clarity, in this Figure we multiplied (divided) the \Planck\ (SPT-SZ) points by a factor of two. The Compton parameter profile corresponding to PIPV is plotted in green. We also plot the PIPV profile multiplied (divided) by two as long dashed (short dashed) green lines. The \Planck\ data only and SPT-SZ data only profiles are compatible. The bottom panel shows the ratio between the \Planck, SPT-SZ, \Planck+SPT-SZ measurements and the PIPV profile on a  logarithmic scale. It is clear from the Figure that \Planck\ provides better constraints than SPT-SZ on large scales ($l \thetaf<2500 \, {\rm arcmin}$). Conversely, on small scales ($l \thetaf> 6500 \, {\rm arcmin}$), SPT-SZ outperforms \Planck. The joint analysis therefore allows us to combine the two data sets optimally at all scales. Removing the fourteen clusters overlapping within $3 \times \thetaf$ from the average modifies only weakly the intermediate and external part of the profile ($l \thetaf< 10^4 \, {\rm arcmin}$), and always within the statistical errors.

\subsection{Best-fitting gNFW profile}
\label{sec:bestfit}

We now discuss the best-fitting parameters and associated errors obtained from the likelihood described in Section~\ref{sec:gnfwfits}. As found in other studies, leaving the five gNFW parameters free led to strong degeneracies between the three slope parameters $\gamma$, $\alpha$, $\beta$, and did not allow meaningful constraints to be put on all five parameters. We therefore fixed the intermediate slope to the UPP value ($\alpha=1.05$) and left the other four parameters (amplitude $\log(P_0)$, concentration parameter $c_{500}$, internal slope $\gamma$ and external slope $\beta$) free. Results are shown in Figure~\ref{fig:gnfw_parameters500}. Fixing $\alpha$ allowed the chains to converge to a small region of the parameter space and led to closed contours for all four free parameters. The best-fitting parameters are displayed with the dotted line; their numerical values (which maximise the likelihood given by the MCMC chains) and associated errors (at 68\% confidence) are provided in the first row of Table~\ref{tab:bestfits500}. The $\chi^2$ of the fit is good, with $\chi^2=92.96$ and 68 degrees of freedom (72 data points minus four parameters), leading to a reduced $\chi^2/{\rm d.o.f}=1.37$.

In the left hand panel of Fig.~\ref{fig:press_prof_fit}, we show the joint measurement (blue diamonds, identical to Fig.~\ref{fig:press_prof_all}) with the best-fitting gNFW model overplotted (solid blue line). As in Fig.~\ref{fig:press_prof_all}, the bottom panel shows the ratio with respect to the PIPV profile, but this time in a linear instead of a logarithmic $y$-scale.
The best gNFW fit is similar to the UPP in the inner part and to the PIPV profile in the outer part. We will further discuss the comparison to previous work in Section~\ref{sec:previous} and~\ref{sec:cores}. The right hand panel of Fig.~\ref{fig:press_prof_fit} shows the 3D pressure profiles in real space corresponding to the 2D Compton parameter profiles shown in the left hand panel of the figure. The best-fitting gNFW parameters provided in Table~\ref{tab:bestfits500} are sufficient to reproduce the average gNFW profile but are strongly correlated, as can be seen in Fig.~\ref{fig:gnfw_parameters500}. They are therefore not sufficient to fully describe the uncertainties on the best-fitting profile (that is, the error envelope). To compute the error envelope, we used the MCMC chains. At each $r$ (in real space) and each ${l\thetaf}$ (in harmonic space), we used the 16 and 84 percentiles of the distribution of the profiles provided by the MCMC chains. The resulting 68\% C.L. envelope based on the MCMC chains is delimited with dashed blue lines in both panels of Fig.~\ref{fig:press_prof_fit}. We provide the tabulated best-fitting profile and associated error envelope in harmonic and real space for the full sample in Table~\ref{tab:prof_full}. We graphically compare the error envelopes of the various subsamples in Fig.~\ref{fig:errsubsamp}.

\subsection{Intrinsic scatter around the average profile}
\label{sec:intrinsic}

The average profile is computed as the inverse variance weighted mean of the individual profiles.  The error bars are statistical. We computed 1,000 bootstraps of the full analysis, where a single bootstrap corresponded to a draw with replacement of a sample of 461 clusters from the original sample of 461 clusters. We performed the full analysis again for each bootstrap sample. The bootstrap errors were  then computed as the standard deviation of the 1,000 recovered profiles. The result is shown in the bottom panel of Fig.~\ref{fig:bootstrap}, where the bootstrap errors are displayed in green. They are of the same order of magnitude as the statistical inverse variance weighted errors (shown in red) at low and high $l\thetaf$, but they are larger than the statistical errors in the intermediate $l\thetaf$ range ($2 \times 10^3 \, {\rm arcmin} < l\thetaf < 2 \times 10^4 \, {\rm arcmin} $). This effect is an indication of some intrinsic scatter ($\sigma_{\rm int}$) in the profiles in the intermediate $l\thetaf$ range. We tried to measure it using the approach described in Section~5.3 of \cite{piffaretti2011}. We computed the raw scatter $\sigma_{\rm raw}$ as the standard deviation of the 461 individual profiles. We show this quantity divided by $\sqrt{N}$ (the number of clusters contributing to the $l\thetaf$ bin under consideration) as the dashed blue line in the bottom panel of Fig.~\ref{fig:bootstrap}. We also computed $\sigma_{\rm stat}$ from the square root of the mean of the variances of the individual profiles. Unfortunately, as this figure shows, $\sigma_{\rm raw}$ is of the same order of magnitude as $\sigma_{\rm stat}$ across the full $l\thetaf$ range, and the estimated difference $\sigma_{\rm int}^2=\sigma_{\rm raw}^2-\sigma_{\rm stat}^2$ is compatible with zero. This estimator of the intrinsic scatter $\sigma_{\rm int}$ is simple, robust, and has well-characterised uncertainties, but it is suboptimal when the variances of the individual profiles are very different with respect to each other, which is the case here.

\subsection{Impact of positional uncertainty}
\label{sec:posuncert}

Before quantitative comparison to X-ray selected samples, it is prudent to consider what effect the  significantly reduced positional uncertainty of the SZ experiments has on accuracy of the recovery of the inner slope of the gNFW profile. We estimated the impact of cluster positional uncertainty on the inner slope of the profile by injecting simulated SPT clusters at random locations in the SPT footprint. We adopted the UPP parameters to describe the pressure profile of the simulated objects. We first extracted clusters at the exact location of injection and ran the gNFW fitting analysis including the MCMC chains. We recovered the parameters of the UPP. We then extracted the clusters but with randomly shifted positions, with the positional uncertainty described by a Gaussian distribution with FWHM given by \citep[arXiv v1 of][]{story2011,song2012,zenteno2020}
\begin{equation}
\sqrt{\theta_{\rm beam}^2+ (k \theta_c) ^2}/\xi .
\end{equation}
Here, $\theta_{\rm beam}=(1.1+1.6)/2 \, {\rm arcmin}$ is the mean of the FWHM of the 95 and 150\,GHz original beams, $k=2$, $\theta_c$ is the cluster core radius, and $\xi$ is the signal-to-noise ratio. The positional uncertainty varies between 0 and about 1~arcmin, with a median around 0.1~arcmin. Posterior contours for the best-fitting gNFW models are presented in Appendix~\ref{app:pos_unc}, and show no significant difference between the average profile extracted from the true and the randomly shifted positions. We therefore conclude that positional uncertainty has a negligible effect on our result.

\section{Discussion}
\label{sec:discussion}

%------
\begin{figure}
\centering
\includegraphics[width=\hsize]{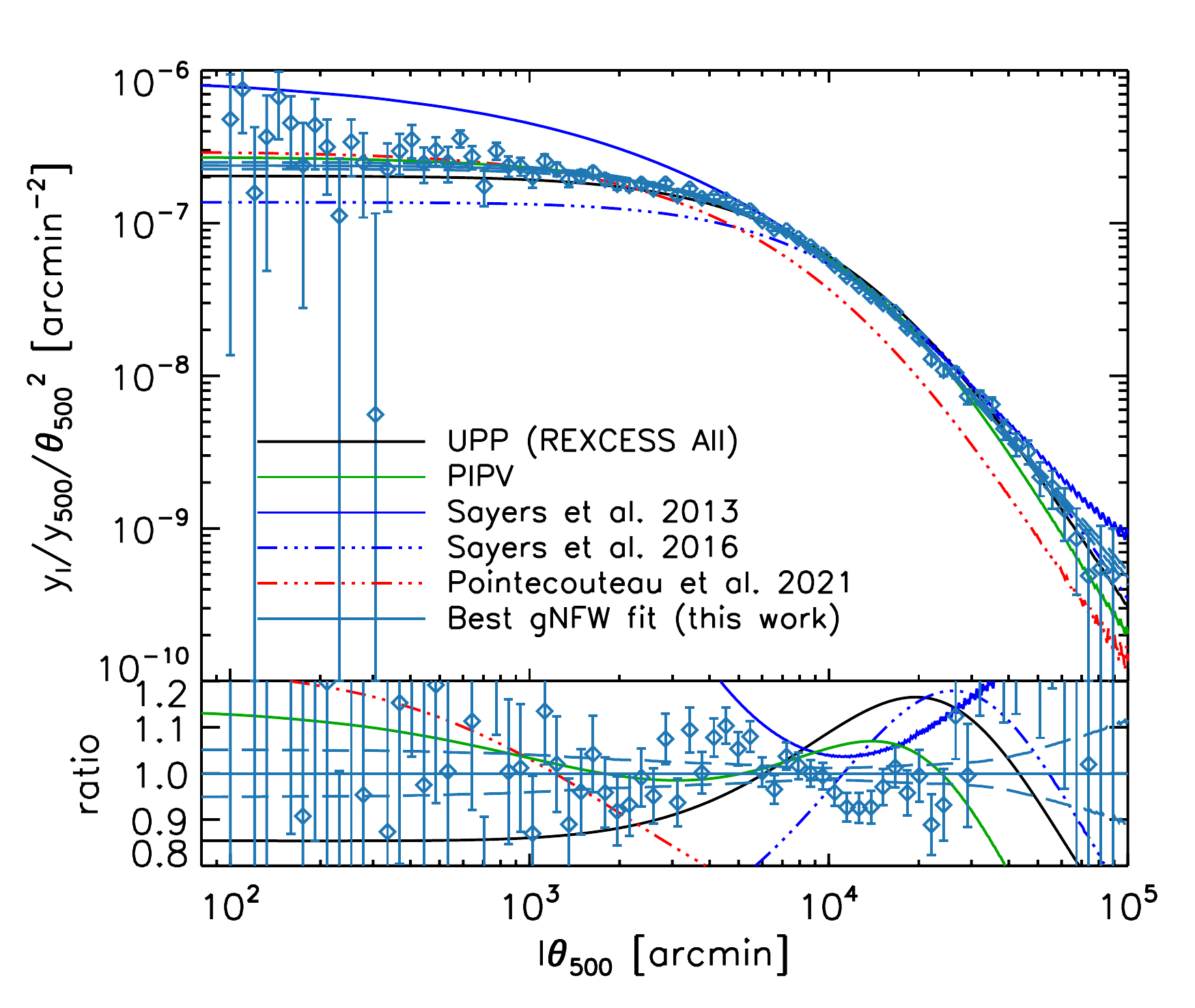}
\caption{\footnotesize Average rescaled harmonic transform Compton parameter joint measurement and its best fitting gNFW model compared to other gNFW profile fits published in other works. See text for discussion of the differences.}
\label{fig:press_prof_oworks}
\end{figure}
%------

We now compare our results to previous work (Section~\ref{sec:previous}). We discuss the consequence of our result on cluster cores in Section~\ref{sec:cores}. In Section~\ref{sec:evolz} and \ref{sec:evolmass}, we study the evolution of the profile with redshift and mass. We discuss the outskirts in Section~\ref{sec:outskirts}.

\subsection{Comparison to previous work}
\label{sec:previous}

In Fig.~\ref{fig:press_prof_oworks}, we show the joint measurement of the average rescaled harmonic transform Compton parameter  (blue diamonds, similar to Fig.~\ref{fig:press_prof_all} and \ref{fig:press_prof_fit}), with gNFW fits from various previous works overplotted. As already noted, the UPP~\citep{arnaud2010}, corresponding to the average profile of all 31 X-ray selected  \rexcess\ clusters, is in excellent agreement with our measurement in the inner part of the profile (at high $l\thetaf$) but is slightly below in the outer part (at low $l\thetaf$). The second and third columns of Fig.~\ref{fig:illu_P0} in Appendix~\ref{app:realfour} help the reader to pass from harmonic to real space for the inner and outer part of the profile. Conversely, the PIPV profile~\citep{pipv}, based on 62 nearby massive clusters observed by \Planck, is in excellent agreement with our measurement in the outer part, but is below in the inner part. Our measurement is thus very well reproduced by the UPP in the inner part and by the PIPV profile in the outer part.

\cite{sayers2013} averaged the signal of 45 massive clusters observed with Bolocam. The corresponding gNFW fit (solid blue line) is above our data points in the inner and in the outer parts but passes through our data points in the central range $7 \times 10^3 \, {\rm arcmin} < l\thetaf < 4 \times 10^4 \, {\rm arcmin} $.
\cite{sayers2016} used \Planck\ and Bolocam data for 47 massive clusters to constrain the amplitude $P_0$ and outer slope $\beta$ of the profile, fixing the other shape parameters to the UPP. This profile (triple dot dashed blue line) is in excellent agreement with our data in the inner region, but has an outer slope that is considerably steeper than our measurement. \cite{pointecouteau2021} measured the profile of 31 clusters in \Planck\ and ACT data. The gNFW fit is shown as the triple dot dashed red line in the Figure. In contrast to \cite{sayers2016}, their profile is in excellent agreement with our measurement in the outer part, but lies significantly below in the inner part.

\begin{figure*}
\centering
\includegraphics[width=0.49\hsize]{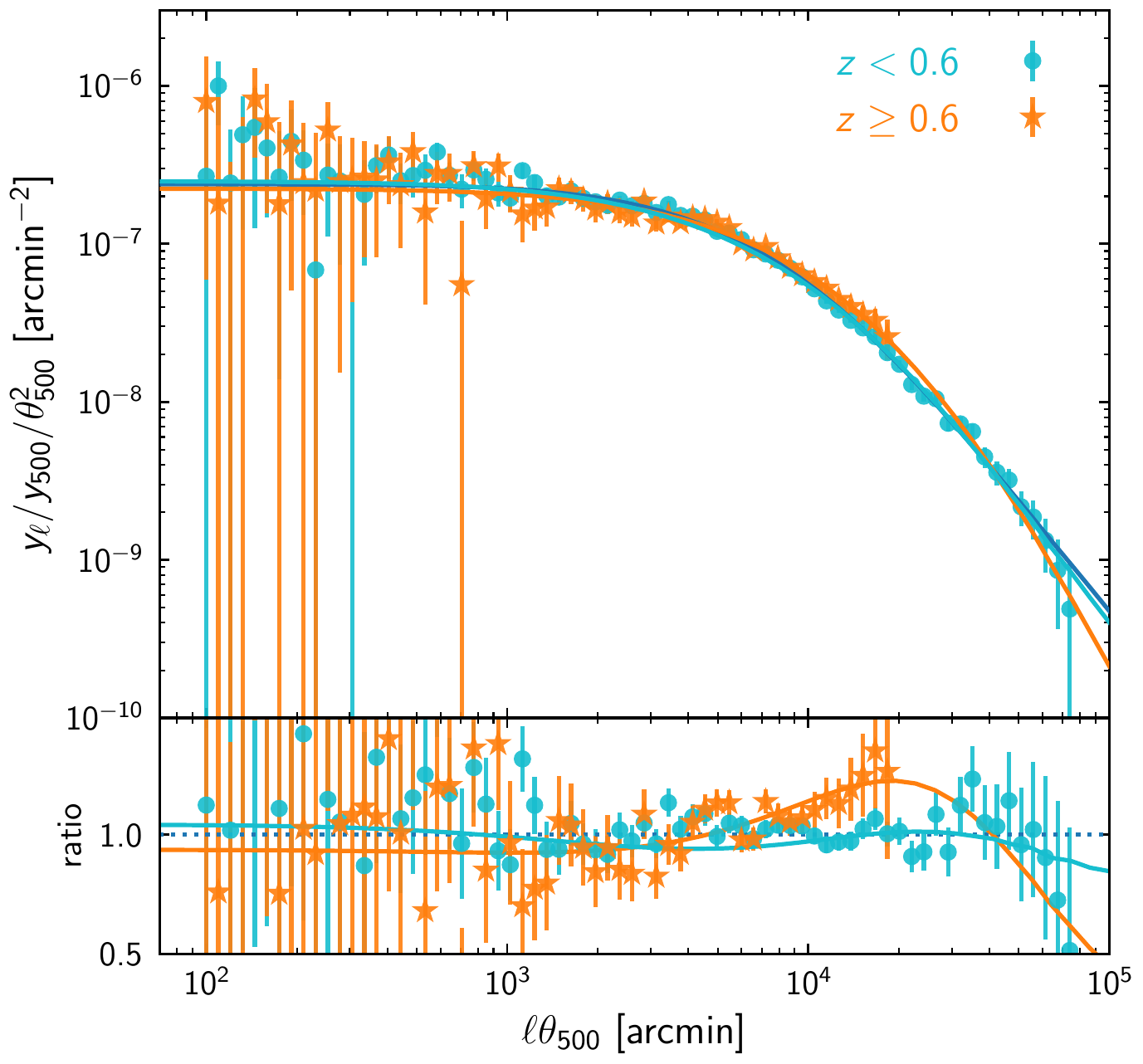}
\hfill
\includegraphics[width=0.455\hsize]{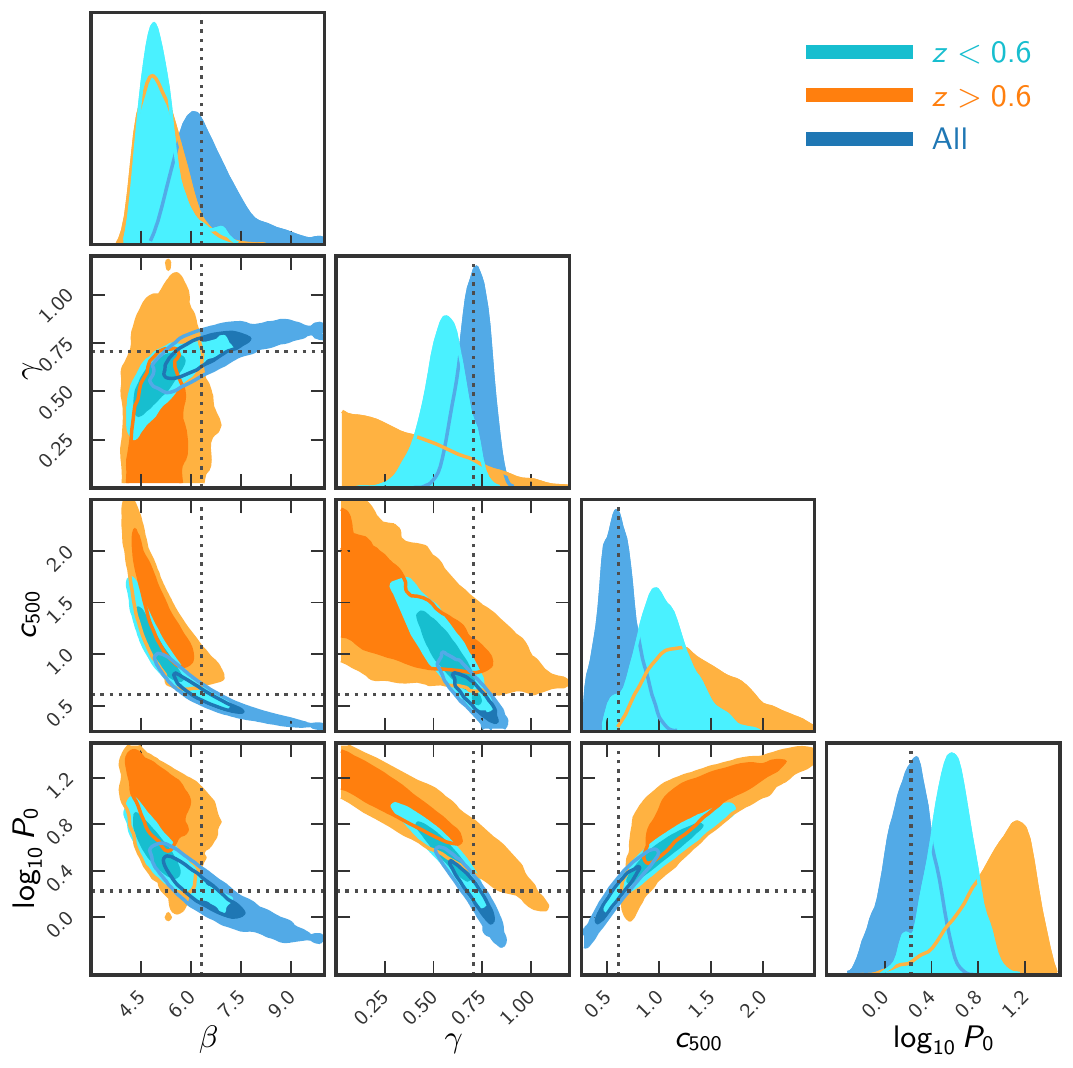}

\caption{\footnotesize Redshift subsamples. {\it Left}: Average pressure profile for $z < 0.6$ (cyan circles), and $z \geqslant 0.6$ (orange stars). The solid blue line is the best-fitting gNFW model for the full sample (Table~\ref{tab:bestfits500}); the ratio plot on the bottom panel is calculated with respect to the best fit for the full sample. {\it Right}: Marginalised posterior likelihood for the parameters of the best-fitting gNFW model of the the low z (cyan) and high z (orange) subsamples, compared to the results for the full sample (blue). Contours represent the 68\% and 95\% confidence regions.}
\label{fig:press_prof_evolz}
\end{figure*}

Our study comprises about a factor of ten more clusters than these previous works, and used an optimal treatment of the PSF and cluster rescaling thanks to the novel analysis in harmonic space. This allows us to significantly improve the constraints on the average pressure profile based on SZ measurements only. In particular, we only fixed the intermediate shape ($\alpha$) and determined the other four parameters ($\beta$, $\gamma$, $c_{500}$, $\log(P_0)$). Our contours (Fig.~\ref{fig:gnfw_parameters500}) are consequently reduced with respect to Fig.~5 of~\cite{sayers2013}, for which one parameter was also fixed. Contours from Fig.~6 of~\cite{pointecouteau2021} are smaller than ours, but were obtained by fixing two parameters instead of just one.
\cite{anbajagane2022} also used the SPT-SZ cluster sample for their study. They used the $y$-map from~\cite{bleem2022}, constructed from \Planck\ and SPT-SZ data. They stacked the SPT-SZ clusters but did not provide a gNFW fit to the average profile. They focused instead on the outer regions to search for pressure shocks. One important missing piece in their analysis is the selection bias of the SPT-SZ sample in SPT-SZ data (Section~\ref{sec:malmquist}), which is expected to impact their result.

\subsection{Cluster cores}
\label{sec:cores}

We now focus on the inner part of the average profile in Fig.~\ref{fig:press_prof_fit}. We recall that the positional uncertainty has a negligible impact on the recovered gNFW parameters, as discussed in Sect.~\ref{sec:posuncert}. We also note that the constraint on the inner part of the profile originates primarily from the low redshift objects, as can be seen in Fig.~\ref{fig:press_prof_evolz}. The joint measurement is in good agreement with the UPP (solid black line, based on all \rexcess\ clusters) in the inner part of the profile ($l \thetaf > 10^4  \, {\rm arcmin}$). It is lower than the \rexcess\ cool-core profile and higher than the \rexcess\ morphologically disturbed and the PIPV profiles in this high $l \thetaf$ region. This indicates that the central slope in real space is closer to that of the UPP,  shallower than that of the \rexcess\  cool-cores, and steeper than that of the \rexcess\  morphologically disturbed and the PIPV samples. This is an indication that the \rexcess\ and the SPT-SZ samples may contain similar fractions of cool-core and morphologically disturbed clusters. \cite{rossetti2016} found that the \Planck\ sample contains more disturbed clusters than three X-ray selected samples, including \rexcess. If the SPT-SZ selection were similar to that of \Planck, we would have expected the SPT-SZ sample to contain more disturbed clusters than \rexcess. This would lead to us finding a shallower inner slope than the UPP, closer to that seen for the \rexcess\ morphologically disturbed profile, such as the PIPV profile. This is not what we find. \cite{zenteno2020} also found no evidence that SPT-SZ selected clusters include a higher fraction of mergers than X-ray selected clusters. The only way to reconcile the results obtained by \cite{rossetti2016} and \cite{zenteno2020} and our work is to consider that the \Planck\ and SPT-SZ selections differ, in the sense that they do not select the same fraction of disturbed clusters. 
 
\begin{figure*}
\centering
\includegraphics[width=0.49\hsize]{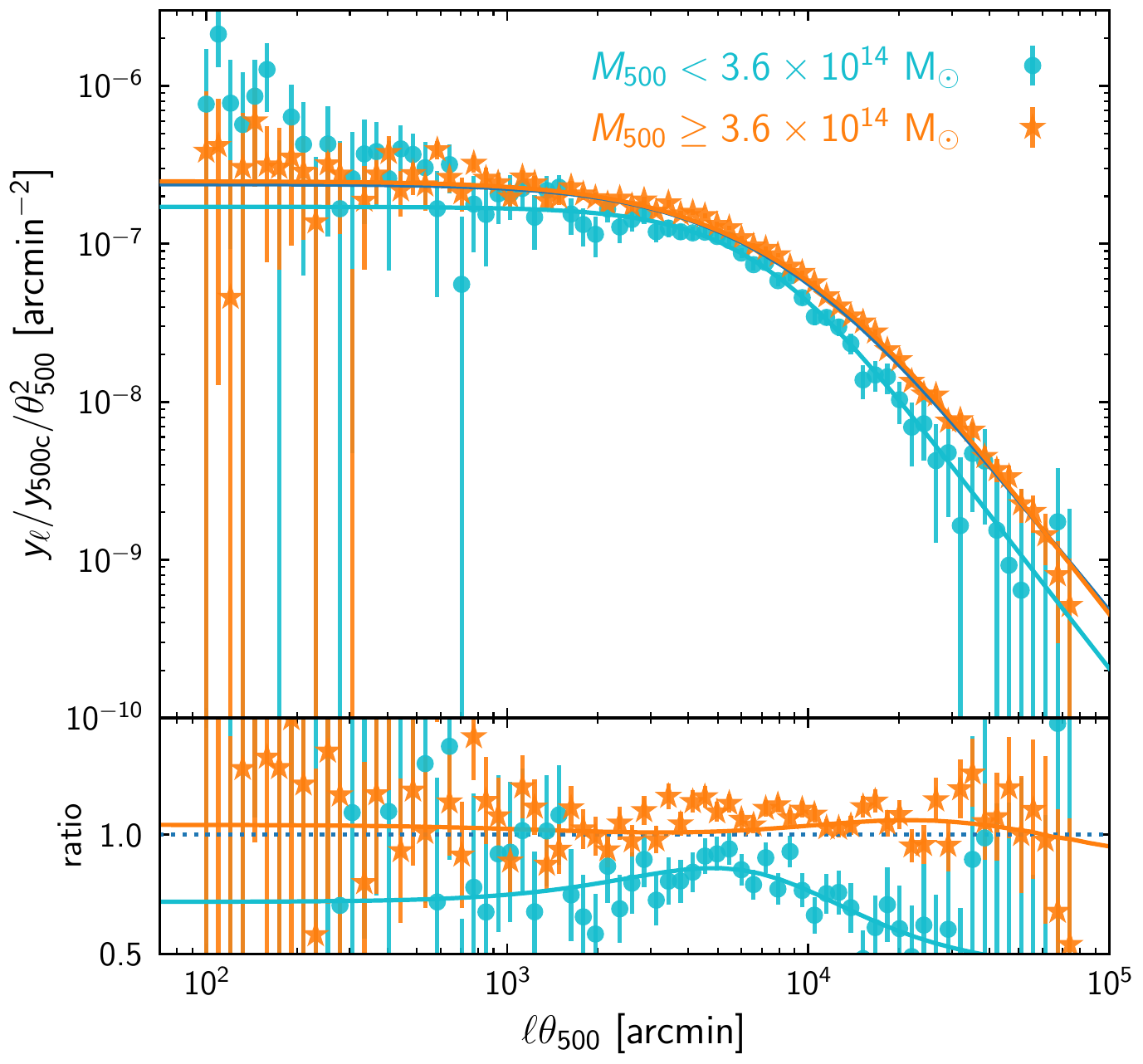}
\hfill
\includegraphics[width=0.455\hsize]{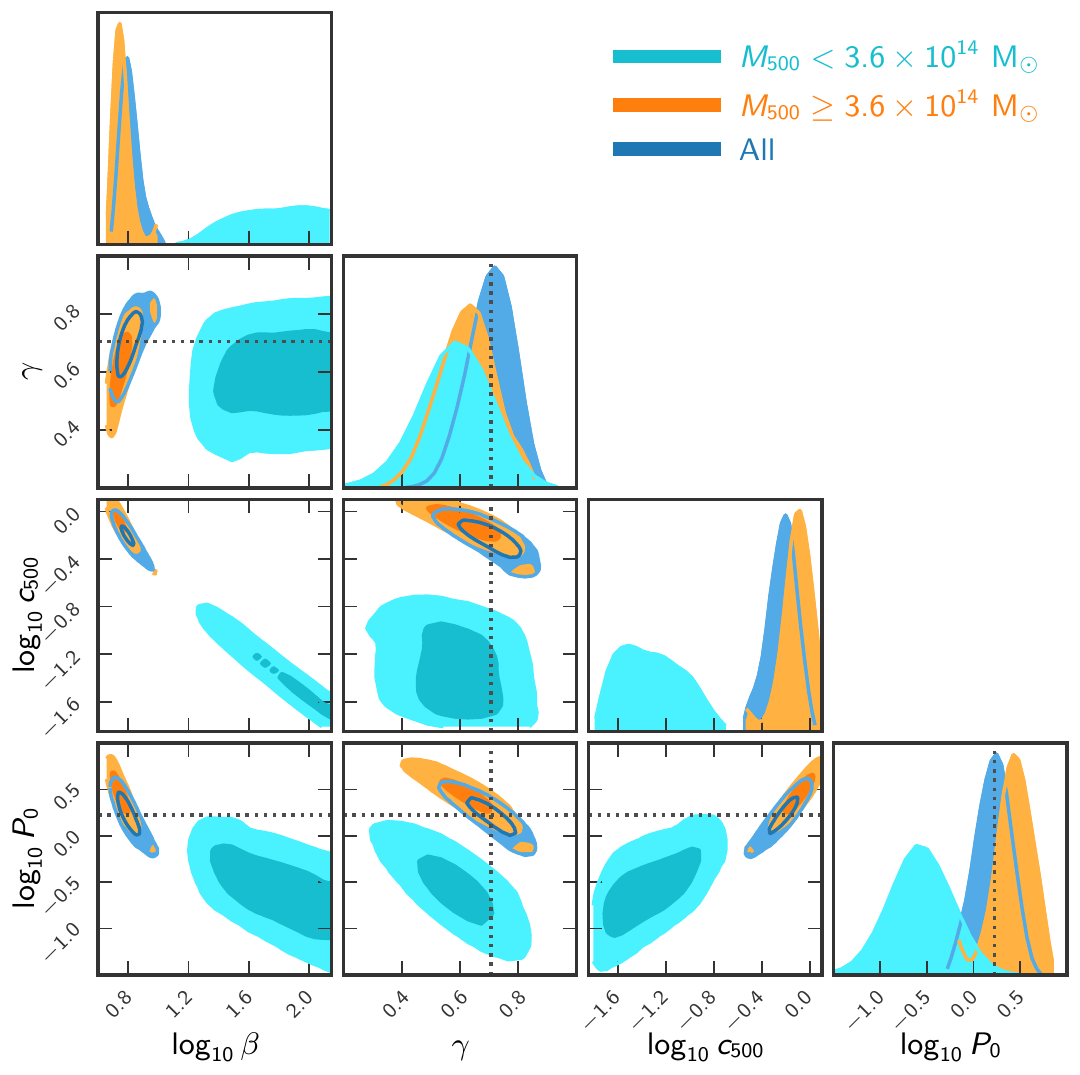}

\caption{\footnotesize Mass subsamples. {\it Left}: Average pressure profile for $\Mfive < 3.6 \times 10^{14} M_\sun$ (cyan circles), and $\Mfive \geqslant 3.6 \times 10^{14} M_\sun$ (orange stars). The solid blue line is the best-fitting gNFW model for the full sample (Table~\ref{tab:bestfits500}); the ratio plot on the bottom panel is calculated with respect to the best fit for the full sample. {\it Right}: Marginalised posterior likelihood for the parameters of the best-fitting gNFW model of the the low mass (cyan) and high mass (orange) subsamples, compared to the results for the full sample (blue). Contours represent the 68\% and 95\% confidence regions. Note that some parameters are plotted with logarithmic axes.}
\label{fig:press_prof_evolmass}
\end{figure*}

\subsection{Trends with redshift}
\label{sec:evolz}

We built low and high redshift subsamples by splitting our sample in two at $z=0.6$, a value close to the median redshift of the sample ($z=0.56$). The low $z$ subsample contains 266 clusters, and the high $z$ subsample 195 objects.
The resulting Compton parameter profile measurements are shown in Fig.~\ref{fig:press_prof_evolz}. The low $z$ subsample is well measured across all $l \thetaf$, while the high $z$ subsample lacks data points at $l \thetaf > 2 \times 10^{4} {\rm arcmin}$ (small scales) due to the small angular size of the high $z$ clusters with respect to the resolution of the two instruments.
 
 We fitted the subsample profiles as described in Sect.~\ref{sec:gnfwfits}, limiting the fit to ${l\thetaf}_{\rm max} < 2 \times 10^4 \, {\rm arcmin}$ for the high $z$ subsample owing to the more limited range of this profile. The corner plot for the parameters of the two samples is shown in Figure~\ref{fig:press_prof_evolz} and the best fit values are given in Table~\ref{tab:bestfits500}. The fits for the low and high z subsamples are both in agreement with the best fit of the full sample. We thus find no strong indication for evolution of the mean cluster pressure profile between $z=0.40$ (median redshift of the low $z$ sample) and $z=0.74$ (median redshift of the high $z$ sample).
 
We provide the tabulated best-fitting profile and associated error envelope in harmonic and real space for the low and high redshift subsamples in Table~\ref{tab:prof_lowz} and Table~\ref{tab:prof_highz}. Furthermore, the low and high redshift subsample error envelopes are graphically compared to that of the full sample in Fig.~\ref{fig:errsubsamp}.

\subsection{Trends with mass}
\label{sec:evolmass}

We next divided our sample into two mass subsamples at $\Mfive=3.6 \times 10^{14} M_\sun$.  This mass is close to the median mass of the sample ($\Mfive=3.58 \times 10^{14} M_\sun$). The low mass subsample contains 233 clusters with a median mass of $\Mfive=3.15 \times 10^{14} M_\sun$. The high mass subsample contains 228 clusters with a median mass of $\Mfive=4.46 \times 10^{14} M_\sun$. The resulting subsample profiles are shown in Fig.~\ref{fig:press_prof_evolmass}. The profile of the low mass sample is significantly below the profile of the high mass sample over the whole $l \thetaf$ range. The profile of the high mass sample is slightly above the full sample gNFW fit.

We fitted the subsample profiles as described in Sect.~\ref{sec:gnfwfits}, limiting the fit to ${l\thetaf}_{\rm max} < 6 \times 10^4 \, {\rm arcmin}$ for the high mass subsample owing to the more limited range of this profile. The best fitting gNFW models for the two subsamples are shown in Fig.~\ref{fig:press_prof_evolmass} and the best fitting parameter values are given in Table~\ref{tab:bestfits500}. These fits confirm the visual impression that the low and high mass samples are not compatible. In Fig.~\ref{fig:press_prof_evolmass} the 95\% C.L. contours of the low mass subsample do not overlap for any of the parameters. In particular,  for a given $\log(P_0)$, the profile of the low mass subsample has a lower $\gamma$ than the profile of the high mass subsample. This shows that the profile is less peaked for the low mass than for the high mass subsample at fixed $\log(P_0)$. This can be physically interpreted as an increase in the impact of non-gravitational processes on the pressure profile at low masses, or alternatively, by the faction of cool-core clusters being smaller for the low mass systems selected by SPT. On the other hand, at fixed $\gamma$, $\log(P_0)$ is lower for the low mass than for the high mass subsample. If one assumes that there is no change of the pressure profile with mass, this result could be due to an overestimated $\Pfive$ value which would drive the fit to a lower $P_0$. This might point towards a possible under-correction of the Malmquist bias of the SPT data for the low mass sample. At this stage, it is not possible to decide between the two interpretations (less peaked profiles for low mass clusters, or residual systematics in the Malmquist bias correction of the SPT subsample).

We provide the tabulated best-fitting profile and associated error envelope in harmonic and real space for the low and high mass subsamples in Table~\ref{tab:prof_lowm} and Table~\ref{tab:prof_highm}. We also compare the low and high mass subsample error envelopes  to that of the full sample in Fig.~\ref{fig:errsubsamp}.

\subsection{Outskirts}
\label{sec:outskirts}

\subsubsection{External slope}
\label{sec:beta}

We now focus on the outer part of the average profile of the full sample shown in Fig.~\ref{fig:press_prof_fit}. For $l \thetaf < 10^3  \, {\rm arcmin}$, we showed that the joint measurement is in good agreement with the PIPV profile, and therefore has a shallower slope than all the profiles based on \rexcess. This result confirms the excess of pressure in the outer part of the cluster ($r>\Rfive$) with respect to~\cite{arnaud2010} found by~\cite{pipv}. For reference, the third row of Fig.~\ref{fig:illu_P0} shows harmonic and real space representations for the outer part.

The best-fitting gNFW model to the average profile of the full sample yields $\beta= 6.32 \substack{+0.88 \\ -0.91}$, a value which is higher than $\beta=5.49$, the value for the UPP, although the slope for our profile is shallower in the outer part than the UPP. This result is due to degeneracies between parameters. In fact, the  external slope is not determined only by $\beta$, but by a combination of $c_{500}$ and $\beta$, as can be seen in the first column of Fig.~\ref{fig:illu_P0}. The degeneracy between $c_{500}$ and $\beta$ for our profile can clearly be seen in Fig.~\ref{fig:gnfw_parameters500}. Imposing the UPP value $c_{500}=1.18$ would lead to a best-fitting value for $\beta$ lower than 5.49, as expected for a shallower slope. 

We now look for a possible dependence of the outer slope with cluster mass, focussing on the $c_{500}$ versus $\beta$ plot in the right hand panel of Fig.~\ref{fig:press_prof_evolmass}. The contours for the low mass and high mass subsamples are not compatible, indicating a dependence of the external slope on mass. This can also be seen to some extent in the left hand panel of Fig.~\ref{fig:press_prof_evolmass}, but the data points of the low mass subsample have large uncertainties at $l \thetaf < 10^3  \, {\rm arcmin}$.

%--------
% R200m
%
\begin{figure*}
\centering
\includegraphics[width=\columnwidth]{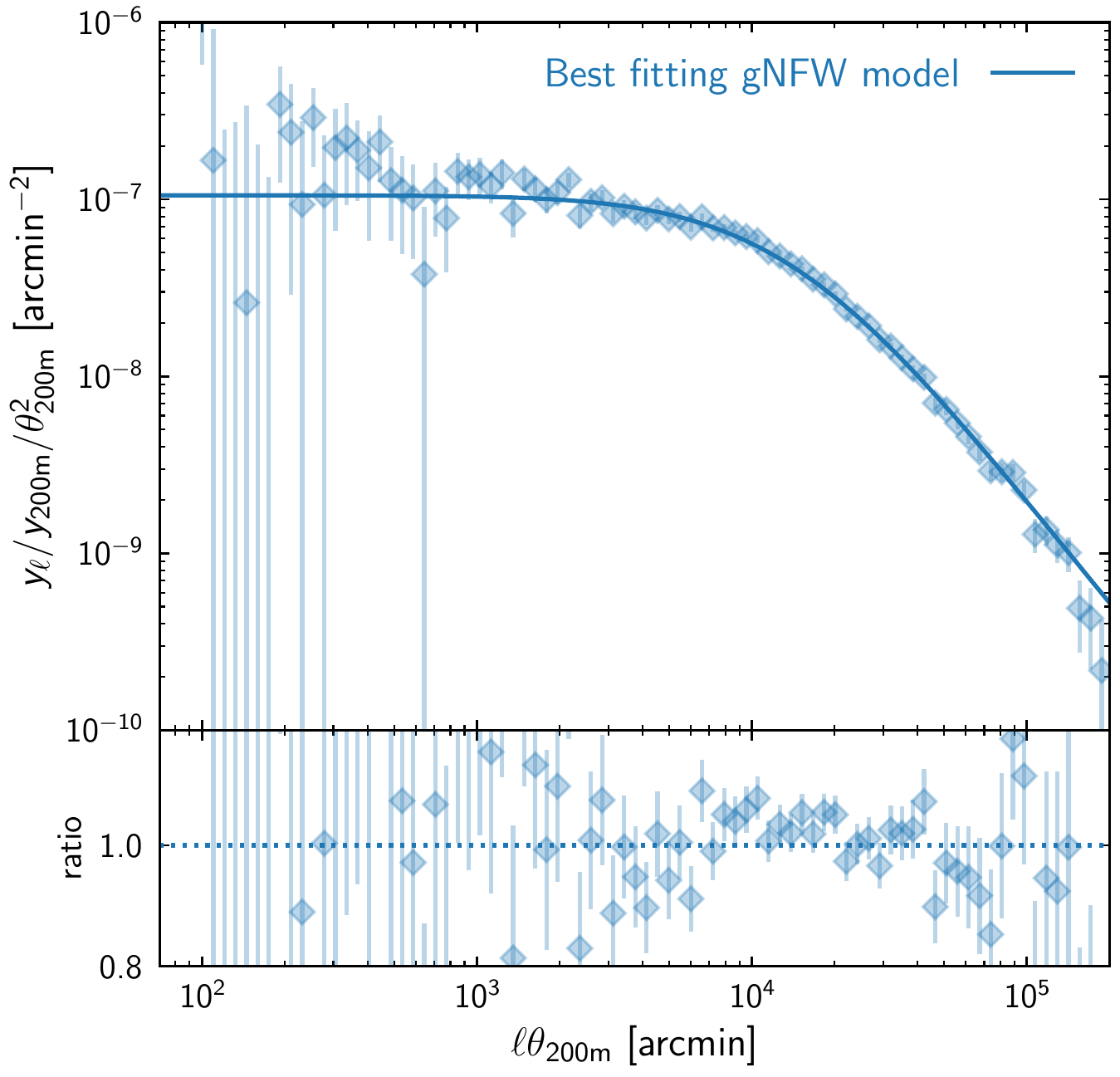}
\hfill
\includegraphics[width=0.955\columnwidth]{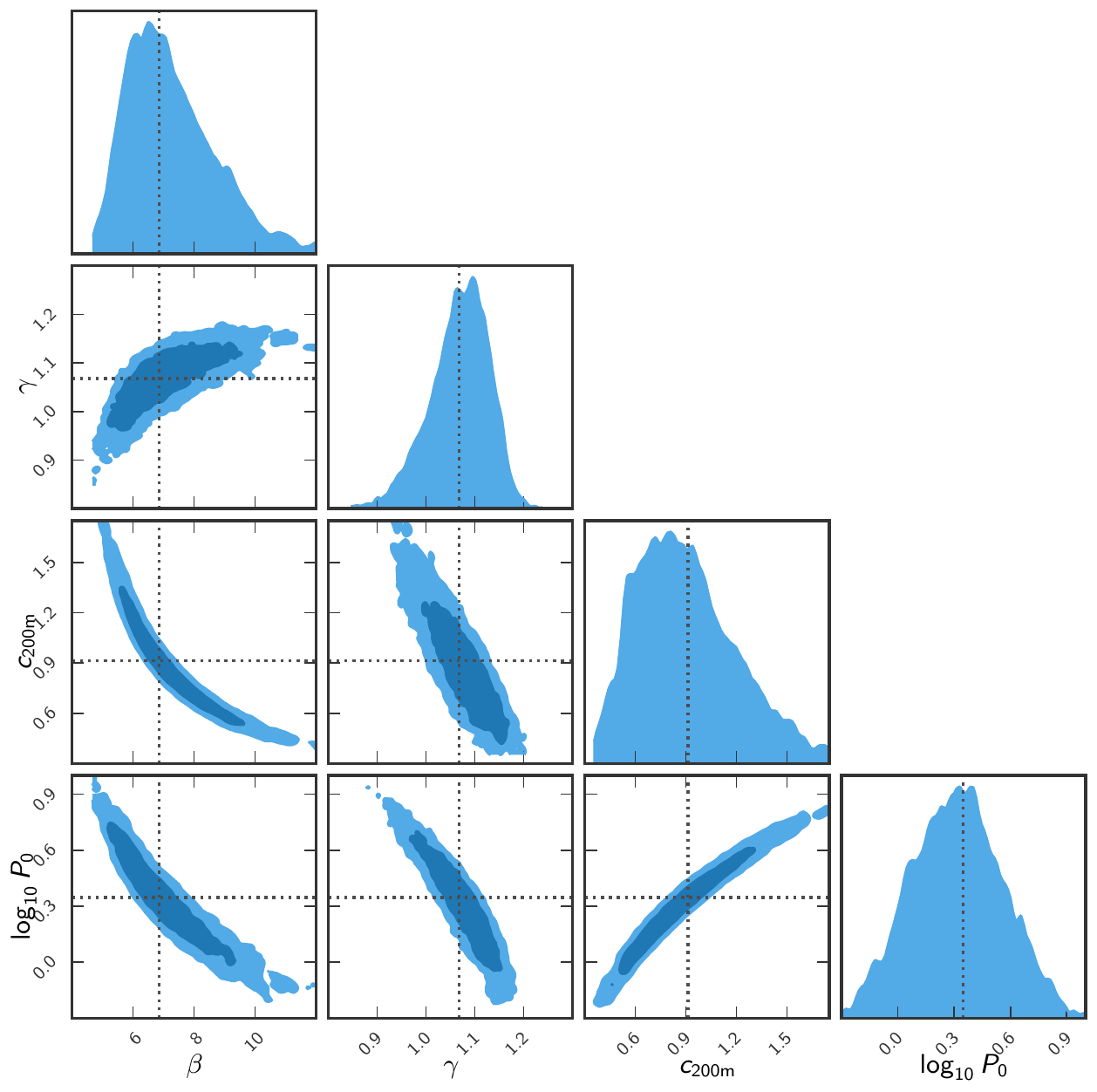}
\caption{\footnotesize Pressure profile scaled by $\theta_{\rm 200m}$ and best fitting model. {\it Left:} Same as Fig.~\ref{fig:press_prof_fit} Left but averaged according to $\theta_{\rm 200m}$ instead of $\thetaf$. {\it Right:} Same as Fig.~\ref{fig:gnfw_parameters500} but for the profile averaged according to $\theta_{\rm 200m}$. Values for the gNFW best-fitting parameters are given in Table~\ref{tab:bestfits200}.}
\label{fig:press_prof_r200m}
\end{figure*}
%-------

\subsubsection{Scaling in $\theta_{\rm 200m}$}
\label{sec:th200m}

\cite{lau2015} found that the self-similarity in the outskirts is better preserved if the profiles are averaged with respect to the mean instead of the critical density of the Universe. We thus performed a new analysis but we rescaled the cluster profiles with $\theta_{\rm 200m}$ instead of $\thetaf$. We converted SPT masses $\Mfive$ to $M_{\rm 200m}$, assuming a NFW profile and using $c_{\rm 200c}$ from \cite{diemer2015}. We did not include the renormalisation factor 0.8 on $\Mfive$ since the profile rescaled with $R_{\rm 200 m}$ will not be compared to other works with different mass definition. We renormalised the profiles by $y_{\rm 200 m}\theta_{\rm 200 m}^2$ instead of $y_{500}\thetaf^2$, and we rescaled the radial quantities by $\theta_{\rm 200 m}$ instead of $\thetaf$. Details of the derivation of $y_{\rm 200 m}$ and associated characteristic pressure $P_{e,\rm 200 m}$ are provided in Appendix~\ref{app:selfsim}. The average rescaled harmonic transform Compton parameter profile is shown in Fig.~\ref{fig:press_prof_r200m} and the values for the best gNFW fit are given in Table~\ref{tab:bestfits200}. The fit is acceptable, with a reduced $\chi^2/{\rm d.o.f} = 1.42$.

We provide the tabulated best-fitting profile and associated error envelope in harmonic and real space for the profile scaled by $\theta_{\rm 200m}$ in Table~\ref{tab:prof_t200m}.

\begin{table}
        \centering
        \caption{\footnotesize Best-fitting parameters (given by the maximum of the likelihood) for the full sample when the profile is rescaled by $\theta_{\rm 200m}$. Errors are the 16 and 84 percentiles. $\alpha=1.05$ is fixed. $\log$ is the decimal logarithm.}
	\label{tab:bestfits200}
	\begin{tabular}{c c c c c c}
\toprule
\toprule
	 $\log(P_0)$ & $c_{\rm 200m}$ & $\gamma$ & $\beta$ & $\alpha$ \\
\midrule
          ${\bf 0.35} \substack{{\bf +0.21}\\ {\bf -0.32}}$ & ${\bf 0.91} \substack{{\bf +0.19} \\ \bf {-0.39}}$ & ${\bf 1.07} \substack{{\bf +0.07} \\ {\bf -0.05}}$ & ${\bf 6.86} \substack{{\bf +1.28} \\ {\bf -1.40}}$   & {\bf 1.05} \\
\bottomrule
       \end{tabular}
\end{table}

\subsubsection{Can we detect an accretion shock?}
\label{sec:shock} 

\begin{figure*}
\centering
\includegraphics[width=0.49\hsize]{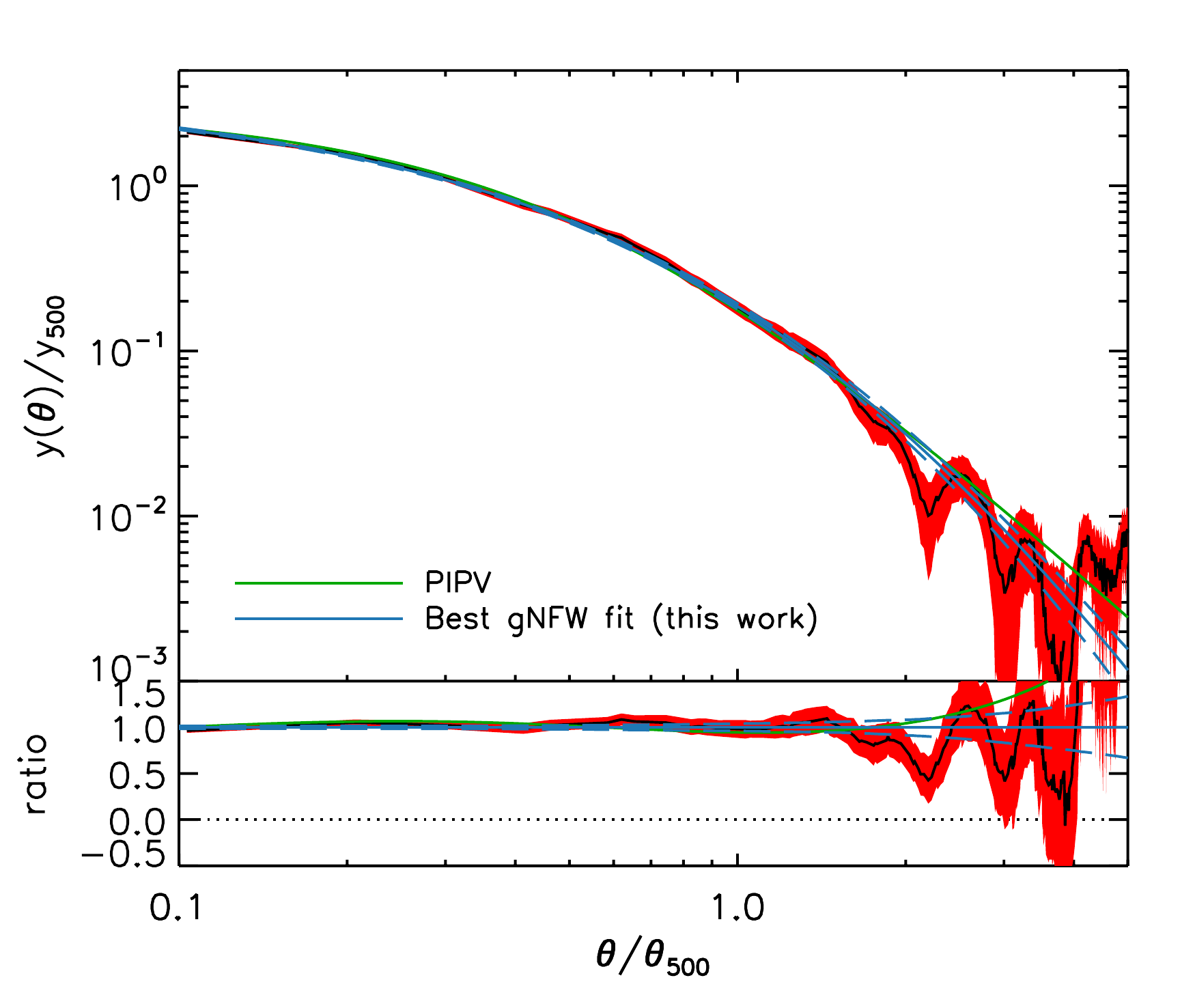}
\includegraphics[width=0.49\hsize]{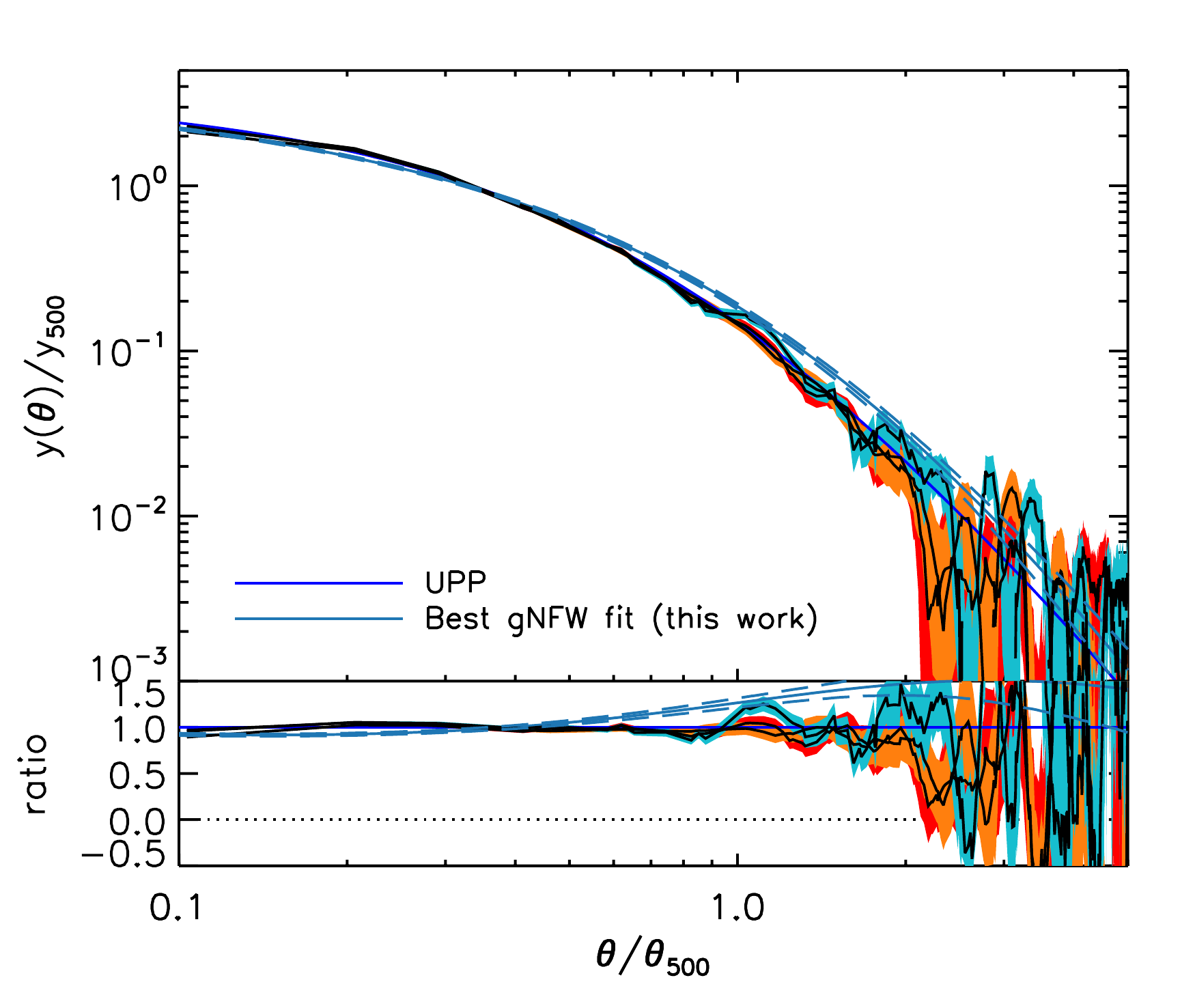}
\caption{\footnotesize Reconstructed Compton profiles (2D real space). {\it Left:} Reconstructed Compton profile (2D real space) from the data points of Fig.~\ref{fig:press_prof_fit} left. The 68\% C.L. envelope is shown in red and the mean in black. The solid green line is the PIPV profile. The solid blue line is the best gNFW fit. The dashed blue lines delineate the 68\% C.L. of the gNFW fit. In the bottom panel, the ratio is computed with respect to the best fit. {\it Right:} Same as the left panel, but with simulated UPP profiles injected into actual data, extracted and reconstructed using the same method as on the real clusters. We performed three injections of simulated SPT clusters at different locations. The reconstructed profiles are in red, orange and cyan. The solid bright blue line is the UPP.}
\label{fig:press_prof_outskirts_recon}
\end{figure*}

To investigate possible accretion features in the outskirts, we reconstructed the 2D real space Compton profile directly from the data points of the left hand panel of Fig.~\ref{fig:press_prof_fit}. This was undertaken by drawing a Gaussian realisation with a mean corresponding to the data points and correlations from the covariance matrix $\mathcal{S}$. We then projected this realisation into a two dimensional map and performed the inverse Fourier transform. This provides one realisation of a possible 2D real space profile. We performed 1,000 such random realisations, projections and inverse Fourier transforms. The mean of the 2D real space profiles is shown as a black line in the left hand panel of  Fig.~\ref{fig:press_prof_outskirts_recon}, and the region encompassing the raw standard deviation of the realisations is shown by the red envelope. The solid green line is the PIPV profile and the blue solid and dashed lines show the best fitting gNFW model and its 68\% confidence envelope. 

The reconstructed profile follows closely the gNFW fit up to $\theta \sim 2 \thetaf$, and does not show a single feature. Instead, multiple pressure drops are visible at $\theta > 2 \thetaf$. To test if these multiple pressure drops are a significant feature of the reconstructed 2D Compton profile, we performed exactly the same reconstruction but for simulated SPT clusters injected at random locations. For this test, we adopted the UPP, and performed three injections of the full simulated SPT sample. Results for the simulated clusters are shown in the right hand panel of  Fig.~\ref{fig:press_prof_outskirts_recon} right, where the three injections are shown in red, orange and cyan. The test shows that the UPP is well reconstructed up to $\theta = 2 \thetaf$. However, at higher $\theta$, the reconstructed profiles also display pressure drops. These features are not simulated in the injected profiles, which are smooth. They are thus most probably due to the noise in the outer part of the profile in the current data sets.

According to recent simulation studies~\citep{aung2021,baxter2021,zhang2021}, a single accretion shock is expected in the average profile of relaxed clusters in the radial range $2<r<2.5 R_{\rm 200m}$. \cite{zenteno2020} studied the dynamical state of 288 SPT-SZ clusters in the Dark Energy Survey footprint and provided a list of 43 disturbed and 41 relaxed clusters. Among the 43 disturbed (41 relaxed) clusters, all (39) are included in our sample. We measured the average profile for each of the subsamples. The two profiles are unfortunately compatible within their error bars due to the small size of the subsamples.

In conclusion, we do not see any obvious single accretion shock in the outskirts of the average pressure profile. This is most probably due to the fact that we are close to the noise floor in the outer parts of the profile with the current data sets. The fact that we cannot separate the clusters in our sample according to their dynamical state makes any such study even more difficult.
In future studies, it will thus be crucial to work with larger samples, including clusters with well characterised dynamical sates, and with deeper data.

\section{Summary and conclusions}
\label{sec:sum}

We measured the average pressure profile of 461 SPT clusters in the \Planck\  and SPT-SZ data jointly. The size of the sample is about ten times larger than in previous studies~\citep{pipv,sayers2013,sayers2016,pointecouteau2021}, which used tens of clusters. The optimal combination of space- and ground-based data allowed to constrain simultaneously the pressure profile at small and large scales. For the first time, we proposed to work in Fourier space, and provided results on the average profile in harmonic instead of real space. This approach greatly simplifies PSF deconvolutions and the rescaling of individual profiles before averaging.
We obtained very competitive constraints on the cluster pressure profile based on SZ data only, without the need for including X-ray data to constrain the inner part or simulations to constrain the outer part. Our main results are summarised below.

\begin{itemize}

\item The Malmquist bias of the SPT clusters in the SPT-SZ maps must be corrected for before combining \Planck\ and SPT-SZ maps. 
\item The average pressure profile measured from the joint \Planck+SPT-SZ dataset is in excellent agreement with the UPP~\citep{arnaud2010} in the inner part and with the PIPV~\citep{pipv} profile in the outer part. A gNFW profile provides a good fit to the data. Fixing  $\alpha$ to the UPP value, we were able to constrain four of the gNFW parameters ($\gamma$, $c_{500}$, $\beta$, $P_0$), for the first time in harmonic space from SZ data only.
\item We split the sample in two low and high redshift subsamples. The resulting profiles are compatible, thus showing no indication of evolution with redshift.
\item Splitting the sample in two low and high mass subsamples, we find inconsistent profiles. This inconsistency may be of physical origin, with lower mass clusters being on average less peaked in the centre, or may be a residual systematic effect of the analysis where the Malmquist bias being underestimated for the low mass subsample.
\item We reconstructed the two dimensional Compton parameter profile in real space to study the outskirts. We found multiple pressure drops at $\theta>\thetaf$. Testing this result by injection of smooth simulated cluster samples into real maps, we conclude that these pressure drops are in fact most probably due our being close to the noise floor in the outer parts of the profile with the current data sets. We lack both sensitivity and information on the dynamical state of the clusters in the SPT sample to probe this further.
\end{itemize}

For practical applications, the average scaled pressure profile from the combination of \Planck\ and SPT-SZ observations in units of $\Rfive$ is given by Eq.~\ref{eq:gnfw500}, with best-fitting parameters given in bold in Table~\ref{tab:bestfits500}. For clusters of given mass $\Mfive$ and redshift $z$, the physical pressure profile can then be derived from Eq.~\ref{eq:p500g}. The equivalent gNFW parameters for the scaled pressure profile in units of $R_{\rm 200m}$ are given in Table~\ref{tab:bestfits200}, which can be used in conjunction with Eq.~\ref{eq:p200m} to obtain the physical pressure profile as a function of $M_{\rm200m}$ and $z$. The MCMC chains corresponding to the best fitting model to the average profiles, scaled as a function of $\Rfive$ and $R_{\rm 200m}$, are available on request to the authors. Finally, the tabulated best-fitting profiles and associated error envelopes are provided in Appendix~\ref{app:envel}.

The harmonic method presented in the work is promising. Upcoming and future ground-based experiments, such as the Simons Observatory~\citep[SO,][]{ade2019} and CMB-S4~\citep{abazajian2019}, will detect around 15,000 and 100,000 objects, respectively. Averaging the cluster profiles in bins of redshift and/or mass in SO/CMB-S4+\Planck\ will allow the study of the cluster pressure profile and its evolution to  unprecedented precision.

\begin{acknowledgements}
We thank the A\&A referee, Shutaro Ueda, for useful suggestions and corrections, which helped to clarify and improve this article. We thank Monique Arnaud for valuable discussions at the beginning of this work, and for comments on the final draft of this article. We thank Rémi Adam, Stefano Ettori, Fabio Gastaldello and Mariachiara Rossetti for useful suggestions while presenting a first version of this work at the NIKA2 conference in June 2021, and Colin Hill, Nick Battaglia and Daisuke Nagai for useful suggestions while presenting a second version of this work at the SZ workshop (Center for Computational Astrophysics, Flatiron Institute) in June 2022. We also thank Etienne Pointecouteau for clarifications and discussions about the PACT article. We acknowledge the use of the Legacy Archive for Microwave Background Data Analysis (LAMBDA), part of the High Energy Astrophysics Science Archive Center (HEASARC). HEASARC/LAMBDA is a service of the Astrophysics Science Division at the NASA Goddard Space Flight Center. We also acknowledge the use of the \Planck\ Legacy Archive. We used the HEALPix software~\citep{gorski2005} available at \url{https://healpix.sourceforge.io} and the WebPlotDigitizer by Ankit Rohatgi. GWP acknowledges financial support from CNES, the French space agency.
\end{acknowledgements}

\bibliographystyle{aa}
\bibliography{sptszprofile}

\begin{thebibliography}{58}
\expandafter\ifx\csname natexlab\endcsname\relax\def\natexlab#1{#1}\fi

\bibitem[{{Abazajian} {et~al.}(2019){Abazajian}, {Addison}, {Adshead}, {Ahmed},
  {Allen}, {Alonso}, {Alvarez}, {Anderson}, {Arnold}, {Baccigalupi}, {Bailey},
  {Barkats}, {Barron}, {Barry}, {Bartlett}, {Basu Thakur}, {Battaglia},
  {Baxter}, {Bean}, {Bebek}, {Bender}, {Benson}, {Berger}, {Bhimani},
  {Bischoff}, {Bleem}, {Bocquet}, {Boddy}, {Bonato}, {Bond}, {Borrill},
  {Bouchet}, {Brown}, {Bryan}, {Burkhart}, {Buza}, {Byrum}, {Calabrese},
  {Calafut}, {Caldwell}, {Carlstrom}, {Carron}, {Cecil}, {Challinor}, {Chang},
  {Chinone}, {Cho}, {Cooray}, {Crawford}, {Crites}, {Cukierman}, {Cyr-Racine},
  {de Haan}, {de Zotti}, {Delabrouille}, {Demarteau}, {Devlin}, {Di Valentino},
  {Dobbs}, {Duff}, {Duivenvoorden}, {Dvorkin}, {Edwards}, {Eimer}, {Errard},
  {Essinger-Hileman}, {Fabbian}, {Feng}, {Ferraro}, {Filippini}, {Flauger},
  {Flaugher}, {Fraisse}, {Frolov}, {Galitzki}, {Galli}, {Ganga}, {Gerbino},
  {Gilchriese}, {Gluscevic}, {Green}, {Grin}, {Grohs}, {Gualtieri}, {Guarino},
  {Gudmundsson}, {Habib}, {Haller}, {Halpern}, {Halverson}, {Hanany},
  {Harrington}, {Hasegawa}, {Hasselfield}, {Hazumi}, {Heitmann}, {Henderson},
  {Henning}, {Hill}, {Hlozek}, {Holder}, {Holzapfel}, {Hubmayr},
  {Huffenberger}, {Huffer}, {Hui}, {Irwin}, {Johnson}, {Johnstone}, {Jones},
  {Karkare}, {Katayama}, {Kerby}, {Kernovsky}, {Keskitalo}, {Kisner}, {Knox},
  {Kosowsky}, {Kovac}, {Kovetz}, {Kuhlmann}, {Kuo}, {Kurita}, {Kusaka},
  {Lahteenmaki}, {Lawrence}, {Lee}, {Lewis}, {Li}, {Linder}, {Loverde},
  {Lowitz}, {Madhavacheril}, {Mantz}, {Matsuda}, {Mauskopf}, {McMahon},
  {McQuinn}, {Meerburg}, {Melin}, {Meyers}, {Millea}, {Mohr}, {Moncelsi},
  {Mroczkowski}, {Mukherjee}, {M{\"u}nchmeyer}, {Nagai}, {Nagy}, {Namikawa},
  {Nati}, {Natoli}, {Negrello}, {Newburgh}, {Niemack}, {Nishino}, {Nordby},
  {Novosad}, {O'Connor}, {Obied}, {Padin}, {Pandey}, {Partridge}, {Pierpaoli},
  {Pogosian}, {Pryke}, {Puglisi}, {Racine}, {Raghunathan}, {Rahlin},
  {Rajagopalan}, {Raveri}, {Reichanadter}, {Reichardt}, {Remazeilles}, {Rocha},
  {Roe}, {Roy}, {Ruhl}, {Salatino}, {Saliwanchik}, {Schaan}, {Schillaci},
  {Schmittfull}, {Scott}, {Sehgal}, {Shandera}, {Sheehy}, {Sherwin},
  {Shirokoff}, {Simon}, {Slosar}, {Somerville}, {Spergel}, {Staggs}, {Stark},
  {Stompor}, {Story}, {Stoughton}, {Suzuki}, {Tajima}, {Teply}, {Thompson},
  {Timbie}, {Tomasi}, {Treu}, {Tristram}, {Tucker}, {Umilt{\`a}}, {van
  Engelen}, {Vieira}, {Vieregg}, {Vogelsberger}, {Wang}, {Watson}, {White},
  {Whitehorn}, {Wollack}, {Kimmy Wu}, {Xu}, {Yasini}, {Yeck}, {Yoon}, {Young},
  \& {Zonca}}]{abazajian2019}
{Abazajian}, K., {Addison}, G., {Adshead}, P., {et~al.} 2019, arXiv e-prints,
  arXiv:1907.04473

\bibitem[{{Ade} {et~al.}(2019){Ade}, {Aguirre}, {Ahmed}, {Aiola}, {Ali},
  {Alonso}, {Alvarez}, {Arnold}, {Ashton}, {Austermann}, {Awan}, {Baccigalupi},
  {Baildon}, {Barron}, {Battaglia}, {Battye}, {Baxter}, {Bazarko}, {Beall},
  {Bean}, {Beck}, {Beckman}, {Beringue}, {Bianchini}, {Boada}, {Boettger},
  {Bond}, {Borrill}, {Brown}, {Bruno}, {Bryan}, {Calabrese}, {Calafut},
  {Calisse}, {Carron}, {Challinor}, {Chesmore}, {Chinone}, {Chluba}, {Cho},
  {Choi}, {Coppi}, {Cothard}, {Coughlin}, {Crichton}, {Crowley}, {Crowley},
  {Cukierman}, {D'Ewart}, {D{\"u}nner}, {de Haan}, {Devlin}, {Dicker},
  {Didier}, {Dobbs}, {Dober}, {Duell}, {Duff}, {Duivenvoorden}, {Dunkley},
  {Dusatko}, {Errard}, {Fabbian}, {Feeney}, {Ferraro}, {Flux{\`a}}, {Freese},
  {Frisch}, {Frolov}, {Fuller}, {Fuzia}, {Galitzki}, {Gallardo}, {Tomas Galvez
  Ghersi}, {Gao}, {Gawiser}, {Gerbino}, {Gluscevic}, {Goeckner-Wald}, {Golec},
  {Gordon}, {Gralla}, {Green}, {Grigorian}, {Groh}, {Groppi}, {Guan},
  {Gudmundsson}, {Han}, {Hargrave}, {Hasegawa}, {Hasselfield}, {Hattori},
  {Haynes}, {Hazumi}, {He}, {Healy}, {Henderson}, {Hervias-Caimapo}, {Hill},
  {Hill}, {Hilton}, {Hilton}, {Hincks}, {Hinshaw}, {Hlo{\v{z}}ek}, {Ho}, {Ho},
  {Howe}, {Huang}, {Hubmayr}, {Huffenberger}, {Hughes}, {Ijjas}, {Ikape},
  {Irwin}, {Jaffe}, {Jain}, {Jeong}, {Kaneko}, {Karpel}, {Katayama}, {Keating},
  {Kernasovskiy}, {Keskitalo}, {Kisner}, {Kiuchi}, {Klein}, {Knowles},
  {Koopman}, {Kosowsky}, {Krachmalnicoff}, {Kuenstner}, {Kuo}, {Kusaka},
  {Lashner}, {Lee}, {Lee}, {Leon}, {Leung}, {Lewis}, {Li}, {Li}, {Limon},
  {Linder}, {Lopez-Caraballo}, {Louis}, {Lowry}, {Lungu}, {Madhavacheril},
  {Mak}, {Maldonado}, {Mani}, {Mates}, {Matsuda}, {Maurin}, {Mauskopf}, {May},
  {McCallum}, {McKenney}, {McMahon}, {Meerburg}, {Meyers}, {Miller},
  {Mirmelstein}, {Moodley}, {Munchmeyer}, {Munson}, {Naess}, {Nati},
  {Navaroli}, {Newburgh}, {Nguyen}, {Niemack}, {Nishino}, {Orlowski-Scherer},
  {Page}, {Partridge}, {Peloton}, {Perrotta}, {Piccirillo}, {Pisano},
  {Poletti}, {Puddu}, {Puglisi}, {Raum}, {Reichardt}, {Remazeilles},
  {Rephaeli}, {Riechers}, {Rojas}, {Roy}, {Sadeh}, {Sakurai}, {Salatino},
  {Sathyanarayana Rao}, {Schaan}, {Schmittfull}, {Sehgal}, {Seibert}, {Seljak},
  {Sherwin}, {Shimon}, {Sierra}, {Sievers}, {Sikhosana}, {Silva-Feaver},
  {Simon}, {Sinclair}, {Siritanasak}, {Smith}, {Smith}, {Spergel}, {Staggs},
  {Stein}, {Stevens}, {Stompor}, {Suzuki}, {Tajima}, {Takakura}, {Teply},
  {Thomas}, {Thorne}, {Thornton}, {Trac}, {Tsai}, {Tucker}, {Ullom},
  {Vagnozzi}, {van Engelen}, {Van Lanen}, {Van Winkle}, {Vavagiakis},
  {Verg{\`e}s}, {Vissers}, {Wagoner}, {Walker}, {Ward}, {Westbrook},
  {Whitehorn}, {Williams}, {Williams}, {Wollack}, {Xu}, {Yu}, {Yu}, {Zago},
  {Zhang}, {Zhu}, \& {Simons Observatory Collaboration}}]{ade2019}
{Ade}, P., {Aguirre}, J., {Ahmed}, Z., {et~al.} 2019, \jcap, 2019, 056

\bibitem[{{Anbajagane} {et~al.}(2022){Anbajagane}, {Chang}, {Jain}, {Adhikari},
  {Baxter}, {Benson}, {Bleem}, {Bocquet}, {Calzadilla}, {Carlstrom}, {Chang},
  {Chown}, {Crawford}, {Crites}, {Cui}, {de Haan}, {Di Mascolo}, {Dobbs},
  {Everett}, {George}, {Grandis}, {Halverson}, {Holder}, {Holzapfel}, {Hrubes},
  {Lee}, {Luong-Van}, {McDonald}, {McMahon}, {Meyer}, {Millea}, {Mocanu},
  {Mohr}, {Natoli}, {Omori}, {Padin}, {Pryke}, {Reichardt}, {Ruhl}, {Saro},
  {Schaffer}, {Shirokoff}, {Staniszewski}, {Stark}, {Vieira}, \&
  {Williamson}}]{anbajagane2022}
{Anbajagane}, D., {Chang}, C., {Jain}, B., {et~al.} 2022, \mnras, 514, 1645

\bibitem[{{Andreon} {et~al.}(2023){Andreon}, {Romero}, {Aussel}, {Bhandarkar},
  {Devlin}, {Dicker}, {Ladjelate}, {Lowe}, {Mason}, {Mroczkowski}, {Raichoor},
  {Sarazin}, \& {Trinchieri}}]{andreon2023}
{Andreon}, S., {Romero}, C., {Aussel}, H., {et~al.} 2023, \mnras, 522, 4301

\bibitem[{{Andreon} {et~al.}(2021){Andreon}, {Romero}, {Castagna}, {Ragagnin},
  {Devlin}, {Dicker}, {Mason}, {Mroczkowski}, {Sarazin}, {Sievers}, \&
  {Stanchfield}}]{andreon2021}
{Andreon}, S., {Romero}, C., {Castagna}, F., {et~al.} 2021, \mnras, 505, 5896

\bibitem[{{Arnaud} {et~al.}(2010){Arnaud}, {Pratt}, {Piffaretti},
  {B{\"o}hringer}, {Croston}, \& {Pointecouteau}}]{arnaud2010}
{Arnaud}, M., {Pratt}, G.~W., {Piffaretti}, R., {et~al.} 2010, \aap, 517, A92

\bibitem[{{Aung} {et~al.}(2021){Aung}, {Nagai}, \& {Lau}}]{aung2021}
{Aung}, H., {Nagai}, D., \& {Lau}, E.~T. 2021, \mnras, 508, 2071

\bibitem[{{Battaglia} {et~al.}(2012){Battaglia}, {Bond}, {Pfrommer}, \&
  {Sievers}}]{battaglia2012}
{Battaglia}, N., {Bond}, J.~R., {Pfrommer}, C., \& {Sievers}, J.~L. 2012, \apj,
  758, 75

\bibitem[{{Baxter} {et~al.}(2021){Baxter}, {Adhikari}, {Vega-Ferrero}, {Cui},
  {Chang}, {Jain}, \& {Knebe}}]{baxter2021}
{Baxter}, E.~J., {Adhikari}, S., {Vega-Ferrero}, J., {et~al.} 2021, \mnras,
  508, 1777

\bibitem[{{Bleem} {et~al.}(2022){Bleem}, {Crawford}, {Ansarinejad}, {Benson},
  {Bocquet}, {Carlstrom}, {Chang}, {Chown}, {Crites}, {Haan}, {Dobbs},
  {Everett}, {George}, {Gualtieri}, {Halverson}, {Holder}, {Holzapfel},
  {Hrubes}, {Knox}, {Lee}, {Luong-Van}, {Marrone}, {McMahon}, {Meyer},
  {Millea}, {Mocanu}, {Mohr}, {Natoli}, {Omori}, {Padin}, {Pryke},
  {Raghunathan}, {Reichardt}, {Ruhl}, {Schaffer}, {Shirokoff}, {Staniszewski},
  {Stark}, {Vieira}, \& {Williamson}}]{bleem2022}
{Bleem}, L.~E., {Crawford}, T.~M., {Ansarinejad}, B., {et~al.} 2022, \apjs,
  258, 36

\bibitem[{{Bleem} {et~al.}(2015){Bleem}, {Stalder}, {de Haan}, {Aird}, {Allen},
  {Applegate}, {Ashby}, {Bautz}, {Bayliss}, {Benson}, {Bocquet}, {Brodwin},
  {Carlstrom}, {Chang}, {Chiu}, {Cho}, {Clocchiatti}, {Crawford}, {Crites},
  {Desai}, {Dietrich}, {Dobbs}, {Foley}, {Forman}, {George}, {Gladders},
  {Gonzalez}, {Halverson}, {Hennig}, {Hoekstra}, {Holder}, {Holzapfel},
  {Hrubes}, {Jones}, {Keisler}, {Knox}, {Lee}, {Leitch}, {Liu}, {Lueker},
  {Luong-Van}, {Mantz}, {Marrone}, {McDonald}, {McMahon}, {Meyer}, {Mocanu},
  {Mohr}, {Murray}, {Padin}, {Pryke}, {Reichardt}, {Rest}, {Ruel}, {Ruhl},
  {Saliwanchik}, {Saro}, {Sayre}, {Schaffer}, {Schrabback}, {Shirokoff},
  {Song}, {Spieler}, {Stanford}, {Staniszewski}, {Stark}, {Story}, {Stubbs},
  {Vanderlinde}, {Vieira}, {Vikhlinin}, {Williamson}, {Zahn}, \&
  {Zenteno}}]{bleem2015}
{Bleem}, L.~E., {Stalder}, B., {de Haan}, T., {et~al.} 2015, \apjs, 216, 27

\bibitem[{{B{\"o}hringer} {et~al.}(2007){B{\"o}hringer}, {Schuecker}, {Pratt},
  {Arnaud}, {Ponman}, {Croston}, {Borgani}, {Bower}, {Briel}, {Collins},
  {Donahue}, {Forman}, {Finoguenov}, {Geller}, {Guzzo}, {Henry}, {Kneissl},
  {Mohr}, {Matsushita}, {Mullis}, {Ohashi}, {Pedersen}, {Pierini}, {Quintana},
  {Raychaudhury}, {Reiprich}, {Romer}, {Rosati}, {Sabirli}, {Temple}, {Viana},
  {Vikhlinin}, {Voit}, \& {Zhang}}]{boehringer2007}
{B{\"o}hringer}, H., {Schuecker}, P., {Pratt}, G.~W., {et~al.} 2007, \aap, 469,
  363

\bibitem[{{Bonamente} {et~al.}(2012){Bonamente}, {Hasler}, {Bulbul},
  {Carlstrom}, {Culverhouse}, {Gralla}, {Greer}, {Hawkins}, {Hennessy}, {Joy},
  {Kolodziejczak}, {Lamb}, {Landry}, {Leitch}, {Marrone}, {Miller},
  {Mroczkowski}, {Muchovej}, {Plagge}, {Pryke}, {Sharp}, \&
  {Woody}}]{bonamente2012}
{Bonamente}, M., {Hasler}, N., {Bulbul}, E., {et~al.} 2012, New Journal of
  Physics, 14, 025010

\bibitem[{{Bulbul} {et~al.}(2019){Bulbul}, {Chiu}, {Mohr}, {McDonald},
  {Benson}, {Bautz}, {Bayliss}, {Bleem}, {Brodwin}, {Bocquet}, {Capasso},
  {Dietrich}, {Forman}, {Hlavacek-Larrondo}, {Holzapfel}, {Khullar}, {Klein},
  {Kraft}, {Miller}, {Reichardt}, {Saro}, {Sharon}, {Stalder}, {Schrabback}, \&
  {Stanford}}]{bulbul2019}
{Bulbul}, E., {Chiu}, I.~N., {Mohr}, J.~J., {et~al.} 2019, \apj, 871, 50

\bibitem[{{Chown} {et~al.}(2018){Chown}, {Omori}, {Aylor}, {Benson}, {Bleem},
  {Carlstrom}, {Chang}, {Cho}, {Crawford}, {Crites}, {de Haan}, {Dobbs},
  {Everett}, {George}, {Henning}, {Halverson}, {Harrington}, {Holder},
  {Holzapfel}, {Hou}, {Hrubes}, {Knox}, {Lee}, {Luong-Van}, {Marrone},
  {McMahon}, {Meyer}, {Millea}, {Mocanu}, {Mohr}, {Natoli}, {Padin}, {Pryke},
  {Reichardt}, {Ruhl}, {Sayre}, {Schaffer}, {Shirokoff}, {Simard},
  {Staniszewski}, {Stark}, {Story}, {Vanderlinde}, {Vieira}, {Williamson},
  {Wu}, \& {South Pole Telescope Collaboration}}]{chown2018}
{Chown}, R., {Omori}, Y., {Aylor}, K., {et~al.} 2018, \apjs, 239, 10

\bibitem[{{Diemer} \& {Kravtsov}(2015)}]{diemer2015}
{Diemer}, B. \& {Kravtsov}, A.~V. 2015, \apj, 799, 108

\bibitem[{{Foreman-Mackey} {et~al.}(2013){Foreman-Mackey}, {Hogg}, {Lang}, \&
  {Goodman}}]{foreman2013}
{Foreman-Mackey}, D., {Hogg}, D.~W., {Lang}, D., \& {Goodman}, J. 2013, \pasp,
  125, 306

\bibitem[{{Ghirardini} {et~al.}(2018){Ghirardini}, {Ettori}, {Eckert},
  {Molendi}, {Gastaldello}, {Pointecouteau}, {Hurier}, \&
  {Bourdin}}]{ghirardini2018}
{Ghirardini}, V., {Ettori}, S., {Eckert}, D., {et~al.} 2018, \aap, 614, A7

\bibitem[{{G{\'o}rski} {et~al.}(2005){G{\'o}rski}, {Hivon}, {Banday}, {Wand
  elt}, {Hansen}, {Reinecke}, \& {Bartelmann}}]{gorski2005}
{G{\'o}rski}, K.~M., {Hivon}, E., {Banday}, A.~J., {et~al.} 2005, \apj, 622,
  759

\bibitem[{{He} {et~al.}(2021){He}, {Mansfield}, {Rau}, {Trac}, \&
  {Battaglia}}]{he2021}
{He}, Y., {Mansfield}, P., {Rau}, M.~M., {Trac}, H., \& {Battaglia}, N. 2021,
  \apj, 908, 91

\bibitem[{{Hurier} {et~al.}(2019){Hurier}, {Adam}, \& {Keshet}}]{hurier2019}
{Hurier}, G., {Adam}, R., \& {Keshet}, U. 2019, \aap, 622, A136

\bibitem[{{Itoh} {et~al.}(1998){Itoh}, {Kohyama}, \& {Nozawa}}]{itoh1998}
{Itoh}, N., {Kohyama}, Y., \& {Nozawa}, S. 1998, \apj, 502, 7

\bibitem[{{K{\'e}ruzor{\'e}} {et~al.}(2020){K{\'e}ruzor{\'e}}, {Mayet},
  {Pratt}, {Adam}, {Ade}, {Andr{\'e}}, {Andrianasolo}, {Arnaud}, {Aussel},
  {Bartalucci}, {Beelen}, {Beno{\^\i}t}, {Berta}, {Bourrion}, {Calvo},
  {Catalano}, {De Petris}, {D{\'e}sert}, {Doyle}, {Driessen}, {Gomez}, {Goupy},
  {Kramer}, {Ladjelate}, {Lagache}, {Leclercq}, {Lestrade},
  {Mac{\'\i}as-P{\'e}rez}, {Mauskopf}, {Monfardini}, {Perotto}, {Pisano},
  {Pointecouteau}, {Ponthieu}, {Rev{\'e}ret}, {Ritacco}, {Romero}, {Roussel},
  {Ruppin}, {Schuster}, {Shu}, {Sievers}, \& {Tucker}}]{keruzore2020}
{K{\'e}ruzor{\'e}}, F., {Mayet}, F., {Pratt}, G.~W., {et~al.} 2020, \aap, 644,
  A93

\bibitem[{{Lau} {et~al.}(2015){Lau}, {Nagai}, {Avestruz}, {Nelson}, \&
  {Vikhlinin}}]{lau2015}
{Lau}, E.~T., {Nagai}, D., {Avestruz}, C., {Nelson}, K., \& {Vikhlinin}, A.
  2015, \apj, 806, 68

\bibitem[{{McCarthy} {et~al.}(2014){McCarthy}, {Le Brun}, {Schaye}, \&
  {Holder}}]{mccarthy2014}
{McCarthy}, I.~G., {Le Brun}, A.~M.~C., {Schaye}, J., \& {Holder}, G.~P. 2014,
  \mnras, 440, 3645

\bibitem[{{Melin} {et~al.}(2021){Melin}, {Bartlett}, {Tarr{\'\i}o}, \&
  {Pratt}}]{melin2021}
{Melin}, J.~B., {Bartlett}, J.~G., {Tarr{\'\i}o}, P., \& {Pratt}, G.~W. 2021,
  \aap, 647, A106

\bibitem[{{Molnar} {et~al.}(2009){Molnar}, {Hearn}, {Haiman}, {Bryan},
  {Evrard}, \& {Lake}}]{molnar2009}
{Molnar}, S.~M., {Hearn}, N., {Haiman}, Z., {et~al.} 2009, \apj, 696, 1640

\bibitem[{{Mroczkowski} {et~al.}(2009){Mroczkowski}, {Bonamente}, {Carlstrom},
  {Culverhouse}, {Greer}, {Hawkins}, {Hennessy}, {Joy}, {Lamb}, {Leitch},
  {Loh}, {Maughan}, {Marrone}, {Miller}, {Muchovej}, {Nagai}, {Pryke}, {Sharp},
  \& {Woody}}]{mroczkowski2009}
{Mroczkowski}, T., {Bonamente}, M., {Carlstrom}, J.~E., {et~al.} 2009, \apj,
  694, 1034

\bibitem[{{Nagai} {et~al.}(2007){Nagai}, {Kravtsov}, \&
  {Vikhlinin}}]{nagai2007}
{Nagai}, D., {Kravtsov}, A.~V., \& {Vikhlinin}, A. 2007, \apj, 668, 1

\bibitem[{{Oppizzi} {et~al.}(2022){Oppizzi}, {De Luca}, {Bourdin}, {Mazzotta},
  {Ettori}, {Gastaldello}, {Kay}, {Lovisari}, {Maughan}, {Pointecouteau},
  {Pratt}, {Rossetti}, {Sayers}, \& {Sereno}}]{oppizzi2022}
{Oppizzi}, F., {De Luca}, F., {Bourdin}, H., {et~al.} 2022, arXiv e-prints,
  arXiv:2209.09601

\bibitem[{{Plagge} {et~al.}(2010){Plagge}, {Benson}, {Ade}, {Aird}, {Bleem},
  {Carlstrom}, {Chang}, {Cho}, {Crawford}, {Crites}, {de Haan}, {Dobbs},
  {George}, {Hall}, {Halverson}, {Holder}, {Holzapfel}, {Hrubes}, {Joy},
  {Keisler}, {Knox}, {Lee}, {Leitch}, {Lueker}, {Marrone}, {McMahon}, {Mehl},
  {Meyer}, {Mohr}, {Montroy}, {Padin}, {Pryke}, {Reichardt}, {Ruhl},
  {Schaffer}, {Shaw}, {Shirokoff}, {Spieler}, {Stalder}, {Staniszewski},
  {Stark}, {Vanderlinde}, {Vieira}, {Williamson}, \& {Zahn}}]{plagge2010}
{Plagge}, T., {Benson}, B.~A., {Ade}, P.~A.~R., {et~al.} 2010, \apj, 716, 1118

\bibitem[{{Planck Collaboration Int. V}(2013)}]{pipv}
{Planck Collaboration Int. V}. 2013, \aap, 550, A131

\bibitem[{{Planck Collaboration X}(2011)}]{piffaretti2011}
{Planck Collaboration X}. 2011, \aap, 536, A10

\bibitem[{{Planck Collaboration XXII}(2016)}]{planck_szmap2016}
{Planck Collaboration XXII}. 2016, \aap, 594, A22

\bibitem[{{Planck Collaboration XXVII}(2016)}]{PSZ2}
{Planck Collaboration XXVII}. 2016, \aap, 594, A27

\bibitem[{{Pointecouteau} {et~al.}(2021){Pointecouteau}, {Santiago-Bautista},
  {Douspis}, {Aghanim}, {Crichton}, {Diego}, {Hurier}, {Macias-Perez},
  {Marriage}, {Remazeilles}, {Caretta}, \& {Bravo-Alfaro}}]{pointecouteau2021}
{Pointecouteau}, E., {Santiago-Bautista}, I., {Douspis}, M., {et~al.} 2021,
  \aap, 651, A73

\bibitem[{{Pratt} {et~al.}(2021){Pratt}, {Qu}, \& {Bregman}}]{pratt2021}
{Pratt}, C.~T., {Qu}, Z., \& {Bregman}, J.~N. 2021, \apj, 920, 104

\bibitem[{{Remazeilles} {et~al.}(2011){Remazeilles}, {Delabrouille}, \&
  {Cardoso}}]{remazeilles2011}
{Remazeilles}, M., {Delabrouille}, J., \& {Cardoso}, J.-F. 2011, \mnras, 410,
  2481

\bibitem[{{Romero} {et~al.}(2018){Romero}, {McWilliam},
  {Mac{\'\i}as-P{\'e}rez}, {Adam}, {Ade}, {Andr{\'e}}, {Aussel}, {Beelen},
  {Beno{\^\i}t}, {Bideaud}, {Billot}, {Bourrion}, {Calvo}, {Catalano},
  {Coiffard}, {Comis}, {de Petris}, {D{\'e}sert}, {Doyle}, {Goupy}, {Kramer},
  {Lagache}, {Leclercq}, {Lestrade}, {Mauskopf}, {Mayet}, {Monfardini},
  {Pascale}, {Perotto}, {Pisano}, {Ponthieu}, {Rev{\'e}ret}, {Ritacco},
  {Roussel}, {Ruppin}, {Schuster}, {Sievers}, {Triqueneaux}, {Tucker}, \&
  {Zylka}}]{romero2018}
{Romero}, C., {McWilliam}, M., {Mac{\'\i}as-P{\'e}rez}, J.~F., {et~al.} 2018,
  \aap, 612, A39

\bibitem[{{Romero} {et~al.}(2017){Romero}, {Mason}, {Sayers}, {Mroczkowski},
  {Sarazin}, {Donahue}, {Baldi}, {Clarke}, {Young}, {Sievers}, {Dicker},
  {Reese}, {Czakon}, {Devlin}, {Korngut}, \& {Golwala}}]{romero2017}
{Romero}, C.~E., {Mason}, B.~S., {Sayers}, J., {et~al.} 2017, \apj, 838, 86

\bibitem[{{Rossetti} {et~al.}(2016){Rossetti}, {Gastaldello}, {Ferioli},
  {Bersanelli}, {De Grandi}, {Eckert}, {Ghizzardi}, {Maino}, \&
  {Molendi}}]{rossetti2016}
{Rossetti}, M., {Gastaldello}, F., {Ferioli}, G., {et~al.} 2016, \mnras, 457,
  4515

\bibitem[{{Ruppin} {et~al.}(2017){Ruppin}, {Adam}, {Comis}, {Ade}, {Andr{\'e}},
  {Arnaud}, {Beelen}, {Beno{\^\i}t}, {Bideaud}, {Billot}, {Bourrion}, {Calvo},
  {Catalano}, {Coiffard}, {D'Addabbo}, {De Petris}, {D{\'e}sert}, {Doyle},
  {Goupy}, {Kramer}, {Leclercq}, {Mac{\'\i}as-P{\'e}rez}, {Mauskopf}, {Mayet},
  {Monfardini}, {Pajot}, {Pascale}, {Perotto}, {Pisano}, {Pointecouteau},
  {Ponthieu}, {Pratt}, {Rev{\'e}ret}, {Ritacco}, {Rodriguez}, {Romero},
  {Schuster}, {Sievers}, {Triqueneaux}, {Tucker}, \& {Zylka}}]{ruppin2017}
{Ruppin}, F., {Adam}, R., {Comis}, B., {et~al.} 2017, \aap, 597, A110

\bibitem[{{Ruppin} {et~al.}(2018){Ruppin}, {Mayet}, {Pratt}, {Adam}, {Ade},
  {Andr{\'e}}, {Arnaud}, {Aussel}, {Bartalucci}, {Beelen}, {Beno{\^\i}t},
  {Bideaud}, {Bourrion}, {Calvo}, {Catalano}, {Comis}, {De Petris},
  {D{\'e}sert}, {Doyle}, {Driessen}, {Goupy}, {Kramer}, {Lagache}, {Leclercq},
  {Lestrade}, {Mac{\'\i}as-P{\'e}rez}, {Mauskopf}, {Monfardini}, {Perotto},
  {Pisano}, {Pointecouteau}, {Ponthieu}, {Rev{\'e}ret}, {Ritacco}, {Romero},
  {Roussel}, {Schuster}, {Sievers}, {Tucker}, \& {Zylka}}]{ruppin2018}
{Ruppin}, F., {Mayet}, F., {Pratt}, G.~W., {et~al.} 2018, \aap, 615, A112

\bibitem[{{Ruppin} {et~al.}(2020){Ruppin}, {McDonald}, {Brodwin}, {Adam},
  {Ade}, {Andr{\'e}}, {Andrianasolo}, {Arnaud}, {Aussel}, {Bartalucci},
  {Bautz}, {Beelen}, {Beno{\^\i}t}, {Bideaud}, {Bourrion}, {Calvo}, {Catalano},
  {Comis}, {Decker}, {De Petris}, {D{\'e}sert}, {Doyle}, {Driessen},
  {Eisenhardt}, {Gomez}, {Gonzalez}, {Goupy}, {K{\'e}ruzor{\'e}}, {Kramer},
  {Ladjelate}, {Lagache}, {Leclercq}, {Lestrade}, {Mac{\'\i}as-P{\'e}rez},
  {Mauskopf}, {Mayet}, {Monfardini}, {Moravec}, {Perotto}, {Pisano},
  {Pointecouteau}, {Ponthieu}, {Pratt}, {Rev{\'e}ret}, {Ritacco}, {Romero},
  {Roussel}, {Schuster}, {Shu}, {Sievers}, {Stanford}, {Stern}, {Tucker}, \&
  {Zylka}}]{ruppin2020}
{Ruppin}, F., {McDonald}, M., {Brodwin}, M., {et~al.} 2020, \apj, 893, 74

\bibitem[{{Sayers} {et~al.}(2013){Sayers}, {Czakon}, {Mantz}, {Golwala},
  {Ameglio}, {Downes}, {Koch}, {Lin}, {Maughan}, {Molnar}, {Moustakas},
  {Mroczkowski}, {Pierpaoli}, {Shitanishi}, {Siegel}, {Umetsu}, \& {Van der
  Pyl}}]{sayers2013}
{Sayers}, J., {Czakon}, N.~G., {Mantz}, A., {et~al.} 2013, \apj, 768, 177

\bibitem[{{Sayers} {et~al.}(2016){Sayers}, {Golwala}, {Mantz}, {Merten},
  {Molnar}, {Naka}, {Pailet}, {Pierpaoli}, {Siegel}, \& {Wolman}}]{sayers2016}
{Sayers}, J., {Golwala}, S.~R., {Mantz}, A.~B., {et~al.} 2016, \apj, 832, 26

\bibitem[{{Sayers} {et~al.}(2023){Sayers}, {Mantz}, {Rasia}, {Allen}, {Cui},
  {Golwala}, {Morris}, \& {Wan}}]{sayers2023}
{Sayers}, J., {Mantz}, A.~B., {Rasia}, E., {et~al.} 2023, \apj, 944, 221

\bibitem[{{Sehgal} {et~al.}(2011){Sehgal}, {Trac}, {Acquaviva}, {Ade},
  {Aguirre}, {Amiri}, {Appel}, {Barrientos}, {Battistelli}, {Bond}, {Brown},
  {Burger}, {Chervenak}, {Das}, {Devlin}, {Dicker}, {Bertrand Doriese},
  {Dunkley}, {D{\"u}nner}, {Essinger-Hileman}, {Fisher}, {Fowler}, {Hajian},
  {Halpern}, {Hasselfield}, {Hern{\'a}ndez-Monteagudo}, {Hilton}, {Hilton},
  {Hincks}, {Hlozek}, {Holtz}, {Huffenberger}, {Hughes}, {Hughes}, {Infante},
  {Irwin}, {Jones}, {Baptiste Juin}, {Klein}, {Kosowsky}, {Lau}, {Limon},
  {Lin}, {Lupton}, {Marriage}, {Marsden}, {Martocci}, {Mauskopf}, {Menanteau},
  {Moodley}, {Moseley}, {Netterfield}, {Niemack}, {Nolta}, {Page}, {Parker},
  {Partridge}, {Reid}, {Sherwin}, {Sievers}, {Spergel}, {Staggs}, {Swetz},
  {Switzer}, {Thornton}, {Tucker}, {Warne}, {Wollack}, \& {Zhao}}]{sehgal2011}
{Sehgal}, N., {Trac}, H., {Acquaviva}, V., {et~al.} 2011, \apj, 732, 44

\bibitem[{{Shaw} {et~al.}(2010){Shaw}, {Nagai}, {Bhattacharya}, \&
  {Lau}}]{shaw2010}
{Shaw}, L.~D., {Nagai}, D., {Bhattacharya}, S., \& {Lau}, E.~T. 2010, \apj,
  725, 1452

\bibitem[{{Song} {et~al.}(2012){Song}, {Zenteno}, {Stalder}, {Desai}, {Bleem},
  {Aird}, {Armstrong}, {Ashby}, {Bayliss}, {Bazin}, {Benson}, {Bertin},
  {Brodwin}, {Carlstrom}, {Chang}, {Cho}, {Clocchiatti}, {Crawford}, {Crites},
  {de Haan}, {Dobbs}, {Dudley}, {Foley}, {George}, {Gettings}, {Gladders},
  {Gonzalez}, {Halverson}, {Harrington}, {High}, {Holder}, {Holzapfel},
  {Hoover}, {Hrubes}, {Joy}, {Keisler}, {Knox}, {Lee}, {Leitch}, {Liu},
  {Lueker}, {Luong-Van}, {Marrone}, {McDonald}, {McMahon}, {Mehl}, {Meyer},
  {Mocanu}, {Mohr}, {Montroy}, {Natoli}, {Nurgaliev}, {Padin}, {Plagge},
  {Pryke}, {Reichardt}, {Rest}, {Ruel}, {Ruhl}, {Saliwanchik}, {Saro}, {Sayre},
  {Schaffer}, {Shaw}, {Shirokoff}, {{\v{S}}uhada}, {Spieler}, {Stanford},
  {Staniszewski}, {Stark}, {Story}, {Stubbs}, {van Engelen}, {Vanderlinde},
  {Vieira}, {Williamson}, \& {Zahn}}]{song2012}
{Song}, J., {Zenteno}, A., {Stalder}, B., {et~al.} 2012, \apj, 761, 22

\bibitem[{{Story} {et~al.}(2011){Story}, {Aird}, {Andersson}, {Armstrong},
  {Bazin}, {Benson}, {Bleem}, {Bonamente}, {Brodwin}, {Carlstrom}, {Chang},
  {Clocchiatti}, {Crawford}, {Crites}, {de Haan}, {Desai}, {Dobbs}, {Dudley},
  {Foley}, {George}, {Gladders}, {Gonzalez}, {Halverson}, {High}, {Holder},
  {Holzapfel}, {Hoover}, {Hrubes}, {Joy}, {Keisler}, {Knox}, {Lee}, {Leitch},
  {Lueker}, {Luong-Van}, {Marrone}, {McMahon}, {Mehl}, {Meyer}, {Mohr},
  {Montroy}, {Padin}, {Plagge}, {Pryke}, {Reichardt}, {Rest}, {Ruel}, {Ruhl},
  {Saliwanchik}, {Saro}, {Schaffer}, {Shaw}, {Shirokoff}, {Song}, {Spieler},
  {Stalder}, {Staniszewski}, {Stark}, {Stubbs}, {Vanderlinde}, {Vieira},
  {Williamson}, \& {Zenteno}}]{story2011}
{Story}, K., {Aird}, K.~A., {Andersson}, K., {et~al.} 2011, \apjl, 735, L36

\bibitem[{{Sun} {et~al.}(2011){Sun}, {Sehgal}, {Voit}, {Donahue}, {Jones},
  {Forman}, {Vikhlinin}, \& {Sarazin}}]{sun2011}
{Sun}, M., {Sehgal}, N., {Voit}, G.~M., {et~al.} 2011, \apjl, 727, L49

\bibitem[{{Sunyaev} \& {Zeldovich}(1970)}]{sz1970}
{Sunyaev}, R.~A. \& {Zeldovich}, Y.~B. 1970, Comments on Astrophysics and Space
  Physics, 2, 66

\bibitem[{{Sunyaev} \& {Zeldovich}(1972)}]{sz1972}
{Sunyaev}, R.~A. \& {Zeldovich}, Y.~B. 1972, Comments on Astrophysics and Space
  Physics, 4, 173

\bibitem[{{Tarr{\'\i}o} {et~al.}(2019){Tarr{\'\i}o}, {Melin}, \&
  {Arnaud}}]{tarrio2019}
{Tarr{\'\i}o}, P., {Melin}, J.~B., \& {Arnaud}, M. 2019, \aap, 626, A7

\bibitem[{{Zenteno} {et~al.}(2020){Zenteno}, {Hern{\'a}ndez-Lang}, {Klein},
  {Vergara Cervantes}, {Hollowood}, {Bhargava}, {Palmese}, {Strazzullo},
  {Romer}, {Mohr}, {Jeltema}, {Saro}, {Lidman}, {Gruen}, {Ojeda},
  {Katzenberger}, {Aguena}, {Allam}, {Avila}, {Bayliss}, {Bertin}, {Brooks},
  {Buckley-Geer}, {Burke}, {Capasso}, {Carnero Rosell}, {Carrasco Kind},
  {Carretero}, {Castander}, {Costanzi}, {da Costa}, {De Vicente}, {Desai},
  {Diehl}, {Doel}, {Eifler}, {Evrard}, {Flaugher}, {Floyd}, {Fosalba},
  {Frieman}, {Garc{\'\i}a-Bellido}, {Gerdes}, {Gonzalez}, {Gruendl},
  {Gschwend}, {Gutierrez}, {Hartley}, {Hinton}, {Honscheid}, {James}, {Kuehn},
  {Lahav}, {Lima}, {McDonald}, {Maia}, {March}, {Melchior}, {Menanteau},
  {Miquel}, {Ogando}, {Paz-Chinch{\'o}n}, {Plazas}, {Roodman}, {Rykoff},
  {Sanchez}, {Scarpine}, {Schubnell}, {Serrano}, {Sevilla-Noarbe}, {Smith},
  {Soares-Santos}, {Suchyta}, {Swanson}, {Tarle}, {Thomas}, {Varga}, {Walker},
  {Wilkinson}, \& {DES Collaboration}}]{zenteno2020}
{Zenteno}, A., {Hern{\'a}ndez-Lang}, D., {Klein}, M., {et~al.} 2020, \mnras,
  495, 705

\bibitem[{{Zhang} {et~al.}(2021){Zhang}, {Zhuravleva}, {Kravtsov}, \&
  {Churazov}}]{zhang2021}
{Zhang}, C., {Zhuravleva}, I., {Kravtsov}, A., \& {Churazov}, E. 2021, \mnras,
  506, 839

\bibitem[{{Zhao}(1996)}]{zhao1996}
{Zhao}, H. 1996, \mnras, 278, 488

\end{thebibliography}

\begin{appendix}

\section{Correlated noise in harmonic space}
\label{app:harmonoise}

\begin{figure}[!h]
\centering
\includegraphics[width=\hsize]{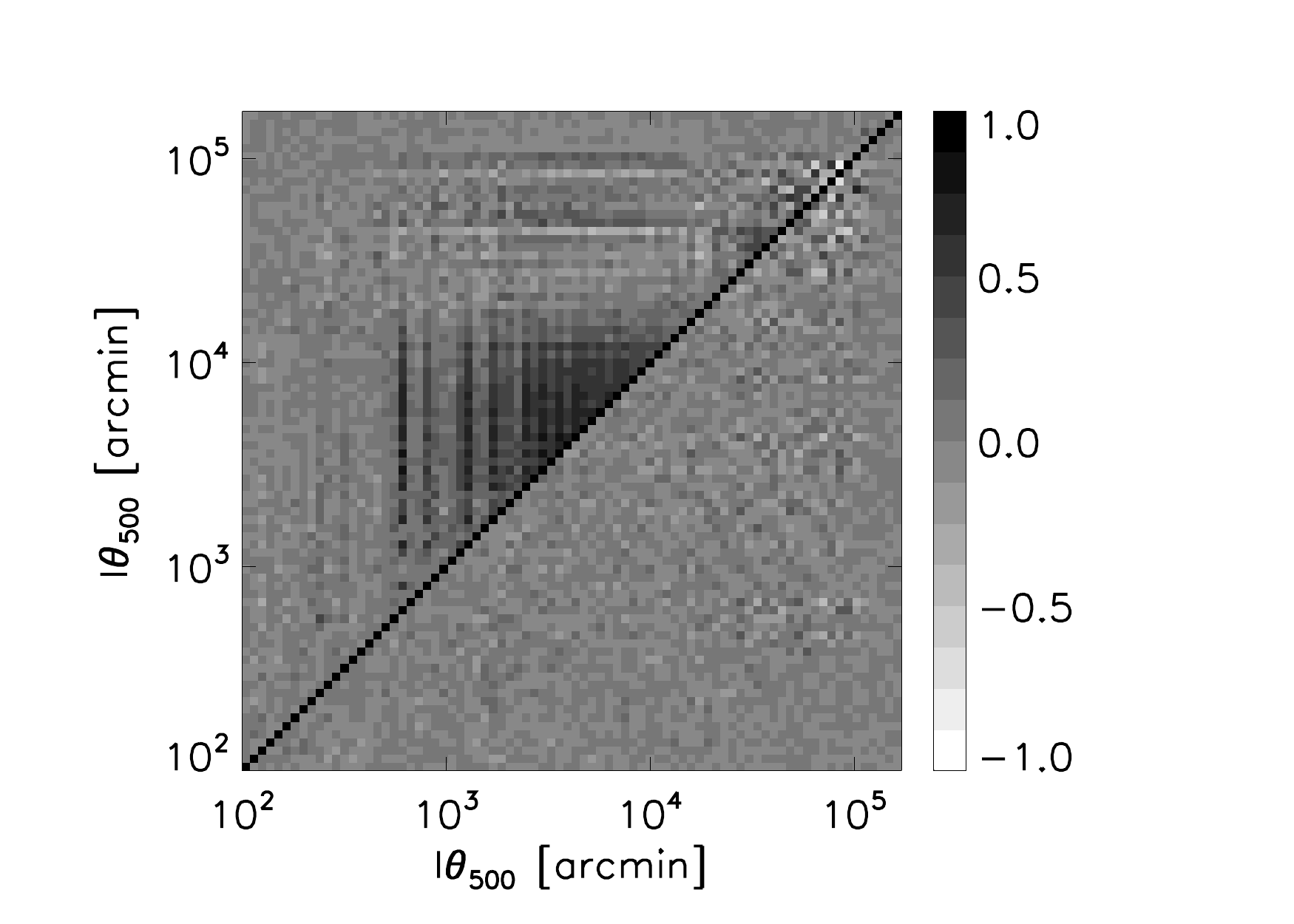}
\caption{\footnotesize Correlation matrix $\mathcal{R}$ between the $l\thetaf$ bins of the average cluster pressure profile estimated by bootstrap resampling. The top left half-matrix is estimated at the location of the clusters, the bottom right half is estimated on blank fields.}
\label{fig:cormat}
\end{figure}

As discussed in Sect.~\ref{sec:av_harmo}, the $l\thetaf$ bins are expected to be uncorrelated to first approximation. We tested this hypothesis by computing the covariance matrix $\mathcal{S}$ between the bins using bootstrap resampling. In practice, we performed 1,000 bootstraps, drawing  with replacement 1,000 SPT-like samples (461 clusters) from the original 461 SPT clusters. For each of the samples, we performed the same analysis as for the original data set. We then estimated the covariance between the $l\thetaf$ bins of the 1,000 samples. The result for the correlation matrix $\mathcal{R}$ is shown as the top left half-matrix in Fig.~\ref{fig:cormat}. We recall the definition
\begin{equation}
 \mathcal{R}_{ll'} = {\mathcal{S}_{ll'} \over \sqrt{\mathcal{S}_{ll}}\sqrt{\mathcal{S}_{l'l'}}}.
\end{equation}
In fact, there is a significant contribution of the off-diagonal terms for $10^{3} < l\thetaf < 10^{4} \, {\rm arcmin}$, with correlation factors larger than 0.5. To further investigate these correlations, we performed the same analysis on blank fields. The results are shown as the bottom right half-matrix in Fig.~\ref{fig:cormat}. The off-diagonal correlations in the range $10^{3} < l\thetaf < 10^{4} \, {\rm arcmin}$ are not visible for the blank fields. We notice however that some correlations exist both on cluster and blank fields in the range $5 \times 10^{4} < l\thetaf < 10^{5} \, {\rm arcmin}$ but are not very significant. As significant correlations in the range $10^{3} < l\thetaf < 10^{4} \, {\rm arcmin}$ are only present at the locations of clusters and not on blank fields, this points toward a cluster-related origin of these correlations, such as lensing of the primary CMB anisotropies, or sources behind the clusters.

\begin{figure}[!h]
\centering
\includegraphics[width=\hsize]{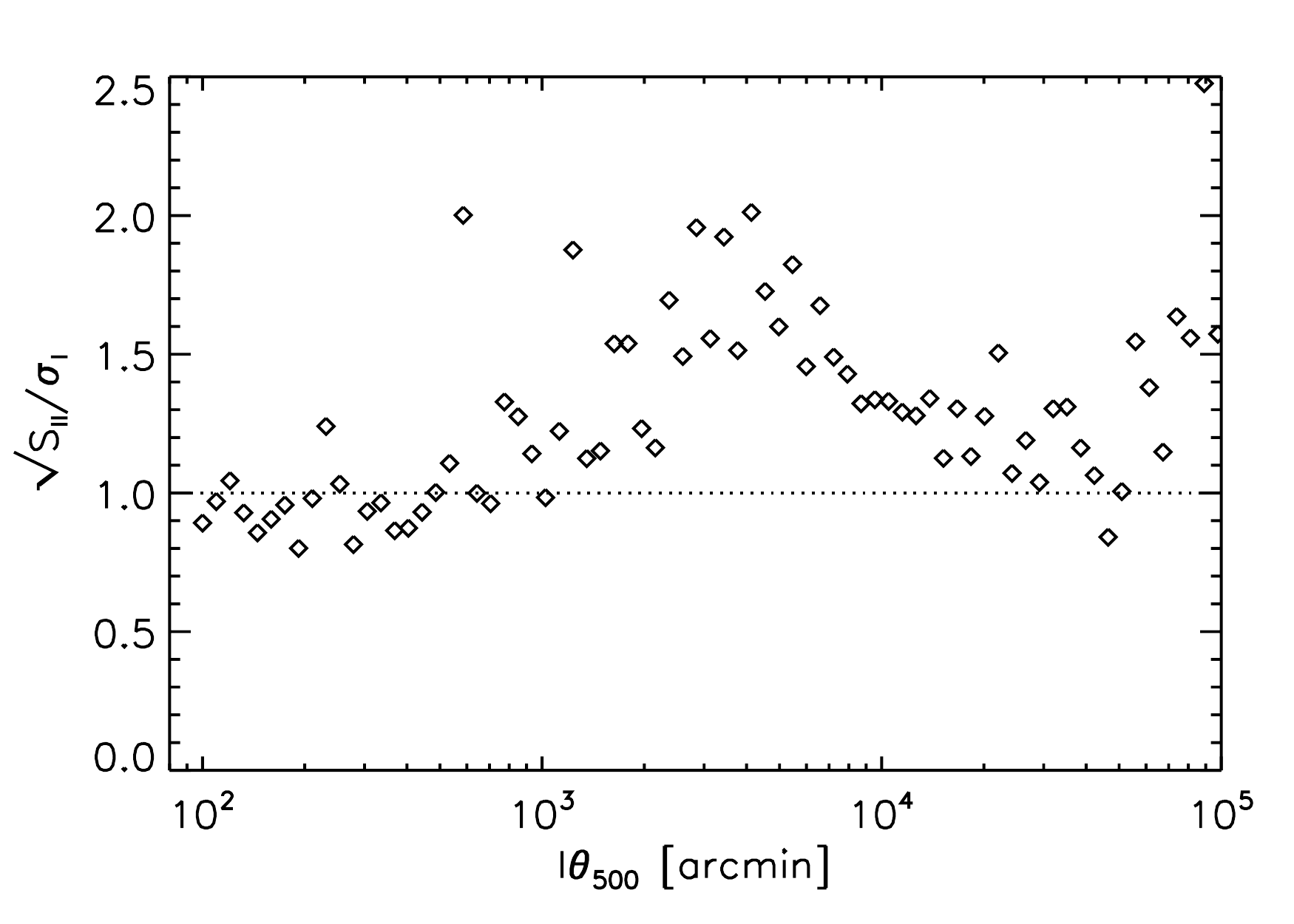}
\caption{\footnotesize Ratio between the diagonal terms of the covariance matrix $\sqrt{\mathcal{S}_{ll}}$ and the statistical error ${\tilde \sigma}_l$ as a function of $l\thetaf$.}
\label{fig:diagterms}
\end{figure}

We also compared the diagonal terms of the covariance matrix $\sqrt{\mathcal{S}_{ll}}$ to the statistical error ${\tilde \sigma}_l$ estimated by our averaging procedure detailed in Sect.~\ref{sec:renorm}. Results are shown in Fig.~\ref{fig:diagterms}. As expected, the $\sqrt{\mathcal{S}_{ll}}$ are in average higher than the statistical ${\tilde \sigma}_l$.

We thus adopted the full covariance matrix $\mathcal{S}$ in the likelihood (Sect.~\ref{sec:gnfwfits}) to be conservative in our gNFW fit.

\FloatBarrier

\section{The gNFW profile in real and harmonic space}
\label{app:realfour}

\begin{figure*}[!h]
\centering
\includegraphics[width=0.31\hsize]{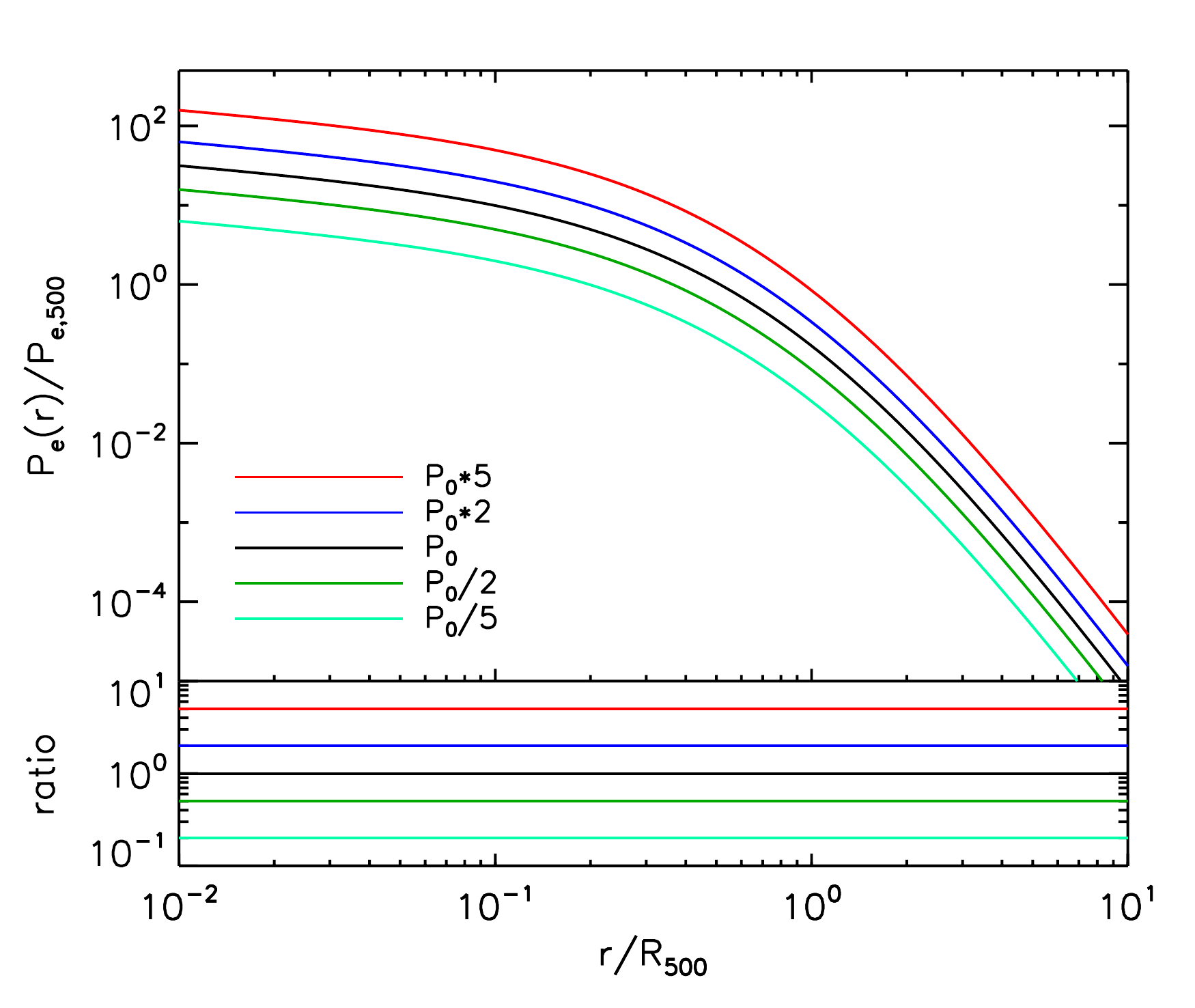}
\includegraphics[width=0.31\hsize]{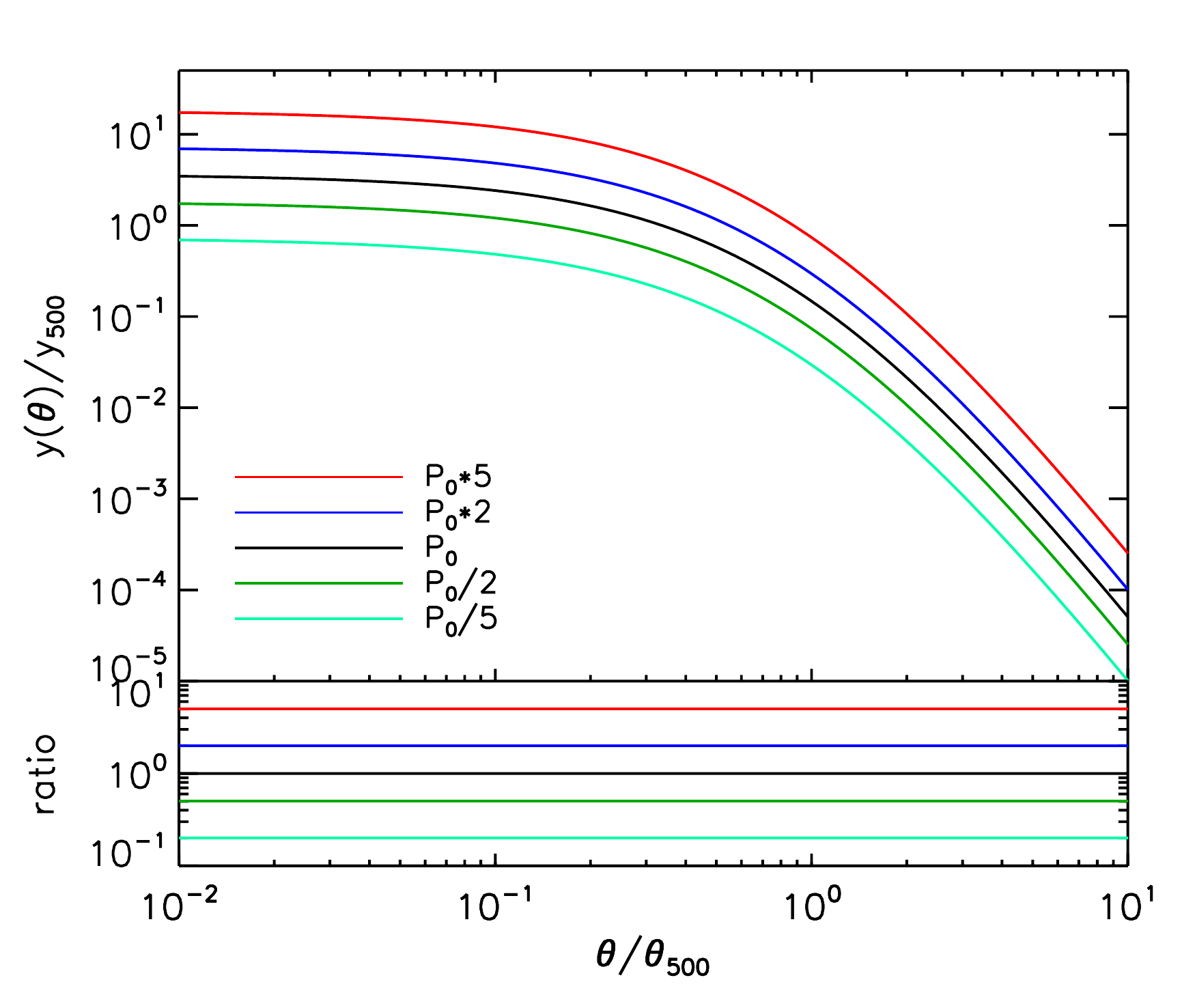}
\includegraphics[width=0.31\hsize]{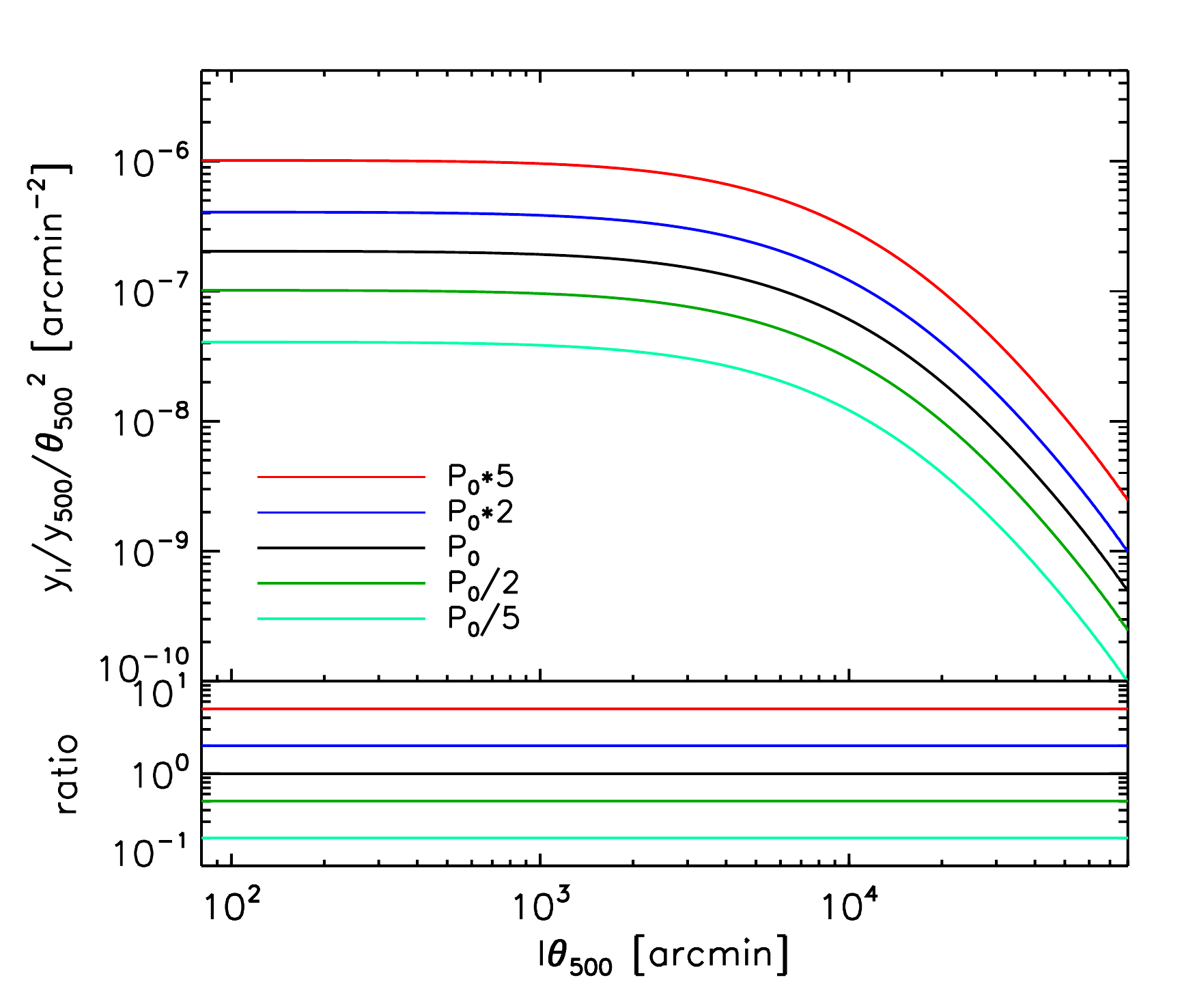}
\includegraphics[width=0.31\hsize]{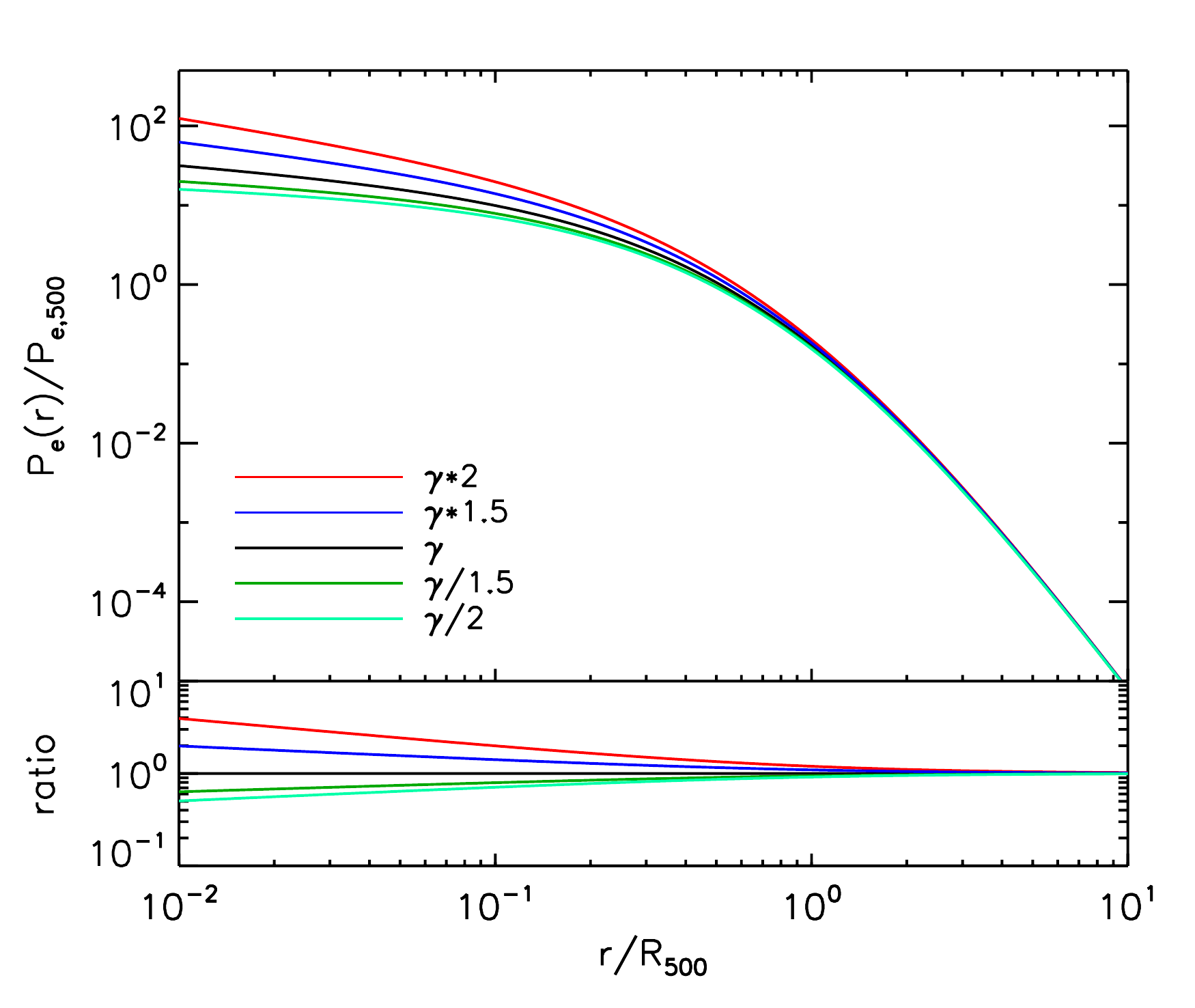}
\includegraphics[width=0.31\hsize]{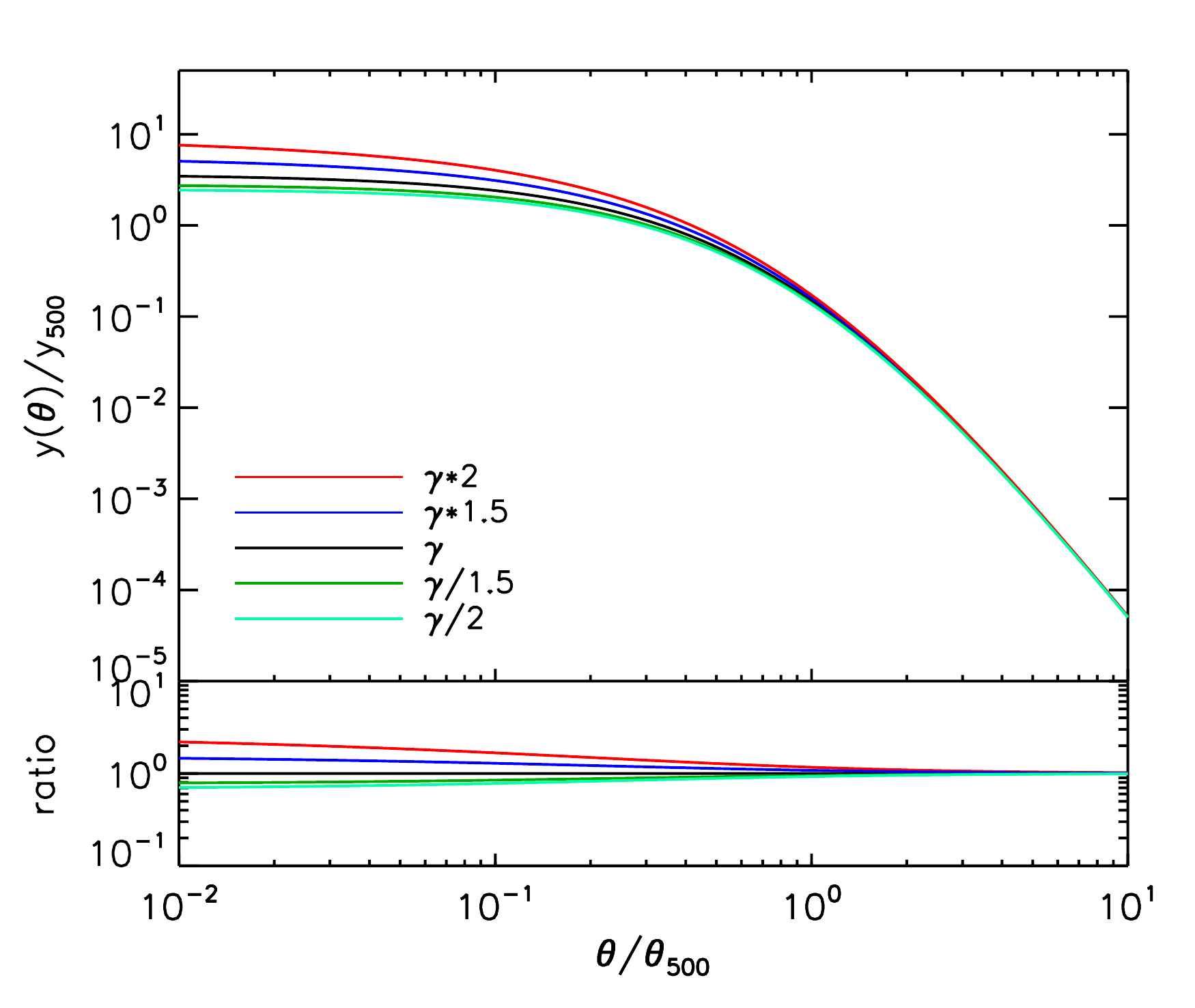}
\includegraphics[width=0.31\hsize]{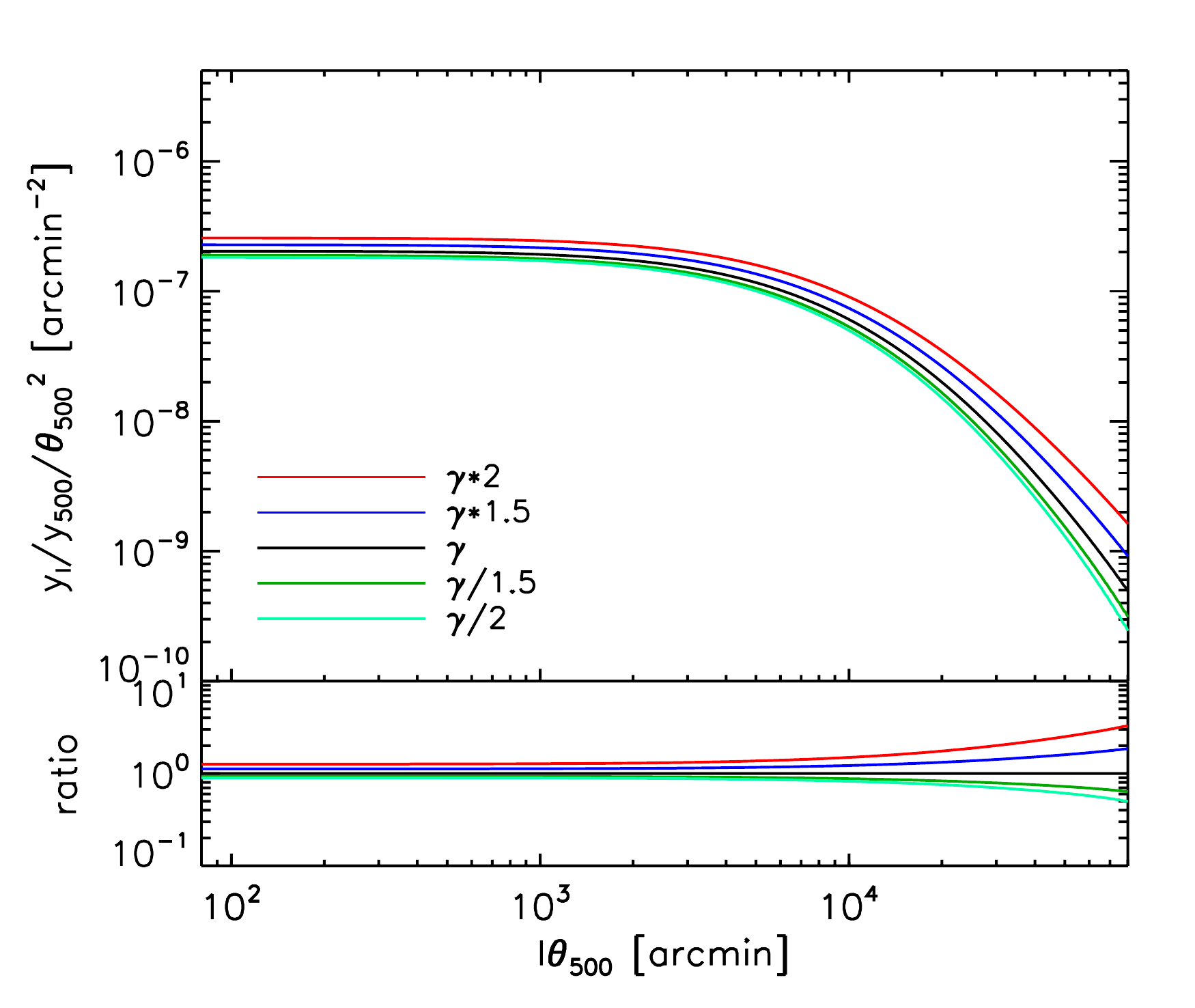}
\includegraphics[width=0.31\hsize]{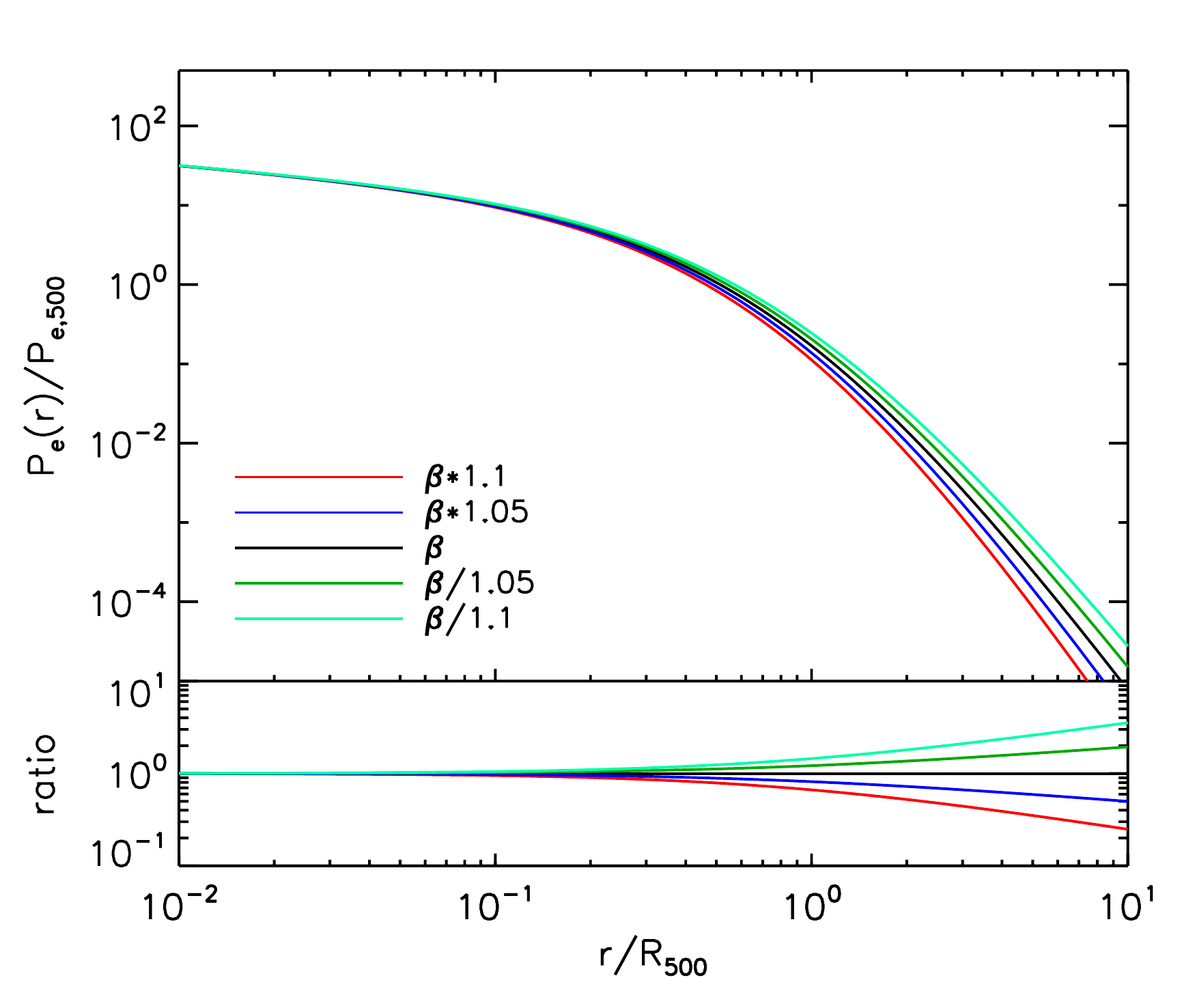}
\includegraphics[width=0.31\hsize]{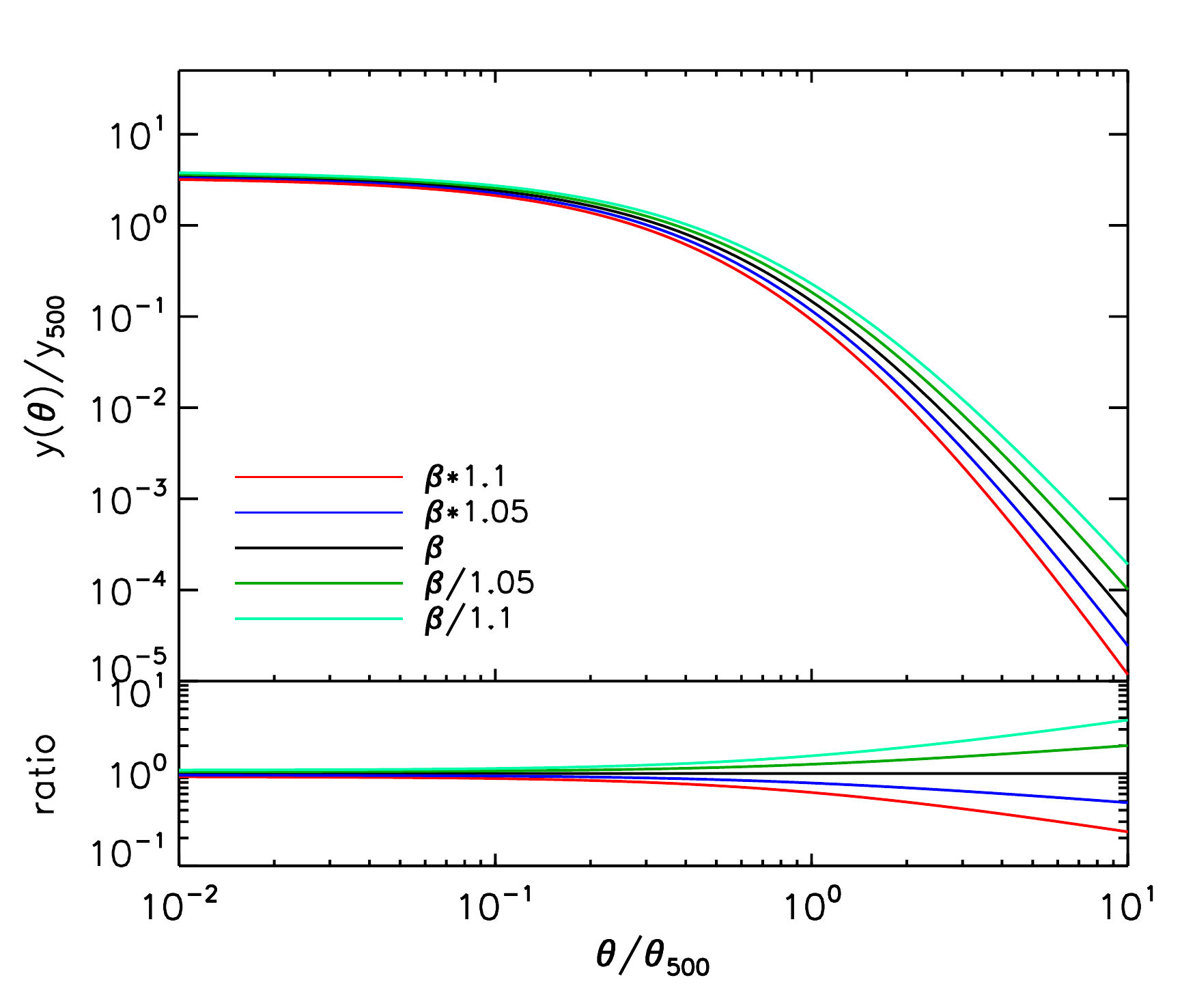}
\includegraphics[width=0.31\hsize]{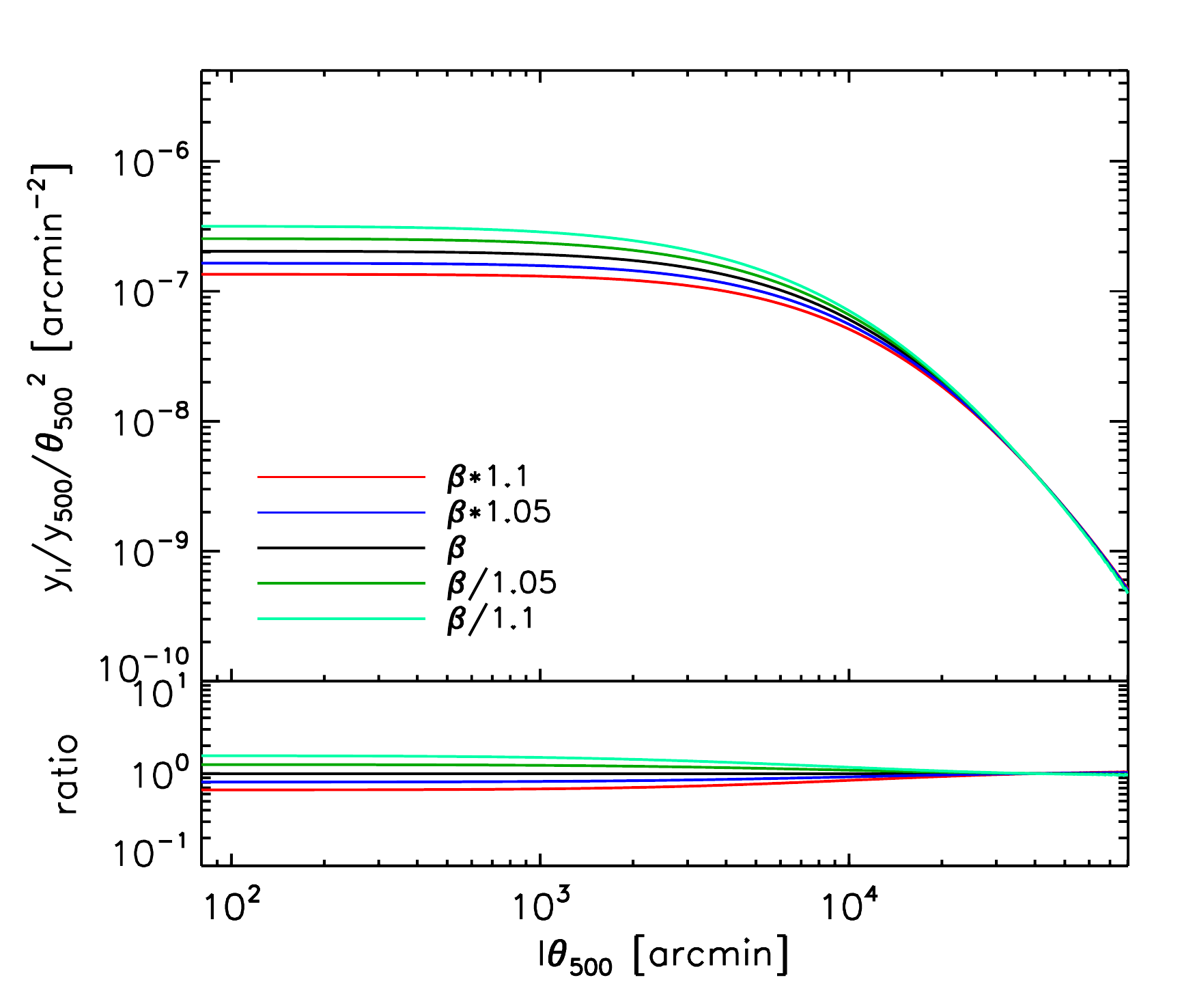}
\includegraphics[width=0.31\hsize]{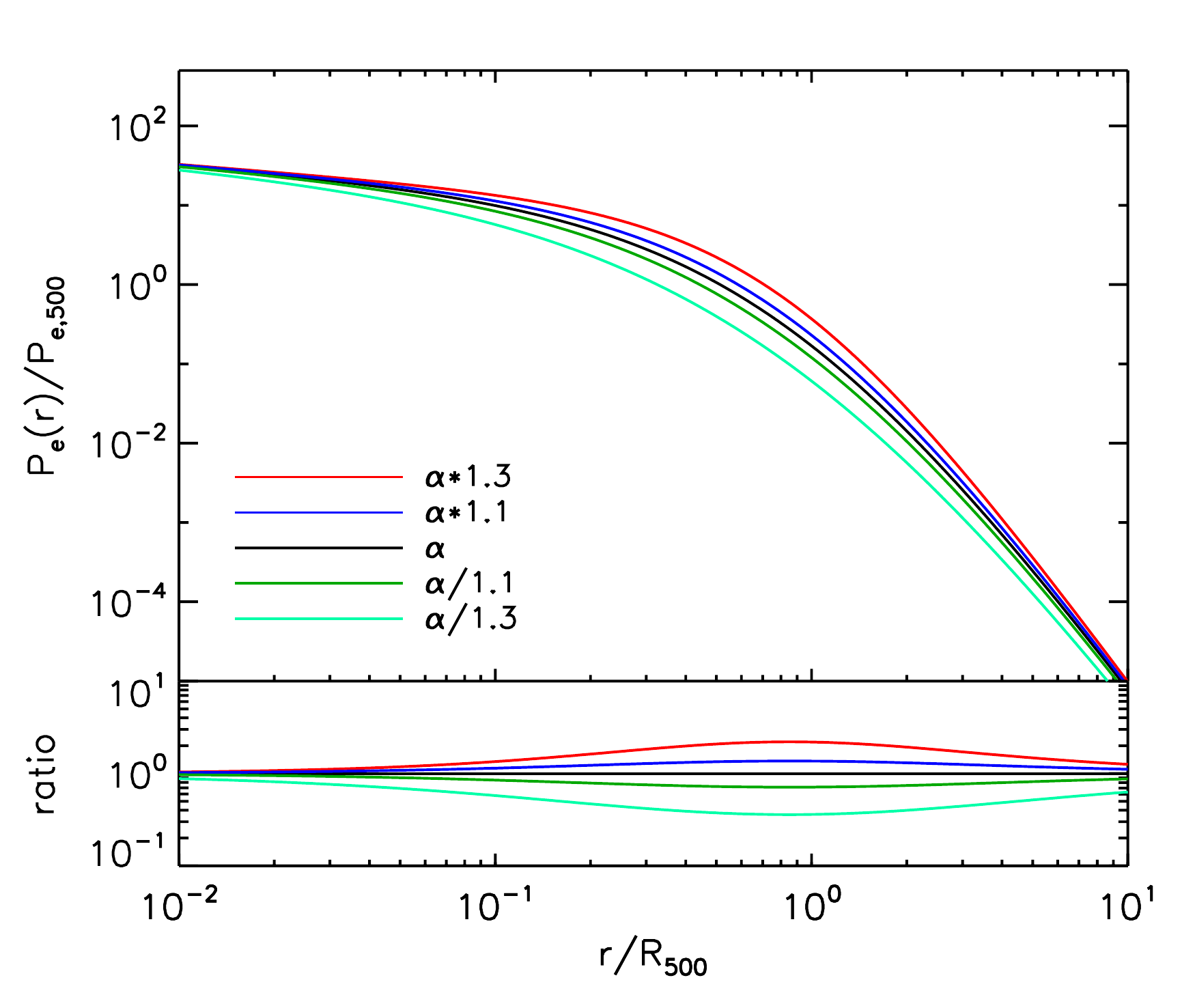}
\includegraphics[width=0.31\hsize]{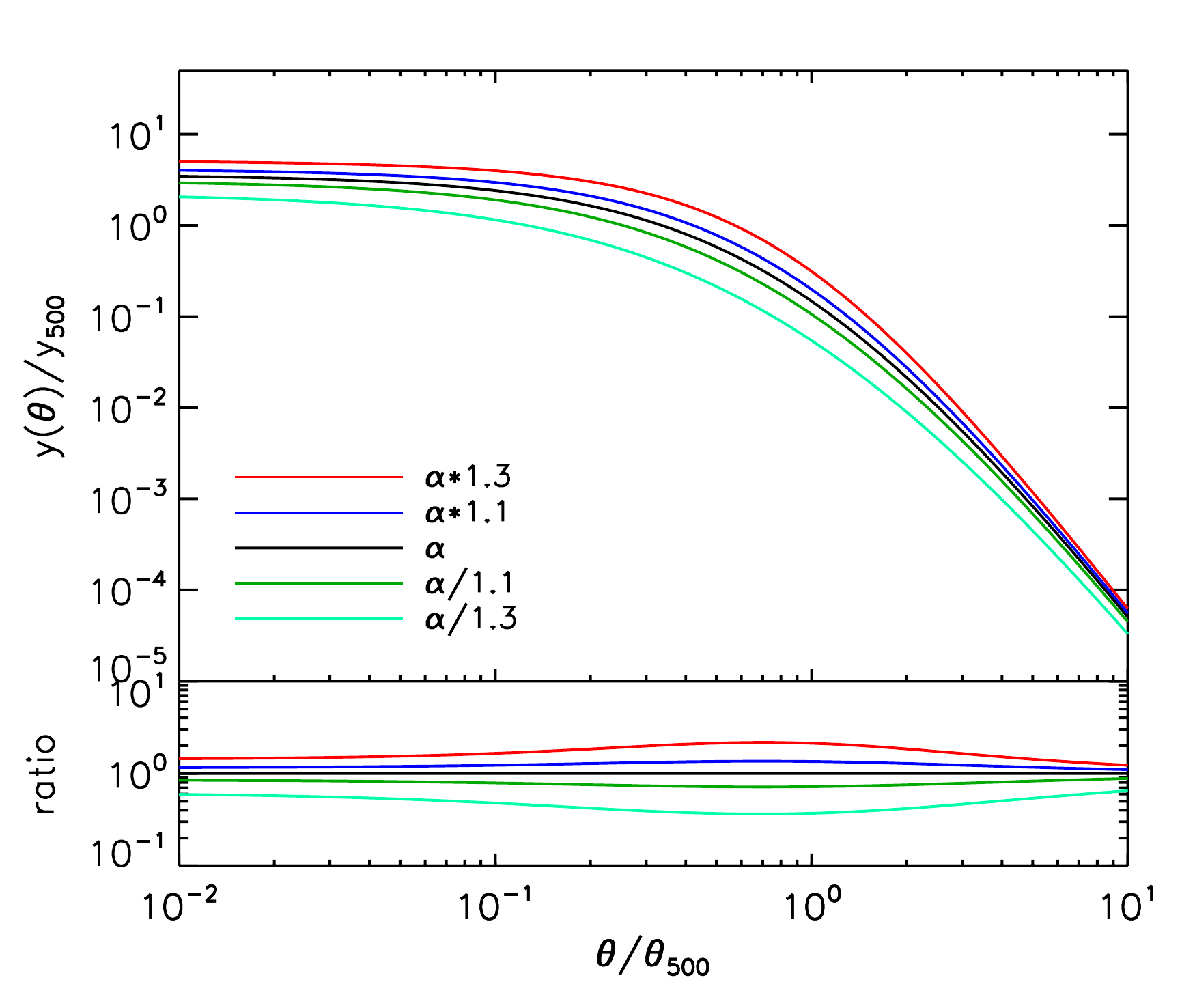}
\includegraphics[width=0.31\hsize]{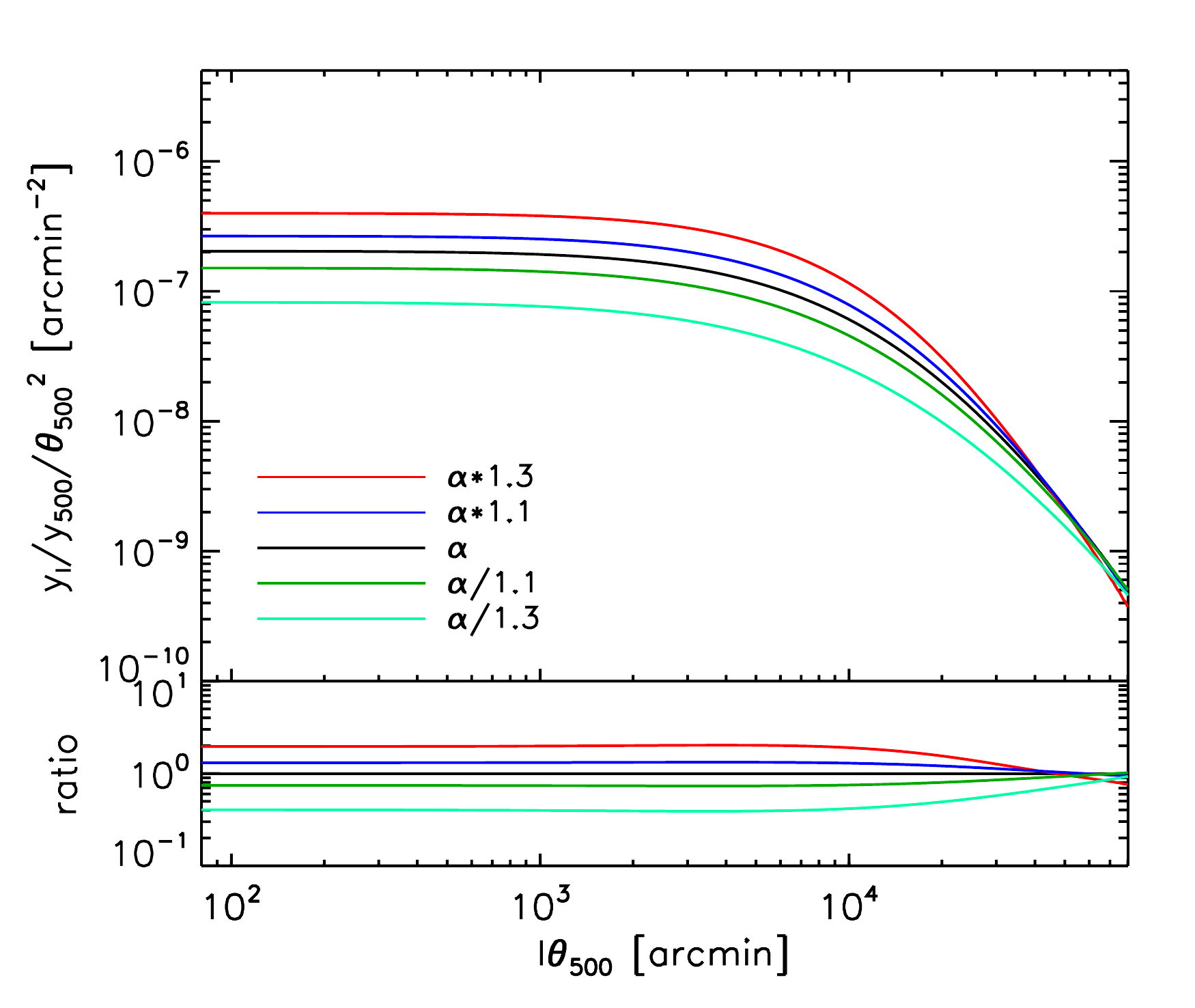}
\includegraphics[width=0.31\hsize]{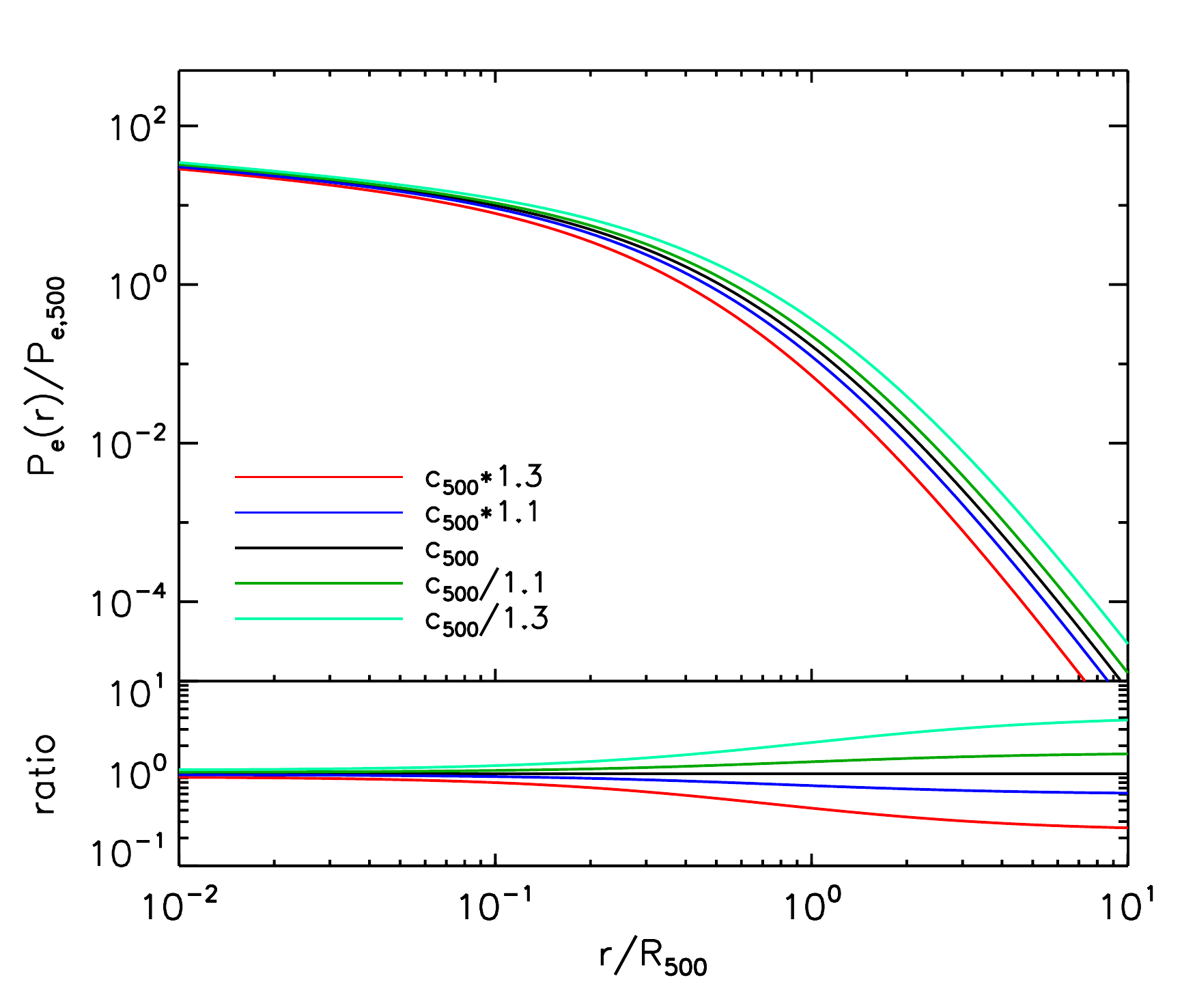}
\includegraphics[width=0.31\hsize]{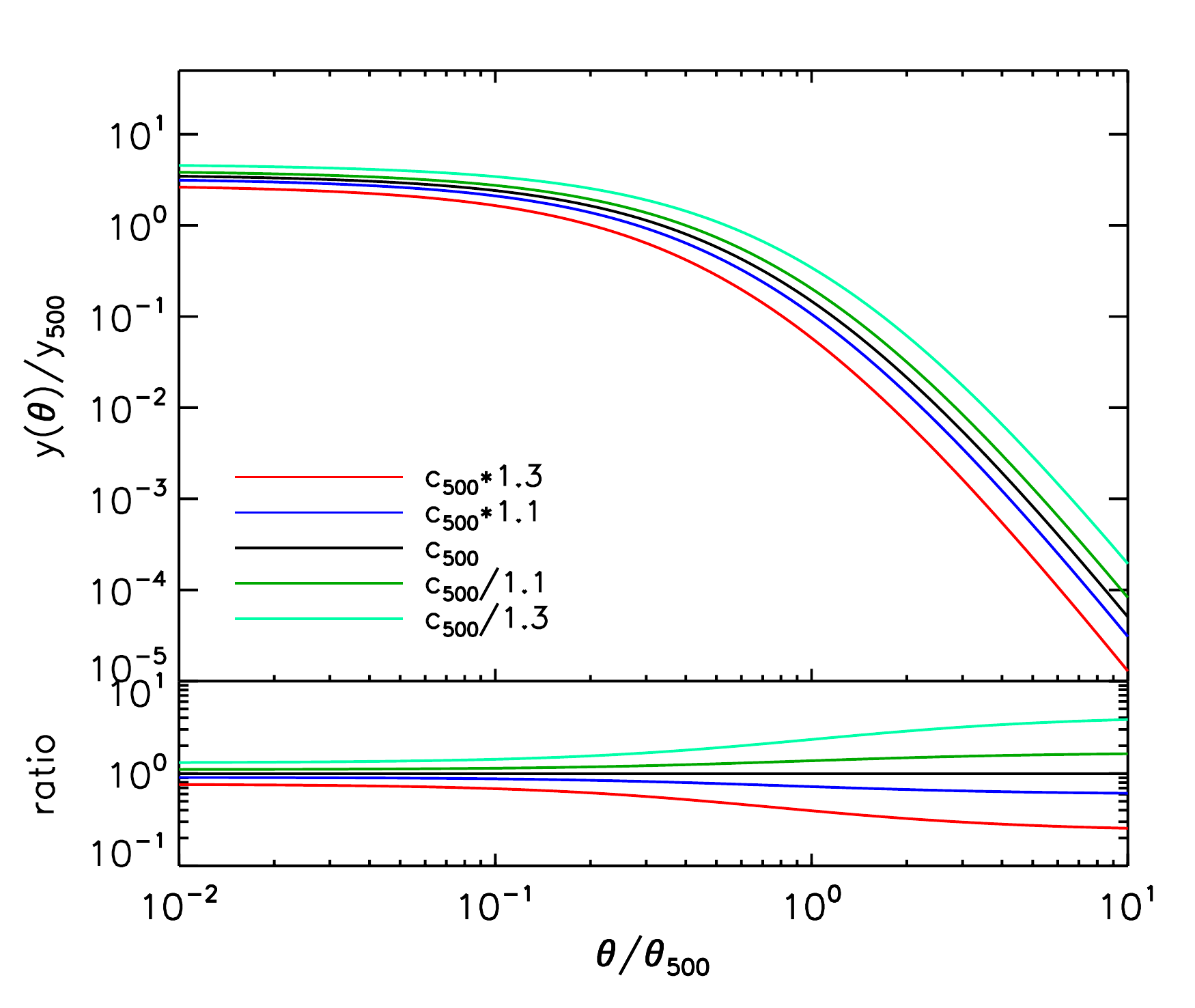}
\includegraphics[width=0.31\hsize]{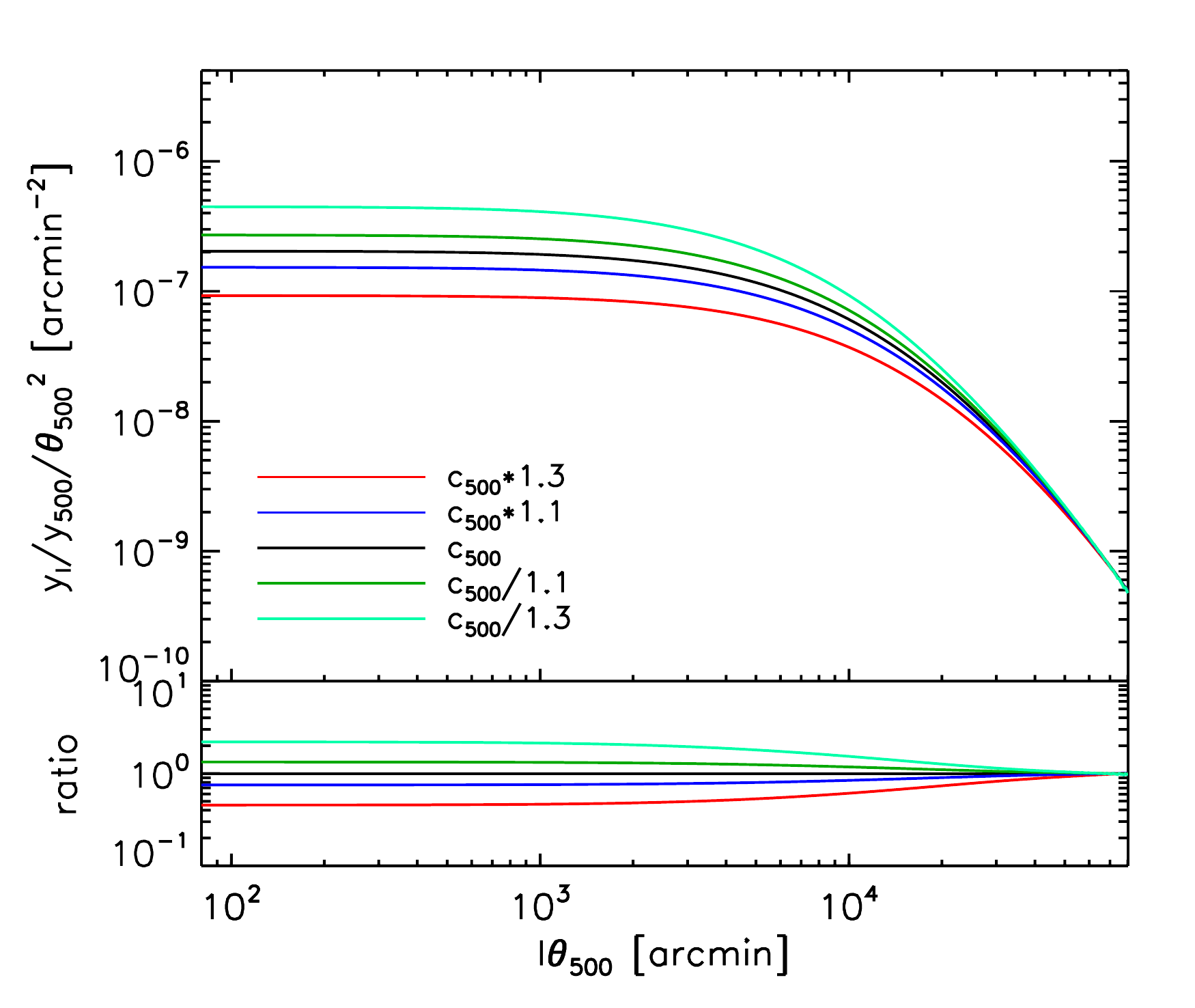}
\caption{\footnotesize Pressure profiles. {\it From left to right:}  3D real space, 2D real space, 2D harmonic space. {\it From top to bottom:} the gNFW parameters $P_0$, $\gamma$, $\beta$, $\alpha$, $c_{500}$ are varied.}
\label{fig:illu_P0}
\end{figure*}

We produced illustrative plots based on variations of the five parameters of the gNFW  around the fiducial values of the UPP~\citep{arnaud2010}. These are shown in Fig.~\ref{fig:illu_P0}. The first column corresponds to the three dimensional pressure profile in real space, the second column to the two dimensional Compton profile in real space that is the three dimensional pressure profile integrated along the line-of-sight, the third column is the two dimensional Fourier transform of the Compton profile expressed in harmonic space. From top to bottom, rows correspond to variations around $P_0$, $\gamma$, $\beta$, $\alpha$ and $c_{500}$.

\FloatBarrier

\section{Self-similar quantities}
\label{app:selfsim}

Following Appendix A of~\cite{arnaud2010}, we define the characteristic temperature $kT_{\rm 200m}=\mu m_p G M_{\rm200m} / (2 R_{\rm 200m})$ where $\mu$ is the mean molecular weight, $m_p$ the proton mass and $G$ the gravitational constant. We note $n_{e,{\rm 200m}}=\rho_{g,{\rm 200m}}/(\mu_e m_p)$ the characteristic electronic density where $\rho_{g,{\rm 200m}}=200 f_{\rm B} \rho_{\rm m}(z)$ is the characteristic gas density with $\mu_e$ the mean molecular weight per free electron, $f_{\rm B}$ the baryon fraction in the Universe and $\rho_{\rm m}(z)$ the mean density of the Universe at redshift $z$. We define the characteristic pressure in ${\rm 200m}$ as
\begin{equation}
P_{e,\rm 200m} = n_{e,{\rm 200m}}  \, kT_{\rm 200m}
\end{equation}
which can be written as a function of $z$ and $M_{\rm 200m}$
\begin{equation}
\label{eq:p200m}
P_{e,\rm 200m} = {3 \over 8 \pi} \left [ 200 \, G^{-1/4} \Omega_{\rm m}(z) H(z)^2 \over 2 \right]^{4/3}  \, {\mu \over \mu_e} \, f_{\rm B} \, M_{\rm 200m}^{2/3}.
\end{equation}
Then we define
\begin{equation}
\label{eq:def_y200m}
y_{\rm 200m} = {\sigma_{\rm T} \over m_{\rm e} c^2}  \times (2 R_{\rm 200m}) \times P_{e,\rm 200m}
\end{equation}
as in Eq.~\ref{eq:def_y500} but we intentionally do not include any mass dependent factor $\Mfive^{0.12}$ because it has not been constrained for overdensity ${\rm 200m}$.

\section{Impact of positional uncertainty on recovered gNFW parameters}
\label{app:pos_unc}

\begin{figure}[!h]
\centering
\includegraphics[width=\hsize]{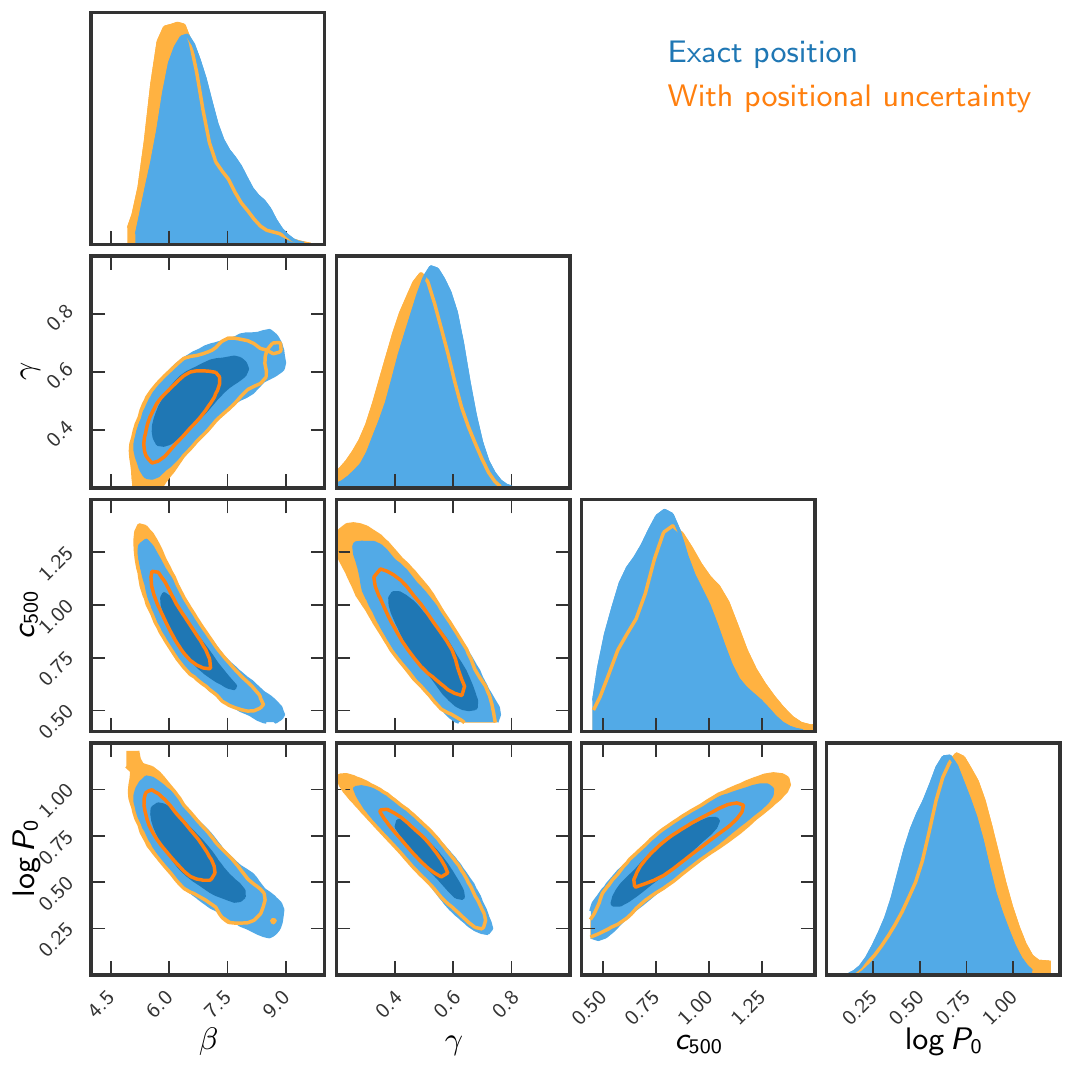}
\caption{\footnotesize Marginalised posterior likelihood for the parameters of the best-fitting gNFW model for clusters injected in the data assuming the universal pressure profile. Blue contours assume the positions of the injected clusters are perfectly known. Orange contours include random positional uncertainties. The shift of the posteriors is negligible with respect to their size.}
\label{fig:pos_unc}
\end{figure}

Fig~\ref{fig:pos_unc} shows the impact of the positional uncertainty on recovered gNFW parameters for clusters injected in the data assuming the universal pressure profile. Details of the procedure are provided in Sect.~\ref{sec:posuncert}. The shift of the posteriors is negligible with respect to their size.

\FloatBarrier

\section{Tabulated best-fitting profiles and error envelopes}
\label{app:envel}

\begin{figure*}[!h]
\centering
\includegraphics[width=0.33\hsize]{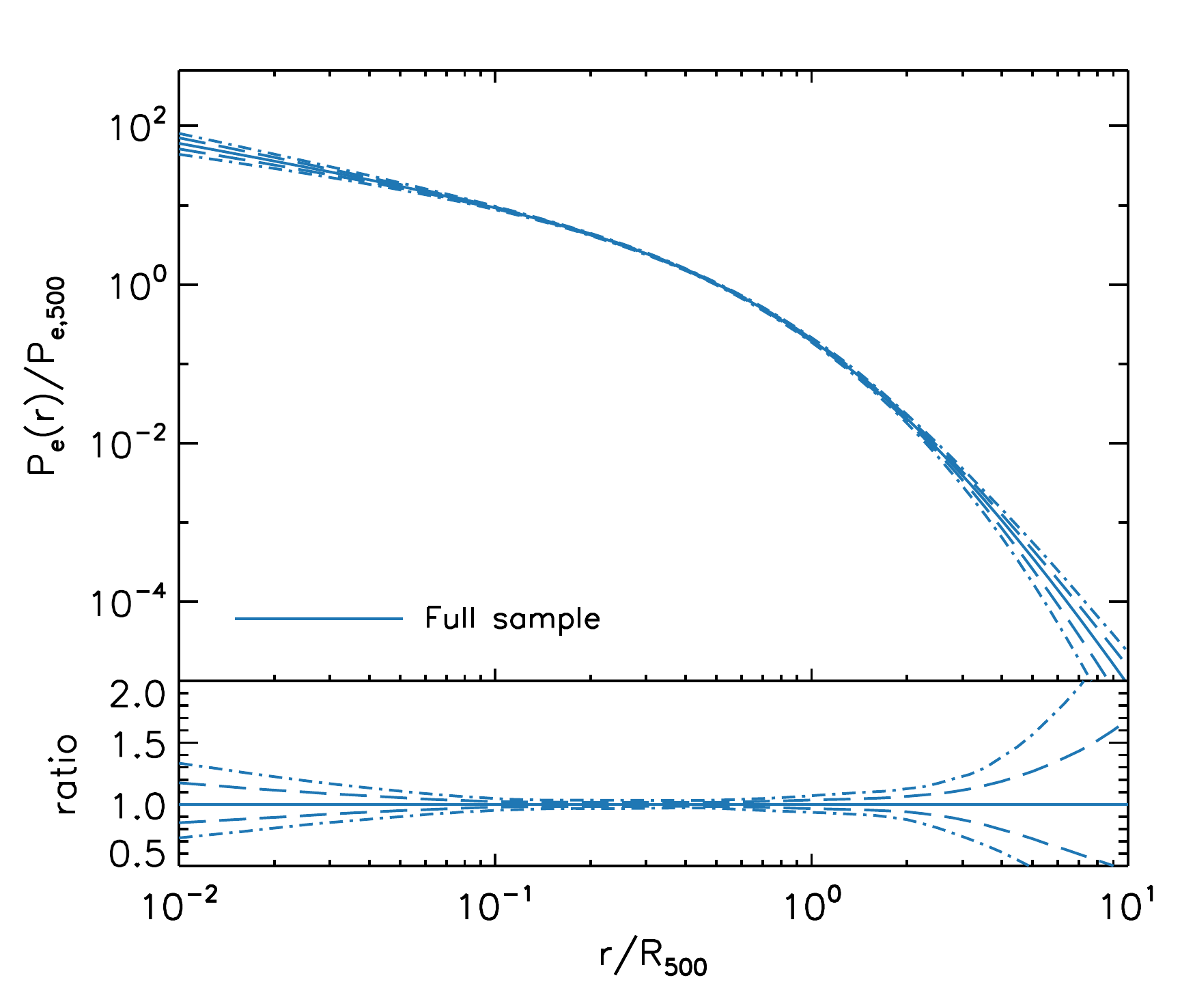}
\includegraphics[width=0.33\hsize]{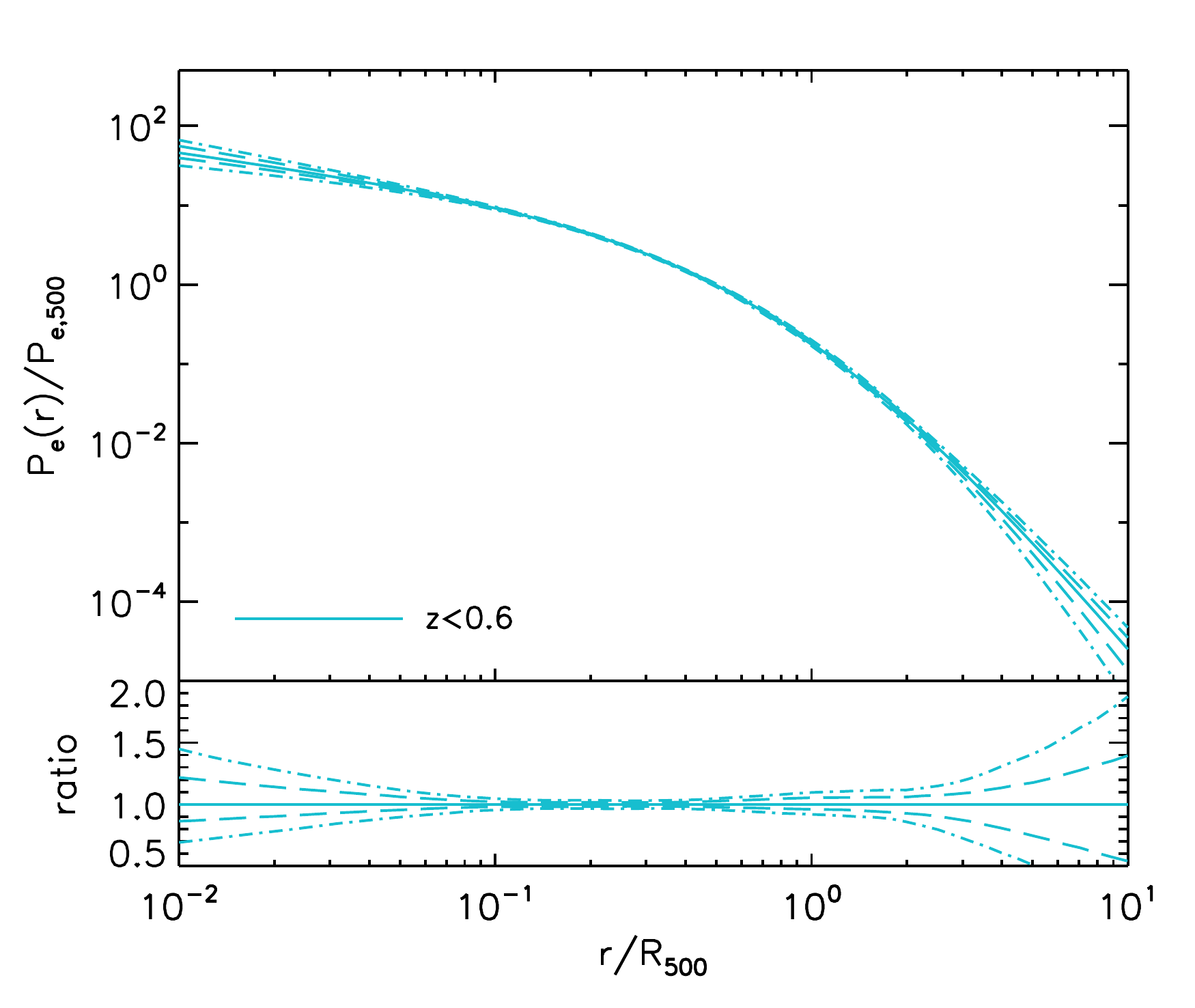}
\includegraphics[width=0.33\hsize]{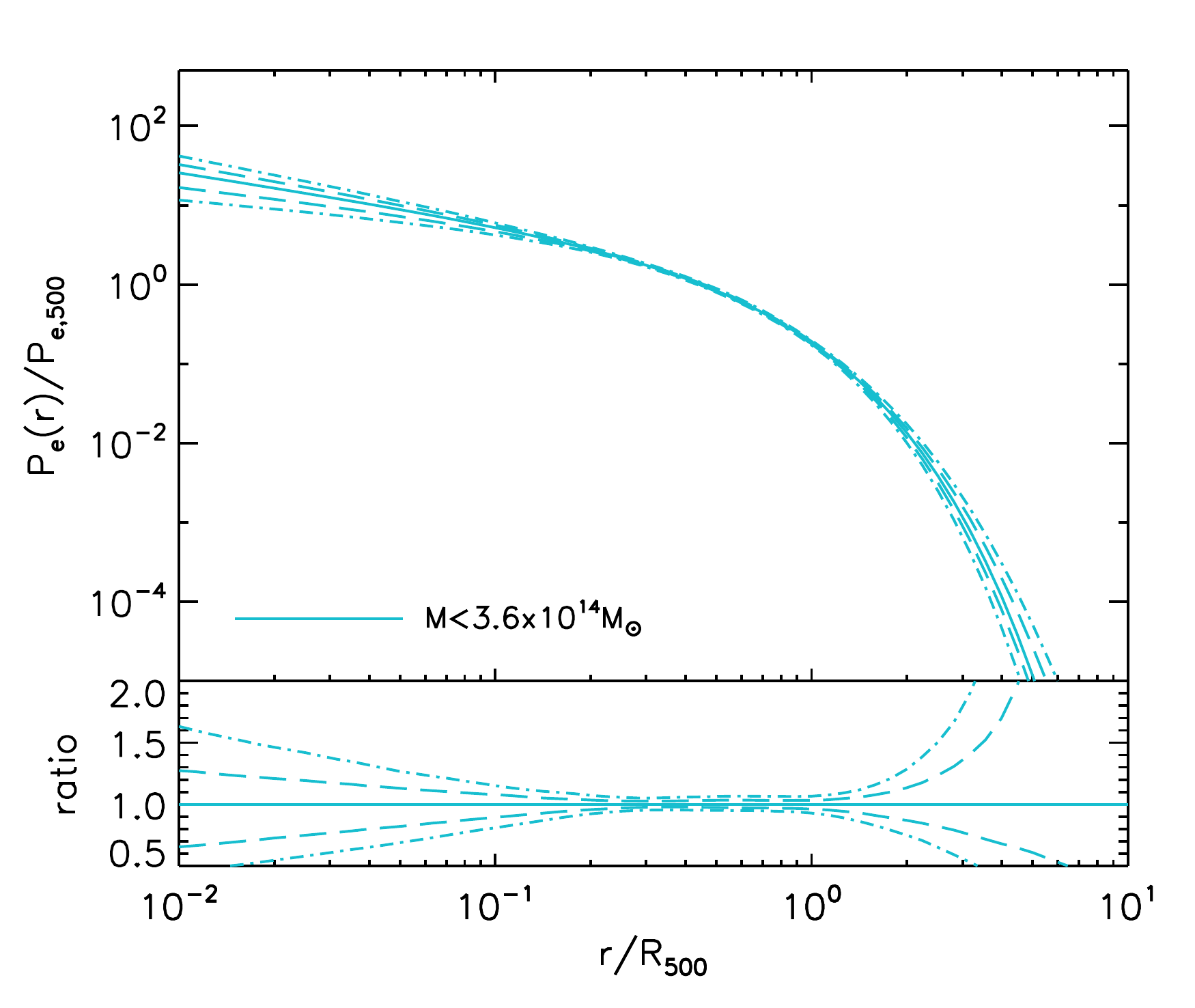}
\includegraphics[width=0.33\hsize]{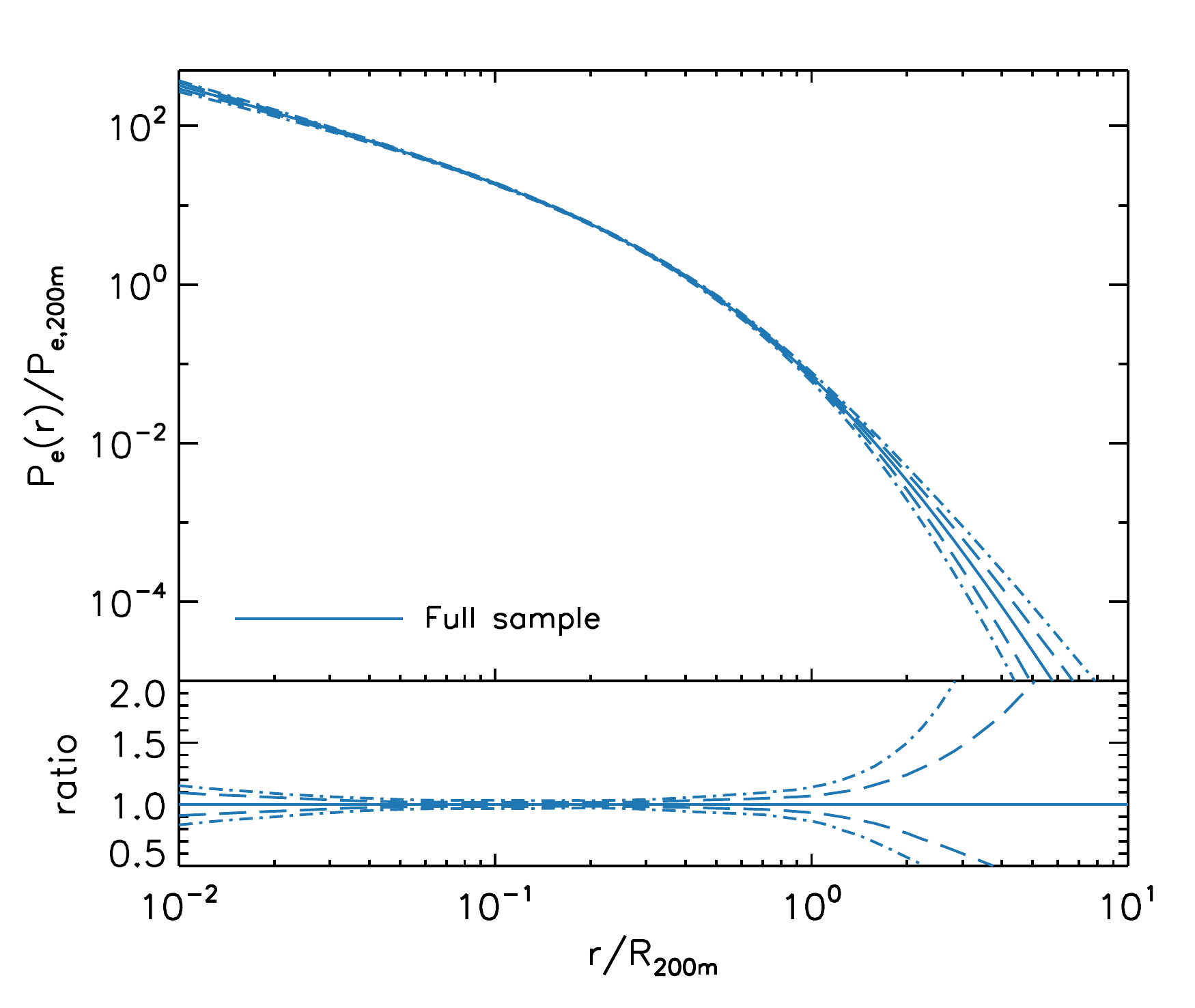}
\includegraphics[width=0.33\hsize]{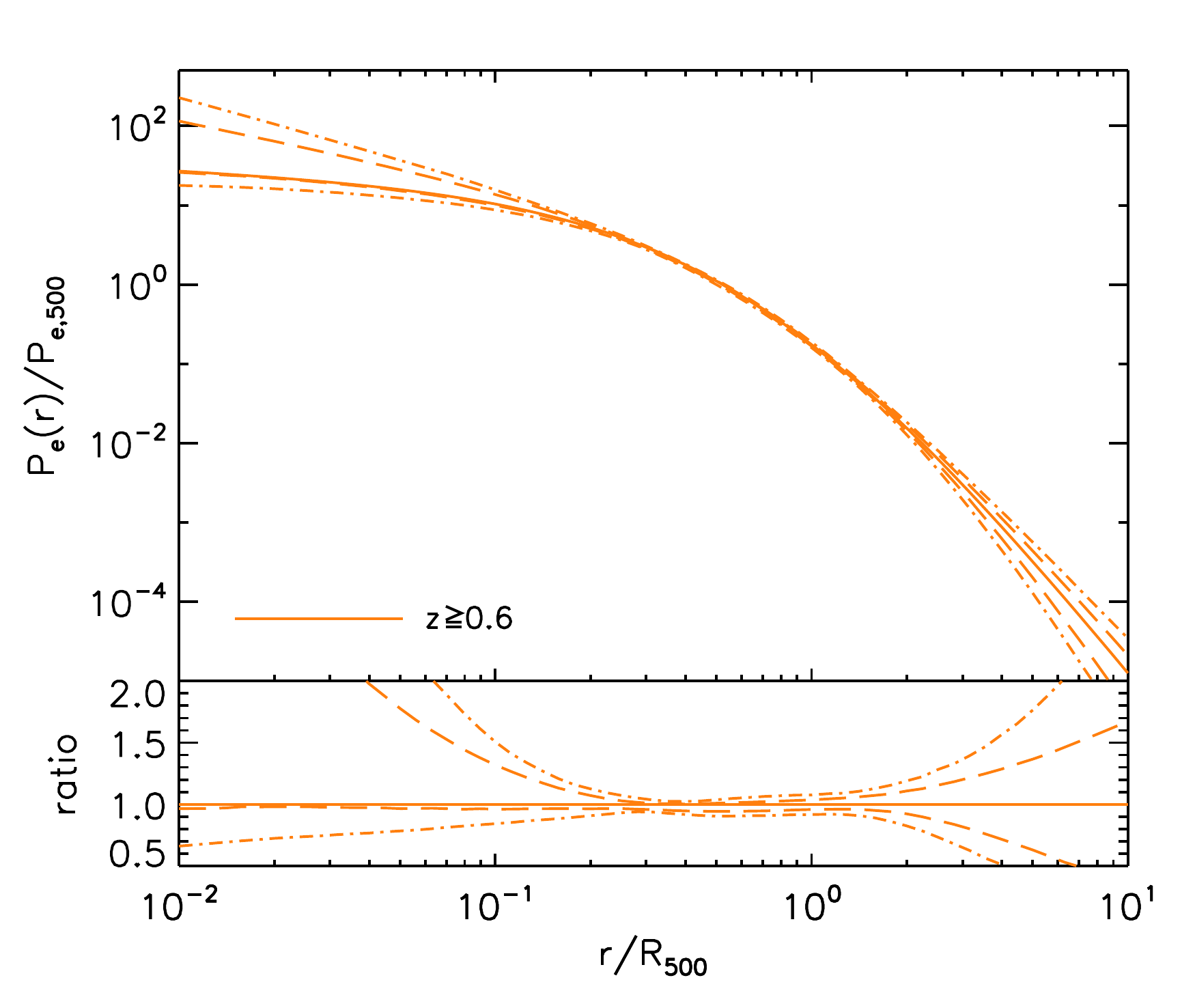}
\includegraphics[width=0.33\hsize]{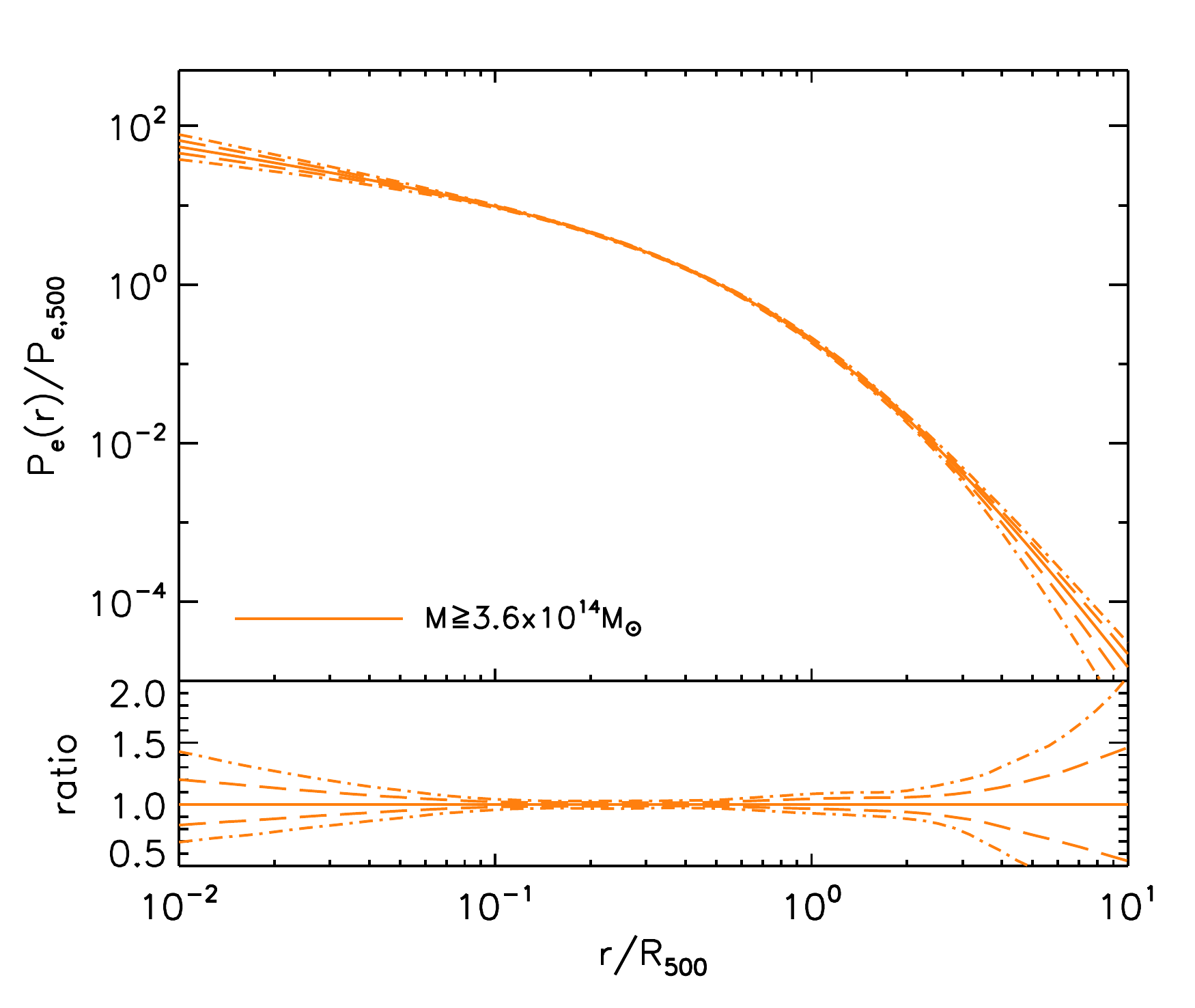}
\caption{\footnotesize Best fitting pressure profiles in real space (solid lines) and associated 68\% (dashed lines) and 95\% (dashed dotted lines) C.L. envelopes. {\it From left to right and top to bottom:}  full sample scaled in $\thetaf$, low redshift subsample, low mass subsample, full sample scaled in $\theta_{\rm 200m}$, high redshift subsample, high mass subsample.}
\label{fig:bestfits_envelopes}
\end{figure*}

\begin{figure*}[!h]
\centering
\includegraphics[width=0.45\hsize]{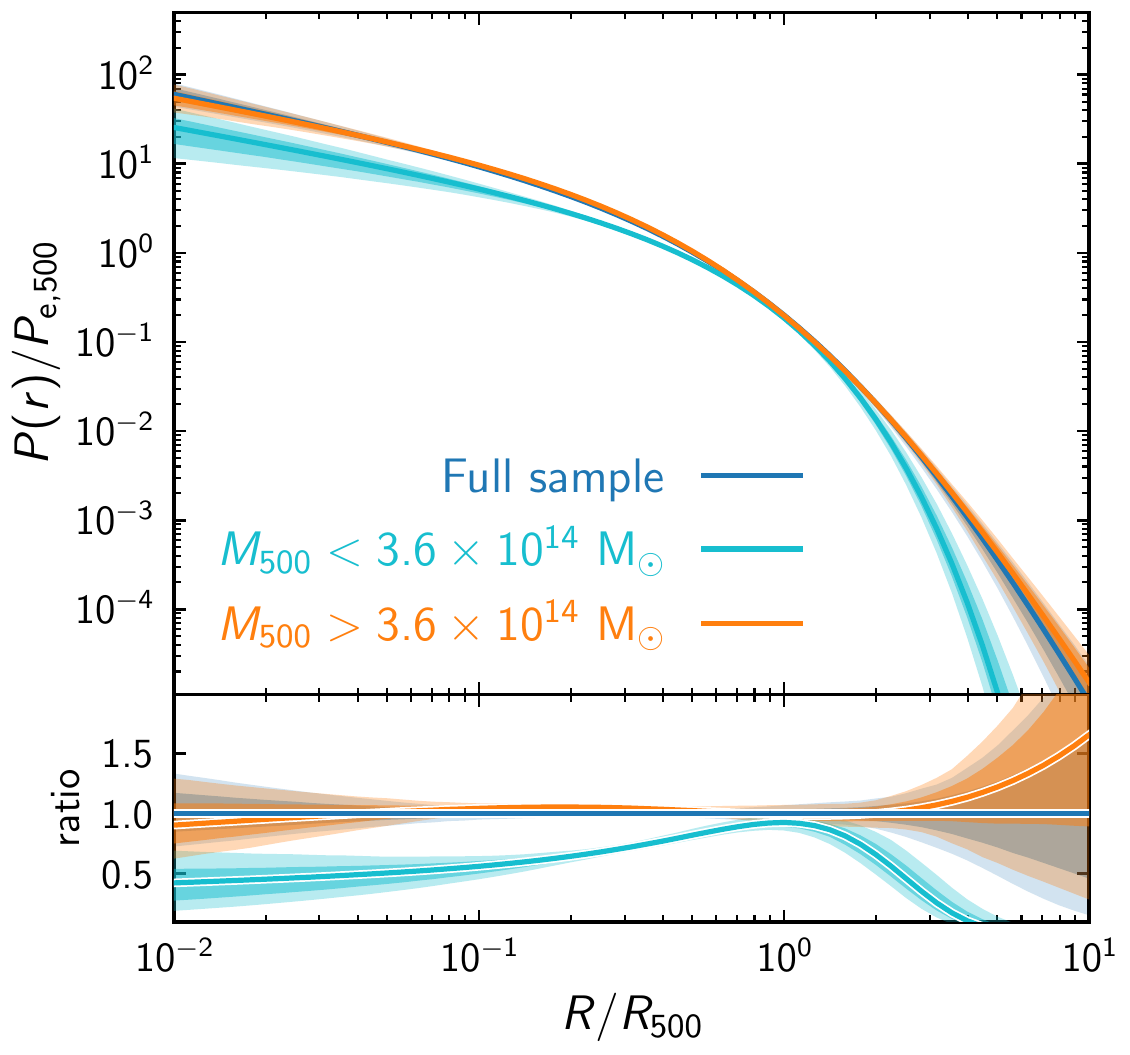}
\hspace{1cm}
\includegraphics[width=0.45\hsize]{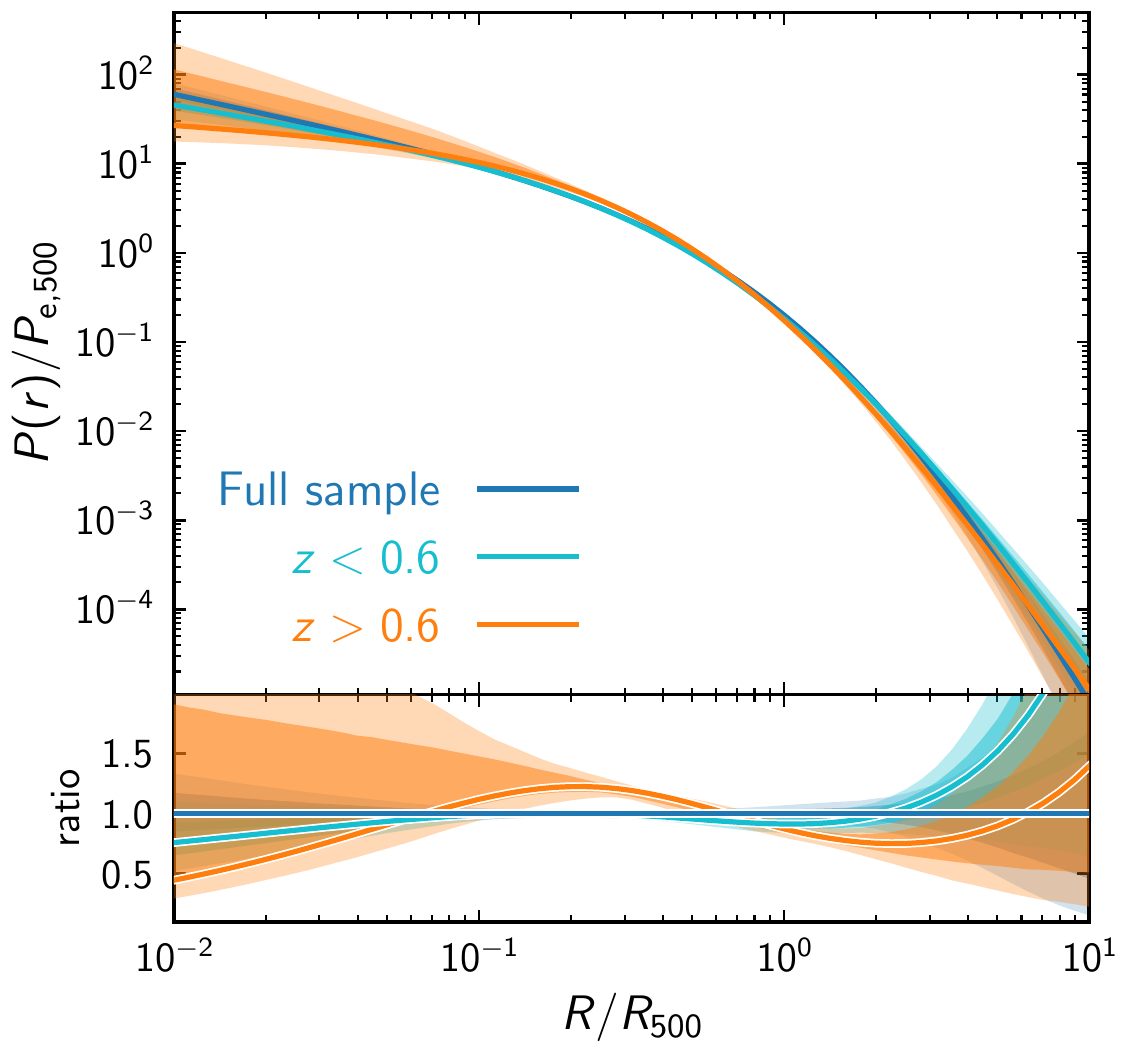}
\caption{\footnotesize Best fitting pressure profiles in real space (solid lines) and associated 68\% (dark) and 95\% (light) C.L. envelopes from the MCMC chains. {\it Left:}  Full sample (dark blue) compared to the low (cyan) and high (orange) mass subsamples. The bottom panel shows the ratio with respect to the full sample. {\it Right:}  Same but for the low (cyan) and high (orange) redshift subsamples.}
\label{fig:errsubsamp}
\end{figure*}

For each sample and subsample, we provide tabulated best-fitting profiles and associated 68\% and 95\% C.L. envelopes (upper and lower bounds). The profiles are provided in 3D real space for the pressure (first six columns of each table), in 2D real space for the Compton parameter (following six columns) and in the 2D harmonic space for the harmonic transform of the Compton parameter (final six columns). The best-fitting pressure profile for each sample and subsample, extracted from the first six columns of each table, is shown in Fig.~\ref{fig:bestfits_envelopes}. The different error envelopes are compared in Fig.~\ref{fig:errsubsamp}.

The complete set of profiles for the full sample are provided in Table~\ref{tab:prof_full}, for the low redshift sample in Table~\ref{tab:prof_lowz}, for the high redshift sample in Table~\ref{tab:prof_highz}, for the low mass sample in Table~\ref{tab:prof_lowm}, for the high mass sample in Table~\ref{tab:prof_highm}, and for the full sample scaled in $\theta_{\rm 200m}$ in Table~\ref{tab:prof_t200m}.

\FloatBarrier

\begin{sidewaystable*}
\caption{\footnotesize Best fitting 3D pressure profile, 2D Compton parameter profile and 2D harmonic transform Compton parameter profile for the full sample, and associated 68\% and 95 \% C.L. envelopes (upper and lower bounds).}
\label{tab:prof_full}
\centering
\resizebox{\textwidth}{!}{%
\begin{tabular}{c c c c c c | c c c c c c | c c c c c c}
\toprule
\toprule
$r/\Rfive$ & $(P_e(r)/P_{e,500})_{\rm best}$  & $(P_e(r)/P_{e,500})_{\rm inf \, 68\%}$  & $(P_e(r)/P_{e,500})_{\rm sup \, 68\%}$ & $(P_e(r)/P_{e,500})_{\rm inf \, 95\%}$  & $(P_e(r)/P_{e,500})_{\rm sup \, 95\%}$  & $\theta/\thetaf$ & $(y(\theta)/y_{500})_{\rm best}$  & $(y(\theta)/y_{500})_{\rm inf \, 68\%}$  & $(y(\theta)/y_{500})_{\rm sup \, 68\%}$  & $(y(\theta)/y_{500})_{\rm inf \, 95\%}$ & $(y(\theta)/y_{500})_{\rm sup \, 95\%}$  & $l\thetaf$  & $(y_l/y_{500}/\thetaf^2)_{\rm best}$  & $(y_l/y_{500}/\thetaf^2)_{\rm inf \, 68\%}$  & $(y_l/y_{500}/\thetaf^2)_{\rm sup \, 68\%}$  & $(y_l/y_{500}/\thetaf^2)_{\rm inf \, 95\%}$  & $(y_l/y_{500}/\thetaf^2)_{\rm sup \, 95\%}$  \\
 -- & -- & -- & -- & -- & --  & -- & -- & --  & -- & -- & -- & arcmin & ${\rm arcmin}^{-2}$ &  ${\rm arcmin}^{-2}$  &  ${\rm arcmin}^{-2}$  &  ${\rm arcmin}^{-2}$  &  ${\rm arcmin}^{-2}$  \\                     
\midrule
1.000e-02 & 6.033e+01 & 5.126e+01 & 7.080e+01 & 4.386e+01 & 8.046e+01 & 1.000e-02 & 3.862e+00 & 3.664e+00 & 4.075e+00 & 3.506e+00 & 4.279e+00 & 1.000e+02 & 2.383e-07 & 2.263e-07 & 2.505e-07 & 2.141e-07 & 2.639e-07 \\
1.122e-02 & 5.544e+01 & 4.752e+01 & 6.451e+01 & 4.101e+01 & 7.290e+01 & 1.122e-02 & 3.805e+00 & 3.617e+00 & 4.002e+00 & 3.468e+00 & 4.196e+00 & 1.122e+02 & 2.382e-07 & 2.262e-07 & 2.504e-07 & 2.140e-07 & 2.637e-07 \\
1.259e-02 & 5.093e+01 & 4.400e+01 & 5.875e+01 & 3.834e+01 & 6.607e+01 & 1.259e-02 & 3.745e+00 & 3.571e+00 & 3.927e+00 & 3.428e+00 & 4.111e+00 & 1.259e+02 & 2.381e-07 & 2.262e-07 & 2.503e-07 & 2.140e-07 & 2.636e-07 \\
1.413e-02 & 4.676e+01 & 4.075e+01 & 5.348e+01 & 3.582e+01 & 5.979e+01 & 1.413e-02 & 3.684e+00 & 3.520e+00 & 3.853e+00 & 3.385e+00 & 4.025e+00 & 1.413e+02 & 2.381e-07 & 2.262e-07 & 2.502e-07 & 2.140e-07 & 2.634e-07 \\
\bottomrule
\end{tabular}}
\end{sidewaystable*}

\begin{sidewaystable*}
\caption{\footnotesize Best fitting 3D pressure profile, 2D Compton parameter profile and 2D harmonic transform Compton parameter profile for the low redshift subsample, and associated 68\% and 95 \% C.L. envelopes (upper and lower bounds).}
\label{tab:prof_lowz}
\centering
\resizebox{\textwidth}{!}{%
\begin{tabular}{c c c c c c | c c c c c c | c c c c c c}
\toprule
\toprule
$r/\Rfive$ & $(P_e(r)/P_{e,500})_{\rm best}$  & $(P_e(r)/P_{e,500})_{\rm inf \, 68\%}$  & $(P_e(r)/P_{e,500})_{\rm sup \, 68\%}$ & $(P_e(r)/P_{e,500})_{\rm inf \, 95\%}$  & $(P_e(r)/P_{e,500})_{\rm sup \, 95\%}$  & $\theta/\thetaf$ & $(y(\theta)/y_{500})_{\rm best}$  & $(y(\theta)/y_{500})_{\rm inf \, 68\%}$  & $(y(\theta)/y_{500})_{\rm sup \, 68\%}$  & $(y(\theta)/y_{500})_{\rm inf \, 95\%}$ & $(y(\theta)/y_{500})_{\rm sup \, 95\%}$  & $l\thetaf$  & $(y_l/y_{500}/\thetaf^2)_{\rm best}$  & $(y_l/y_{500}/\thetaf^2)_{\rm inf \, 68\%}$  & $(y_l/y_{500}/\thetaf^2)_{\rm sup \, 68\%}$  & $(y_l/y_{500}/\thetaf^2)_{\rm inf \, 95\%}$  & $(y_l/y_{500}/\thetaf^2)_{\rm sup \, 95\%}$  \\
 -- & -- & -- & -- & -- & --  & -- & -- & --  & -- & -- & -- & arcmin & ${\rm arcmin}^{-2}$ &  ${\rm arcmin}^{-2}$  &  ${\rm arcmin}^{-2}$  &  ${\rm arcmin}^{-2}$  &  ${\rm arcmin}^{-2}$  \\                     
\midrule
1.000e-02 & 4.581e+01 & 3.950e+01 & 5.577e+01 & 3.161e+01 & 6.638e+01 & 1.000e-02 & 3.548e+00 & 3.395e+00 & 3.771e+00 & 3.210e+00 & 3.976e+00 & 1.000e+02 & 2.478e-07 & 2.311e-07 & 2.618e-07 & 2.146e-07 & 2.804e-07 \\
1.122e-02 & 4.281e+01 & 3.714e+01 & 5.158e+01 & 3.017e+01 & 6.063e+01 & 1.122e-02 & 3.510e+00 & 3.364e+00 & 3.720e+00 & 3.186e+00 & 3.909e+00 & 1.122e+02 & 2.476e-07 & 2.310e-07 & 2.614e-07 & 2.145e-07 & 2.799e-07 \\
1.259e-02 & 3.998e+01 & 3.497e+01 & 4.761e+01 & 2.877e+01 & 5.538e+01 & 1.259e-02 & 3.470e+00 & 3.332e+00 & 3.667e+00 & 3.159e+00 & 3.842e+00 & 1.259e+02 & 2.474e-07 & 2.310e-07 & 2.611e-07 & 2.144e-07 & 2.792e-07 \\
1.413e-02 & 3.732e+01 & 3.290e+01 & 4.396e+01 & 2.742e+01 & 5.065e+01 & 1.413e-02 & 3.428e+00 & 3.299e+00 & 3.611e+00 & 3.135e+00 & 3.776e+00 & 1.413e+02 & 2.472e-07 & 2.308e-07 & 2.608e-07 & 2.142e-07 & 2.782e-07 \\
\bottomrule
\end{tabular}}
\end{sidewaystable*}

\begin{sidewaystable*}
\caption{\footnotesize Best fitting 3D pressure profile, 2D Compton parameter profile and 2D harmonic transform Compton parameter profile for the high redshift subsample, and associated 68\% and 95 \% C.L. envelopes (upper and lower bounds).}
\label{tab:prof_highz}
\centering
\resizebox{\textwidth}{!}{%
\begin{tabular}{c c c c c c | c c c c c c | c c c c c c}
\toprule
\toprule
$r/\Rfive$ & $(P_e(r)/P_{e,500})_{\rm best}$  & $(P_e(r)/P_{e,500})_{\rm inf \, 68\%}$  & $(P_e(r)/P_{e,500})_{\rm sup \, 68\%}$ & $(P_e(r)/P_{e,500})_{\rm inf \, 95\%}$  & $(P_e(r)/P_{e,500})_{\rm sup \, 95\%}$  & $\theta/\thetaf$ & $(y(\theta)/y_{500})_{\rm best}$  & $(y(\theta)/y_{500})_{\rm inf \, 68\%}$  & $(y(\theta)/y_{500})_{\rm sup \, 68\%}$  & $(y(\theta)/y_{500})_{\rm inf \, 95\%}$ & $(y(\theta)/y_{500})_{\rm sup \, 95\%}$  & $l\thetaf$  & $(y_l/y_{500}/\thetaf^2)_{\rm best}$  & $(y_l/y_{500}/\thetaf^2)_{\rm inf \, 68\%}$  & $(y_l/y_{500}/\thetaf^2)_{\rm sup \, 68\%}$  & $(y_l/y_{500}/\thetaf^2)_{\rm inf \, 95\%}$  & $(y_l/y_{500}/\thetaf^2)_{\rm sup \, 95\%}$  \\
 -- & -- & -- & -- & -- & --  & -- & -- & --  & -- & -- & -- & arcmin & ${\rm arcmin}^{-2}$ &  ${\rm arcmin}^{-2}$  &  ${\rm arcmin}^{-2}$  &  ${\rm arcmin}^{-2}$  &  ${\rm arcmin}^{-2}$  \\                     
\midrule
1.000e-02 & 2.696e+01 & 2.599e+01 & 1.152e+02 & 1.779e+01 & 2.270e+02 & 1.000e-02 & 3.491e+00 & 3.429e+00 & 5.788e+00 & 2.992e+00 & 8.007e+00 & 1.000e+02 & 2.209e-07 & 2.064e-07 & 2.366e-07 & 1.949e-07 & 2.555e-07 \\
1.122e-02 & 2.620e+01 & 2.528e+01 & 1.044e+02 & 1.757e+01 & 2.002e+02 & 1.122e-02 & 3.478e+00 & 3.413e+00 & 5.661e+00 & 2.985e+00 & 7.725e+00 & 1.122e+02 & 2.208e-07 & 2.063e-07 & 2.365e-07 & 1.948e-07 & 2.551e-07 \\
1.259e-02 & 2.544e+01 & 2.456e+01 & 9.493e+01 & 1.734e+01 & 1.764e+02 & 1.259e-02 & 3.464e+00 & 3.394e+00 & 5.540e+00 & 2.975e+00 & 7.446e+00 & 1.259e+02 & 2.207e-07 & 2.063e-07 & 2.363e-07 & 1.948e-07 & 2.545e-07 \\
1.413e-02 & 2.467e+01 & 2.394e+01 & 8.585e+01 & 1.709e+01 & 1.554e+02 & 1.413e-02 & 3.448e+00 & 3.373e+00 & 5.415e+00 & 2.966e+00 & 7.169e+00 & 1.413e+02 & 2.206e-07 & 2.063e-07 & 2.362e-07 & 1.948e-07 & 2.539e-07 \\
\bottomrule
\end{tabular}}
\end{sidewaystable*}

\begin{sidewaystable*}
\caption{\footnotesize Best fitting 3D pressure profile, 2D Compton parameter profile and 2D harmonic transform Compton parameter profile for the low mass subsample, and associated 68\% and 95 \% C.L. envelopes (upper and lower bounds).}
\label{tab:prof_lowm}
\centering
\resizebox{\textwidth}{!}{%
\begin{tabular}{c c c c c c | c c c c c c | c c c c c c}
\toprule
\toprule
$r/\Rfive$ & $(P_e(r)/P_{e,500})_{\rm best}$  & $(P_e(r)/P_{e,500})_{\rm inf \, 68\%}$  & $(P_e(r)/P_{e,500})_{\rm sup \, 68\%}$ & $(P_e(r)/P_{e,500})_{\rm inf \, 95\%}$  & $(P_e(r)/P_{e,500})_{\rm sup \, 95\%}$  & $\theta/\thetaf$ & $(y(\theta)/y_{500})_{\rm best}$  & $(y(\theta)/y_{500})_{\rm inf \, 68\%}$  & $(y(\theta)/y_{500})_{\rm sup \, 68\%}$  & $(y(\theta)/y_{500})_{\rm inf \, 95\%}$ & $(y(\theta)/y_{500})_{\rm sup \, 95\%}$  & $l\thetaf$  & $(y_l/y_{500}/\thetaf^2)_{\rm best}$  & $(y_l/y_{500}/\thetaf^2)_{\rm inf \, 68\%}$  & $(y_l/y_{500}/\thetaf^2)_{\rm sup \, 68\%}$  & $(y_l/y_{500}/\thetaf^2)_{\rm inf \, 95\%}$  & $(y_l/y_{500}/\thetaf^2)_{\rm sup \, 95\%}$  \\
 -- & -- & -- & -- & -- & --  & -- & -- & --  & -- & -- & -- & arcmin & ${\rm arcmin}^{-2}$ &  ${\rm arcmin}^{-2}$  &  ${\rm arcmin}^{-2}$  &  ${\rm arcmin}^{-2}$  &  ${\rm arcmin}^{-2}$  \\                     
\midrule
1.000e-02 & 2.561e+01 & 1.674e+01 & 3.260e+01 & 1.161e+01 & 4.179e+01 & 1.000e-02 & 2.206e+00 & 1.934e+00 & 2.412e+00 & 1.737e+00 & 2.667e+00 & 1.000e+02 & 1.660e-07 & 1.587e-07 & 1.742e-07 & 1.511e-07 & 1.830e-07 \\
1.122e-02 & 2.381e+01 & 1.580e+01 & 3.011e+01 & 1.113e+01 & 3.810e+01 & 1.122e-02 & 2.184e+00 & 1.919e+00 & 2.384e+00 & 1.730e+00 & 2.626e+00 & 1.122e+02 & 1.660e-07 & 1.587e-07 & 1.742e-07 & 1.511e-07 & 1.829e-07 \\
1.259e-02 & 2.212e+01 & 1.493e+01 & 2.769e+01 & 1.067e+01 & 3.473e+01 & 1.259e-02 & 2.161e+00 & 1.904e+00 & 2.350e+00 & 1.721e+00 & 2.581e+00 & 1.259e+02 & 1.660e-07 & 1.587e-07 & 1.741e-07 & 1.511e-07 & 1.829e-07 \\
1.413e-02 & 2.055e+01 & 1.412e+01 & 2.546e+01 & 1.021e+01 & 3.170e+01 & 1.413e-02 & 2.137e+00 & 1.891e+00 & 2.319e+00 & 1.712e+00 & 2.536e+00 & 1.413e+02 & 1.660e-07 & 1.587e-07 & 1.741e-07 & 1.511e-07 & 1.829e-07 \\
\bottomrule
\end{tabular}}
\end{sidewaystable*}

\begin{sidewaystable*}
\caption{\footnotesize Best fitting 3D pressure profile, 2D Compton parameter profile and 2D harmonic transform Compton parameter profile for the high mass subsample, and associated 68\% and 95 \% C.L. envelopes (upper and lower bounds).}
\label{tab:prof_highm}
\centering
\resizebox{\textwidth}{!}{%
\begin{tabular}{c c c c c c | c c c c c c | c c c c c c}
\toprule
\toprule
$r/\Rfive$ & $(P_e(r)/P_{e,500})_{\rm best}$  & $(P_e(r)/P_{e,500})_{\rm inf \, 68\%}$  & $(P_e(r)/P_{e,500})_{\rm sup \, 68\%}$ & $(P_e(r)/P_{e,500})_{\rm inf \, 95\%}$  & $(P_e(r)/P_{e,500})_{\rm sup \, 95\%}$  & $\theta/\thetaf$ & $(y(\theta)/y_{500})_{\rm best}$  & $(y(\theta)/y_{500})_{\rm inf \, 68\%}$  & $(y(\theta)/y_{500})_{\rm sup \, 68\%}$  & $(y(\theta)/y_{500})_{\rm inf \, 95\%}$ & $(y(\theta)/y_{500})_{\rm sup \, 95\%}$  & $l\thetaf$  & $(y_l/y_{500}/\thetaf^2)_{\rm best}$  & $(y_l/y_{500}/\thetaf^2)_{\rm inf \, 68\%}$  & $(y_l/y_{500}/\thetaf^2)_{\rm sup \, 68\%}$  & $(y_l/y_{500}/\thetaf^2)_{\rm inf \, 95\%}$  & $(y_l/y_{500}/\thetaf^2)_{\rm sup \, 95\%}$  \\
 -- & -- & -- & -- & -- & --  & -- & -- & --  & -- & -- & -- & arcmin & ${\rm arcmin}^{-2}$ &  ${\rm arcmin}^{-2}$  &  ${\rm arcmin}^{-2}$  &  ${\rm arcmin}^{-2}$  &  ${\rm arcmin}^{-2}$  \\                     
\midrule
1.000e-02 & 5.459e+01 & 4.535e+01 & 6.555e+01 & 3.777e+01 & 7.791e+01 & 1.000e-02 & 3.867e+00 & 3.663e+00 & 4.094e+00 & 3.459e+00 & 4.335e+00 & 1.000e+02 & 2.476e-07 & 2.350e-07 & 2.591e-07 & 2.258e-07 & 2.739e-07 \\
1.122e-02 & 5.061e+01 & 4.251e+01 & 6.019e+01 & 3.578e+01 & 7.095e+01 & 1.122e-02 & 3.819e+00 & 3.629e+00 & 4.032e+00 & 3.431e+00 & 4.256e+00 & 1.122e+02 & 2.475e-07 & 2.349e-07 & 2.590e-07 & 2.257e-07 & 2.736e-07 \\
1.259e-02 & 4.690e+01 & 3.977e+01 & 5.527e+01 & 3.388e+01 & 6.438e+01 & 1.259e-02 & 3.768e+00 & 3.591e+00 & 3.969e+00 & 3.403e+00 & 4.176e+00 & 1.259e+02 & 2.474e-07 & 2.348e-07 & 2.589e-07 & 2.256e-07 & 2.734e-07 \\
1.413e-02 & 4.344e+01 & 3.725e+01 & 5.074e+01 & 3.191e+01 & 5.841e+01 & 1.413e-02 & 3.715e+00 & 3.550e+00 & 3.902e+00 & 3.375e+00 & 4.094e+00 & 1.413e+02 & 2.473e-07 & 2.347e-07 & 2.587e-07 & 2.256e-07 & 2.732e-07 \\
\bottomrule
\end{tabular}}
\end{sidewaystable*}

\begin{sidewaystable*}
\caption{\footnotesize Best fitting 3D pressure profile, 2D Compton parameter profile and 2D harmonic transform Compton parameter profile for the full sample scaled in $\theta_{\rm 200m}$, and associated 68\% and 95 \% C.L. envelopes (upper and lower bounds).}
\label{tab:prof_t200m}
\centering
\resizebox{\textwidth}{!}{%
\begin{tabular}{c c c c c c | c c c c c c | c c c c c c}
\toprule
\toprule
$r/R_{\rm 200m}$ & $(P_e(r)/P_{e,\rm 200m})_{\rm best}$  & $(P_e(r)/P_{e,\rm 200m})_{\rm inf \, 68\%}$  & $(P_e(r)/P_{e,\rm 200m})_{\rm sup \, 68\%}$ & $(P_e(r)/P_{e,\rm 200m})_{\rm inf \, 95\%}$  & $(P_e(r)/P_{e,\rm 200m})_{\rm sup \, 95\%}$  & $\theta/\theta_{\rm 200m}$ & $(y(\theta)/y_{\rm 200m})_{\rm best}$  & $(y(\theta)/y_{\rm 200m})_{\rm inf \, 68\%}$  & $(y(\theta)/y_{\rm 200m})_{\rm sup \, 68\%}$  & $(y(\theta)/y_{\rm 200m})_{\rm inf \, 95\%}$ & $(y(\theta)/y_{\rm 200m})_{\rm sup \, 95\%}$  & $l \theta_{\rm 200m}$  & $(y_l/y_{\rm 200m}/\theta_{\rm 200m}^2)_{\rm best}$  & $(y_l/y_{\rm 200m}/\theta_{\rm 200m}^2)_{\rm inf \, 68\%}$  & $(y_l/y_{\rm 200m}/\theta_{\rm 200m}^2)_{\rm sup \, 68\%}$  & $(y_l/y_{\rm 200m}/\theta_{\rm 200m}^2)_{\rm inf \, 95\%}$  & $(y_l/y_{\rm 200m}/\theta_{\rm 200m}^2)_{\rm sup \, 95\%}$  \\
 -- & -- & -- & -- & -- & --  & -- & -- & --  & -- & -- & -- & arcmin & ${\rm arcmin}^{-2}$ &  ${\rm arcmin}^{-2}$  &  ${\rm arcmin}^{-2}$  &  ${\rm arcmin}^{-2}$  &  ${\rm arcmin}^{-2}$  \\                     
\midrule
1.000e-02 & 3.230e+02 & 2.932e+02 & 3.529e+02 & 2.691e+02 & 3.720e+02 & 1.000e-02 & 1.016e+01 & 9.722e+00 & 1.059e+01 & 9.373e+00 & 1.086e+01 & 1.000e+02 & 1.052e-07 & 9.963e-08 & 1.113e-07 & 9.457e-08 & 1.183e-07 \\
1.122e-02 & 2.842e+02 & 2.598e+02 & 3.087e+02 & 2.402e+02 & 3.236e+02 & 1.122e-02 & 9.760e+00 & 9.377e+00 & 1.014e+01 & 9.047e+00 & 1.037e+01 & 1.122e+02 & 1.052e-07 & 9.963e-08 & 1.113e-07 & 9.457e-08 & 1.182e-07 \\
1.259e-02 & 2.499e+02 & 2.299e+02 & 2.696e+02 & 2.146e+02 & 2.816e+02 & 1.259e-02 & 9.362e+00 & 9.023e+00 & 9.689e+00 & 8.723e+00 & 9.919e+00 & 1.259e+02 & 1.051e-07 & 9.962e-08 & 1.113e-07 & 9.456e-08 & 1.182e-07 \\
1.413e-02 & 2.195e+02 & 2.033e+02 & 2.356e+02 & 1.910e+02 & 2.452e+02 & 1.413e-02 & 8.968e+00 & 8.674e+00 & 9.257e+00 & 8.407e+00 & 9.461e+00 & 1.413e+02 & 1.051e-07 & 9.962e-08 & 1.113e-07 & 9.456e-08 & 1.182e-07 \\
\bottomrule
\end{tabular}}
\end{sidewaystable*}

\end{appendix}

\end{document}